\documentclass[aps,amssymb,amsmath, onecolumn, pra, superscriptaddress]{revtex4}
\usepackage{graphicx}
\usepackage{epsfig}
\usepackage{dcolumn}
\usepackage{amsmath}
\usepackage{bm}
\usepackage{bbm}
\usepackage{ulem}
\usepackage[colorlinks, citecolor=red]{hyperref}
\usepackage{mathtools}
\usepackage{color}
\usepackage{xcolor}
\usepackage{soul}
\usepackage{centernot}
\usepackage[up]{subfigure}
\usepackage{dsfont}	
\usepackage{relsize}

\usepackage{epstopdf}

\usepackage{mathrsfs}
\newcommand{\be}{\begin{equation}}
\newcommand{\ee}{\end{equation}}
\newcommand{\bc}{\begin{center}}
\newcommand{\ec}{\end{center}}
\newcommand{\bea}{\begin{eqnarray}}
\newcommand{\eea}{\end{eqnarray}}
\newcommand{\ba}{\begin{array}}
\newcommand{\ea}{\end{array}}

\def\bra#1{\mathinner{\langle{#1}|}}
\def\ket#1{\mathinner{|{#1}\rangle}}
\def\braket#1{\mathinner{\langle{#1}\rangle}}

\makeatletter
\newcommand{\ovast}{\bBigg@{3}}
\newcommand{\vast}{\bBigg@{4}}
\newcommand{\Vast}{\bBigg@{5}}
\makeatother
\begin{document}
\title{Simulating Dirac Hamiltonian in Curved Space-time by Split-step Quantum Walk }
\author{Arindam Mallick}
\email{marindam@imsc.res.in}
\author{Sanjoy Mandal}
\author{Anirban Karan}
\author{C. M. Chandrashekar}
\email{chandru@imsc.res.in}
\affiliation{Optics and Quantum Information Group, The Institute of Mathematical
Sciences, C. I. T. Campus, Taramani, Chennai 600113, India}
\affiliation{Homi Bhabha National Institute, Training School Complex, Anushakti Nagar, Mumbai 400094,  India}

\begin{abstract}
Dirac particle represents a fundamental constituent of our nature. Simulation of Dirac particle dynamics by a controllable quantum system using quantum walks will allow us to investigate the non-classical nature of dynamics in its discrete form. In this work, starting from a modified version of one-spatial dimensional general inhomogeneous split-step discrete quantum walk we derive an effective Hamiltonian which mimics
a single massive Dirac particle dynamics in curved $(1+1)$ space-time dimension coupled to $U(1)$ gauge potential---which is a forward step towards the simulation of the unification of electromagnetic and gravitational forces in lower dimension and at the single particle level. Implementation of this simulation
scheme in simple qubit-system has been demonstrated. We show that the same Hamiltonian can represent $(2+1)$ space-time dimensional Dirac particle dynamics when one of the spatial momenta remains fixed. 
We also discuss how we can include $U(N)$ gauge potential in our scheme, in order to capture other fundamental 
force effects on the Dirac particle. The emergence of curvature in the two-particle 
split-step quantum walk has also been investigated while the particles are interacting 
through their entangled coin operations.   
\end{abstract}

\maketitle

\section{Introduction}\label{intro}

Quantum walk, an effective algorithmic tool for simulating quantum physical phenomena where 
classical simulator fails or when the computational task is hard to realize via classical algorithm, has been shown to be very useful 
for realization of universal quantum computation \cite{childs1, lovett, childs}. The similarity between discrete quantum walk (DQW) and the dynamics of Dirac particles 
\cite{strauch, bracken, bisio, chandra, di Molfetta, ariano, nesme, sato, chandra2}, at the continuum limit, elevates the DQW as 
a potential candidate to simulate various phenomena where the Dirac fermions play a crucial role \cite{neutrino, neutrino2,fermcon}.
With advancement in field of quantum simulations where many quantum phenomena are mimicked in table-top experiments, algorithmic schemes 
which can simulate Dirac particle dynamics in quantum field theory has garnered considerable interest in recent days. Simulation of 
Dirac particle dynamics in the presence of the external abelian and nonabelian gauge field by DQW has been recently reported\,\cite{arnault,arnault2}. 
Other recent works\,\cite{molfetta, molfettacurve} investigated the inhomogeneous DQW that mimics the Dirac particle dynamics under the influence of 
external gauge-potential and curved space-time as a background. Two-step stroboscopic DQW with space-time dependent $U(2)$ coin operator was used to produce gravitational 
and gauge potential effect in single Dirac fermion, but their approach was unable to capture mass,
gravity and gauge potential in one Hamiltonian\,\cite{molfetta,molfettacurve}.
A generalized single particle Dirac equation in curved space-time was derived from a special DQW---grouped quantum
walk (GQW)---which needs prior unitary encoding and decoding at last\,\cite{qwcur, qwcur3, qwicur}.
DQW with $SU(2)$ coin operator parameters which are spatially independent but depend randomly on time-steps, 
has also been studied in the context of random artificial gauge fields~\cite{randomqw}. 
The randomized coin parameters which mimic random gravitational and gauge 
field act as transition knobs from non-classical probability distribution to classical
probability distribution. A DQW with a single evolution step which contains four spatial shift
operations---mimics the Dirac evolution under influence of gravitational waves in $(2+1)$ 
dimension---was also recently reported in ref.~\cite{grawave}.  In ref.~\cite{mallcm},
it is shown that the SS-DQW, where the coin parameters are space and time-step independent, 
can capture properties of the discretized Dirac particle dynamics in flat $(1+1)$ dimension,
while conventional DQW is unable to capture all properties of it.
This motivates us to generalize the SS-DQW operation and study the consequences of it.
  
In this paper starting from a slightly modified version of the single-step split-step DQW (SS-DQW) 
\cite{kitagawa} whose coin operators are time and position-step dependent (inhomogeneous both in time and position space), we derive a SS-DQW version of the $(1+1)$ dimensional massive Dirac particle Hamiltonian under the influence of the $U(1)$ gauge potential in curved space-time. This scheme is realizable in various physical table-top system as the SS-DQW has been proposed and successfully implemented in various systems like cold atoms \cite{cstom}, superconducting qubits \cite{supercon, balu}, photonic systems \cite{photonic, goyal}. Our scheme can also describe the $(2+1)$ dimensional Dirac Hamiltonian in curved space-time when one component of momentum of the particle remains fixed. We provide realization of our simulation 
scheme using qubit systems. This scheme doesn't require any prior encoding or decoding, 
nor it demands extra conditions on the coin parameters in order to satisfy the boundedness 
(well-defined eigenvalues) of the generator which is the effective Hamiltonian in our case, 
i.e.~the unitary operator should start evolution from identity while the parameter 
of the corresponding lie group evolves from zero to a nonzero real value. Our coin operations are general $U(2)$ group elements in coin-space.
After considering all the terms up to first order in time-step size $\tau$ and position-step size $a$ we have derived the effective Hamiltonian. 
While considering higher dimensional quantum system, i.e.,~qudit instead of qubit system, our scheme will capture more general background $U(N)$ gauge potentials as done in ref.~\cite{arnault}.
The modification of the evolution operator from the conventional SS-DQW evolution operation is not contradictory with the result obtained in ref.~\cite{mallcm}. 
In the refs.~\cite{minar, boada} cold-atom implementation of Dirac particle dynamics in curved space-time has been discussed, but their approaches started by
discretizing Hamiltonian or Lagrangian, while in our case we started from unitary evolution operator discretized in space-time.

We extended our study for the two particle case following the same procedure that we 
have taken for the single particle case.  Extension of single-particle DQW with entangled coin operation has been previously studied~\cite{andraca,liu, liu2}.
Two-particle quantum walk under position dependent or independent coin operations which are separable in their coin degrees of freedom, has been investigated~\cite{busch, gabris, omar, berry, carson, wang}. 
Here we study a two-particle coined SS-DQW
whose coin operations are both position and time dependent and entangled in coin degrees of freedom of the individual particle---the interaction comes solely from the coin operations, while no interaction among the particles is present via their spatial shift operations.
We choose a particular kind of entangled coin operators to demonstrate how the curvature and entanglement in coins enter into the two-particle Hamiltonian.
The existence of discrete space-time steps may help us to study the Planck scale 
physics \cite{jack, pikovski}. But in that case simulation of SS-DQW by two-period conventional
DQW \cite{goyal, pradeep} is not feasible, because of the existence of the fundamental (strictly constant) length scale. 

 This paper is organized as follows. In section \ref{standardham}, we describe how an effective Schrodinger like equation can be derived from the standard Dirac equation in curved space-time.
 In the next section\,\ref{ssdqw} we will describe the conventional form of SS-DQW whose 
 coin operation depends on both space and time steps. In section \ref{modham}, we describe our
 modification to the SS-DQW and derive the effective Hamiltonian. Section \ref{schoice} describes a special choice of coin operations that will produce a Hamiltonian in $(1+1)$ dimensional curved space-time with special metric and gauge potential, we have also describe how we can capture 
 $(2+1)$ dimensional Dirac particle dynamics by looking at the $(1+1)$ dimensional version of the derived Hamiltonian. In section \ref{qubitsim} we demonstrate  
 the implementation scheme in qubit system. In section \ref{uN} we discuss one possible way to include $U(N)$ gauge potential effect in our scheme. 
 In section \ref{twopar} we extend our single-particle $(1+1)$ dimensional SS-DQW scheme to a two-particle case. We concluded with remarks in section \ref{conclu}.


\section{Effective Hamiltonian corresponding to the standard Dirac Equation in curved space-time} \label{standardham}

The general curved space-time Dirac equation \cite{olddir,koke} is written as

\begin{align}
 \left( i \hbar e^{\mu}_{(a)}\gamma^{(a)}\nabla_{\mu} -  m c^2 \right)\psi=0,
 \label{Dirac Curved Eq}
\end{align}
where the covariant derivative $\nabla_{\mu}$=$\partial_{\mu}+\Gamma_{\mu}$;
while in presence of $U(1)$ gauge potential, $\nabla_{\mu}$=$\partial_{\mu} + \Gamma_{\mu}-iA_{\mu}$,
$\gamma^{(a)}$ are local $\gamma$ matrices and satisfy the conditions: $\left\{\gamma^{(b)},\gamma^{(d)}\right\}=2\eta^{(b)(d)} \times $ Identity matrix.
In general, for $(1+1)$ and $(2+1)$ dimensions the 
$\gamma$ matrices can be expressed in terms of the conventional Pauli matrices---$\{\sigma_1, \sigma_2, \sigma_3\}$ like:
$\gamma^{(0)}=\vec{n}.\vec{\sigma},\,\gamma^{(1)}=i\vec{n}_{\perp}.\vec{\sigma}$ are chosen in such a way that 
the sign-convention of the flat space-time metric: $\eta^{(0)(0)} = 1, \eta^{(0)(1)} = \eta^{(1)(0)} = 0, \eta^{(1)(1)} = - 1$ will be obeyed,
where $\vec{n}$ and $\vec{n}_{\perp}$ are mutually orthonormal. The identity matrix in this case will be expressed as $\sigma_0$. The torsion-free and metric compatible connection is defined as
\begin{align}
\Gamma_{\mu}=-\frac{i}{4}S_{(c)(d)}e^{(c)\nu}\left(\frac{\partial e^{(d)}_{\nu}}{\partial x^{\mu}}-\Gamma^{\lambda}_{\mu\nu}e^{(d)}_{\lambda}\right),~\text{where}~
\Gamma^{\sigma}_{\lambda\mu}=\frac{1}{2}g^{\nu\sigma}\left(\partial_{\lambda}g_{\mu\nu}+\partial_{\mu}g_{\lambda\nu}-\partial_{\nu}g_{\mu\lambda}\right), 
\end{align}
and $S_{(c)(d)}$ are the flat spinor matrices defined as $S_{(c)(d)} = \frac{i}{2}[\gamma_{(c)},\gamma_{(d)}]$. $A_{\mu}$ are the external $U(1)$ gauge potentials which themselves
carry the coupling strength. $e^{\mu}_{(b)}$ are the vielbeins, which relate the local and global co-ordinates. The metric is 
defined as $g^{\mu \nu} = e^\mu_{(b)} e^{\nu}_{(d)} \eta^{(b)(d)}$.

Now after using some relations of local $\gamma^{(a)}$ matrices (see Appendix \ref{schroder} for detailed derivation),
it is possible to write the above eq.~(\ref{Dirac Curved Eq}) as follows,
\begin{align}
 \frac{i \hbar}{2}\gamma^{(a)} \bigg[ \bigg\{e^{\mu}_{(a)},\bigg(\frac{\partial}{\partial x^{\mu}} - i A_{\mu} \bigg) \bigg\} 
 + e^{\rho}_{(a)}\Gamma^{\mu}_{\mu\rho} \bigg]\psi
 +\frac{i \hbar}{2}\gamma^{(a)}\gamma_{5}\mathcal{B}_{(a)}\psi = m c^2 \psi,
 \label{dirac equation}
\end{align}
where $$\mathcal{B}_{(a)}=\frac{1}{2}\epsilon_{(a)(b)(c)(d)}e^{(b)\mu}e^{(c)\nu}\frac{\partial e^{(d)}_{\nu}}{\partial x^{\mu}}.$$ 
For $(1+1)$ and $(1+2)$ dimensions $\epsilon_{(a)(b)(c)(d)}$ is always zero, so $\mathcal{B}_{(a)}=0$. 
Then it is possible to write the eq.~(\ref{dirac equation}) in standard Schrodinger equation form as,
$i\hbar\frac{\partial\chi}{\partial t}$ = $H\chi$, applying a transformation to eq.~(\ref{dirac equation}),
$\chi = (-g)^{\frac{1}{4}}\Big[ e^{0}_{(0)}\Big]^{\frac{1}{2}}\psi$ as done in ref.~\cite{olddir}, 
where $g = \det(g_{\mu \nu})$ is the determinant of the metric,
and $e^{0}_{(1)}$ = $e^{0}_{(2)}  = 0$. The corresponding effective Hamiltonian takes the form,
\begin{align}
H =  - \hbar A_{0} + c ~ \alpha^{(a)}\Bigg[\frac{e^i_{(a)}}{e^0_{(0)}}\Bigg] \hat{p}_i
- \frac{i \hbar c}{2} \alpha^{(a)}  \frac{\partial}{\partial x^{i}} \Bigg[\frac{e^i_{(a)}}{e^0_{(0)}}\Bigg]
- \hbar \alpha^{(a)}\Bigg[\frac{e^i_{(a)}}{e^0_{(0)}}\Bigg] A_{i} + \beta \frac{m c^2 }{e^0_{(0)}},
\end{align}
where $\alpha^{(b)}=\beta \cdot \gamma^{(b)}$, $\beta = \gamma^{(0)}$, $i=1, 2$ and $\hat{p}_i$ is the momentum operator corresponding to the $i$-th positional coordinate .
For detailed derivation of the above Hamiltonian, see the Appendix \ref{schroder}.

 In proper notation $A_0$, $A_i$, and all the vielbeins are functions of two position coordinates $x$,
 $y$ and time $t$ such that, in place of them $\sum_{x,y} A_0(x,y,t) \ket{x,y}\bra{x,y}, ~~\sum_{x,y} A_i(x,y,t) \ket{x,y}\bra{x,y},~~ \sum_{x,y} e^{i}_{(b)}(x,y,t) \ket{x,y}\bra{x,y}$ 
 should be used respectively. Also, as the $\gamma$ matrices representing coin space which is different from the position Hilbert space, should be 
 written as tensor products. So, the proper form of this Hamiltonian is
 \begin{align}
 \label{hamcurve1}
 H =  - \hbar \sigma_0 \otimes \sum_{x,y} A_{0}(x,y,t) \ket{x,y} \bra{x,y} 
 + c\alpha^{(a)} \otimes \sum_{x,y} \frac{ e^{i}_{(a)}(x,y,t) }{ e^{0}_{(0)}(x,y,t) } \ket{x,y}\bra{x,y}
  \hat{p}_i + c^2 \beta  \otimes \sum_{x,y} \frac{m(x,y,t)}{e^{0}_{(0)}(x,y,t) }\ket{x,y}\bra{x,y}  \nonumber\\ 
- \frac{i \hbar c}{2}\alpha^{(a)}  \otimes \sum_{x,y} \frac{\partial}{\partial x^{i}} 
\Bigg[ \frac{e^{i}_{(a)}(x,y,t)}{e^{0}_{(0)}(x,y,t)} \Bigg] \ket{x,y}\bra{x,y}
- \hbar \alpha^{(a)} \otimes \sum_{x,y} \frac{ e^{i}_{(a)}(x,y,t)}{ e^{0}_{(0)}(x,y,t)}  A_{i}(x,y,t) \ket{x,y}\bra{x,y}.
\end{align}
In case of fundamental particles, the mass $m(x, y, t) = m$ will be independent of position and time, but for emergent particles which appear in condensed matter systems,
the mass can in general be a function of position and time. For $(1+1)$ dimension the variable $y$ will not be present in eq.~(\ref{hamcurve1}). 
 
However, for notational convenience we will express the Hamiltonian given in eq.~(\ref{hamcurve1}) as follows,
\begin{align}\label{hamcurve}
H =  - \hbar \sigma_0 \otimes  A_{0} 
+ c\alpha^{(a)} \otimes \Bigg[\frac{e^i_{(a)}}{e^0_{(0)}}\Bigg]
\hat{p}_i   
- \frac{i \hbar c}{2}\alpha^{(a)}  \otimes \frac{\partial}{\partial x^{i}} \Bigg[ \frac{e^{i}_{(a)}}{e^{0}_{(0)}} \Bigg]
- \hbar \alpha^{(a)} \otimes \Bigg[\frac{e^i_{(a)}}{e^0_{(0)}}\Bigg]  A_{i}
+ c^2 \beta  \otimes \frac{m}{e^{0}_{(0)} } .
\end{align}

 \section{General Split-step DQW}\label{ssdqw}

 As the conventional DQW is a discrete quantum version of the classical random walk~\cite{partha, aharo, mayer}, the 
 SS-DQW is a generalized version of the conventional DQW, first introduced in ref.~\cite{kitagawa}. 
 Multiple evolution parameters give more control over the evolution of the walk to engineer the dynamics at our desire. 
 The general single-particle SS-DQW in $(1+1)$ dimensional space-time can be defined as a unitary evolution operator that evolves a state $\ket{\psi(t)}$
 at time $t$ to a state $\ket{\psi(t + \tau)}$ at time $t + \tau$,
 \begin{align}\label{unissqw}
  U(t, \tau) = S_+ \cdot C_2(t, \tau) \cdot S_- \cdot C_1(t, \tau)~~~~
\end{align} acting on the Hilbert space $\mathcal{H}_c \otimes \mathcal{H}_x$ where 
$\mathcal{H}_c = \text{span}\{(1 ~ 0)^T, (0 ~ 1)^T \}$ is the coin Hilbert space 
and $\mathcal{H}_x = \text{span}\{\ket{x} : x \in a \mathbb{Z}~\text{or}~x \in a \mathbb{Z}_{\mathcal{N}}\}$
is the position Hilbert space. The general state $\ket{\psi(t)} \in \mathcal{H}_c \otimes \mathcal{H}_x$ for all discrete time-step $t \in \tau  \times\big(\{0\} \cup \mathbb{N}\big)$.  

Here the unitary coin operation is defined as \begin{align}\label{coinssqw}
C_j(t,\tau) = \sum_{x} e^{i \xi_j(x,t, \tau)}\left(\begin{array}{cc}
F_j(x,t, \tau) & G_j(x,t, \tau) \\\\
-G^*_j(x,t, \tau) &  F^*_j(x,t, \tau) \\
\end{array} \right) \otimes \ket{x}\bra{x}
\end{align} for $j = 1, 2$, subject to the condition $|F_j(x, t, \tau)|^2 + |G_j(x, t, \tau)|^2  = 1$ and 
$\xi_j(x,t, \tau)$ are real for all $x, t, \tau$. The $x, t$ dependence of the functions
$F_j(x,t,\tau), G_j(x,t, \tau), \xi_j(x, t,\tau)$ implies inhomogeneity of the coin operation both in position and time steps. 
Here $F_j(x,t,\tau), G_j(x,t,\tau)$ represent the elements 
of $SU(2)$ group operation and after including $\xi_j(x,t,\tau)$,
we have a general $U(2)$ group operation on coin space \cite{mallcm}. 
    
The coin state dependent position shift operators are defined as 
\begin{align}\label{shiftssqw}
S_+ = \sum_{x} \left(\begin{array}{cc}
1 &  0 \\
0 &  0 \\
\end{array} \right) \otimes \ket{x+a}\bra{x}
+ \left(\begin{array}{cc}
0 &  0 \\
0 &  1 \\
\end{array} \right) \otimes \ket{x}\bra{x}, ~~
S_- = \sum_{x} \left(\begin{array}{cc}
1 &  0 \\
0 &  0 \\
\end{array} \right) \otimes \ket{x}\bra{x}
+ \left(\begin{array}{cc}
0 &  0 \\
0 &  1 \\
\end{array} \right) \otimes \ket{x-a}\bra{x}.                        
\end{align}
These position shift operators act homogeneously on all positions, at all time steps.
The usual implementation method of the unitary operator 
$U$ which implement one complete step of SS-DQW is in the following order---the coin operation $C_1(t,\tau)$ is followed by the
shift operation $S_-$ which is further followed by the coin operation $C_2(t,\tau)$ and then the shift operation $S_+$. Here $S_\pm$ are by definition unitary operators. The coin operations are generalized $U(2)$ operations on the coin space while they keep the position state of the particle intact, but the parameters of this $U(2)$ operation depend on the position of the coin. $S_+$ shifts the particle one-step further in position points along the direction of increasing $x$ if the coin state of the particle is in the up-state or $(1 ~ 0)^T$ and does nothing if the coin state of the particle is in the down-state or $(0 ~ 1)^T$.
$S_-$  does nothing if the coin state of the particle is in the up-state or $(1 ~ 0)^T$
and shifts the particle one step further in position points 
along the direction of decreasing $x$ if the coin state of the particle is in the down-state or $(0 ~ 1)^T$.  

\vspace{0.2cm}

Using the expressions given in eqs.~(\ref{coinssqw})-(\ref{shiftssqw}),
the whole evolution operator in eq.\,(\ref{unissqw}) can be written in the form:
 \begin{align}\label{unissqwform}
  U(t, \tau) =  \left( \begin{array}{cc} 1 & 0 \\ 0 & 0 \\ \end{array}\right)
  \otimes U_{00}(t, \tau) 
 + \left( \begin{array}{cc} 0 & 1 \\ 0 & 0 \\ \end{array}\right) 
 \otimes U_{01}(t, \tau) 
 +  \left( \begin{array}{cc} 0 & 0 \\ 1 & 0 \\ \end{array}\right)
 \otimes U_{10}(t, \tau) 
 + \left( \begin{array}{cc} 0 & 0 \\ 0 & 1 \\ \end{array}\right)\
 \otimes U_{11}(t, \tau),
\end{align} where
\begin{align}\label{unissqwform1}
 U_{00}(t, \tau) = \sum_x e^{i [\xi_1(x,t,\tau) + \xi_2(x,t,\tau)]} F_2(x,t,\tau)F_1(x,t,\tau) \ket{x+a}\bra{x} 
- e^{i [\xi_1(x,t,\tau) + \xi_2(x-a,t,\tau)]} G_2(x-a,t,\tau)G_1^*(x,t,\tau) \ket{x}\bra{x}, \nonumber\\
U_{01}(t, \tau) =  \sum_x e^{i [\xi_1(x,t,\tau) + \xi_2(x,t,\tau)]} F_2(x,t,\tau) G_1(x,t,\tau) \ket{x+a}\bra{x}
+ e^{i [\xi_1(x,t,\tau) + \xi_2(x-a,t,\tau)]} G_2(x-a,t,\tau)F_1^*(x,t,\tau) \ket{x}\bra{x},\nonumber\\
U_{10}(t, \tau) =  \sum_x - e^{i [\xi_1(x,t,\tau) + \xi_2(x,t,\tau)]} G_2^*(x,t,\tau) F_1(x,t,\tau) \ket{x}\bra{x} 
- e^{i [\xi_1(x,t,\tau) + \xi_2(x-a,t,\tau)]} F^*_2(x-a,t,\tau) G_1^*(x,t,\tau)\ket{x-a}\bra{x},\nonumber\\
U_{11}(t, \tau) =  \sum_x - e^{i [\xi_1(x,t,\tau) + \xi_2(x,t,\tau)]} G_2^*(x,t,\tau)G_1(x,t,\tau) \ket{x}\bra{x}
+ e^{i [\xi_1(x,t,\tau) + \xi_2(x-a,t,\tau)]} F^*_2(x-a,t,\tau)F^*_1(x,t,\tau)\ket{x-a}\bra{x}.
\end{align}

The effective Hamiltonian $H_{\text{eff}}(t): H^\dagger_{\text{eff}}(t) = H_{\text{eff}}(t) $ is defined by  
\begin{align}
U(t, \tau)  =  \exp \bigg( - i \frac{H_{\text{eff}}(t)\tau}{\hbar} \bigg).
\end{align}

Because of the inhomogeneity of the evolution operator in space-time in eq.~(\ref{unissqw}), it is difficult to diagonalize the whole 
operator and derive the effective Hamiltonian as done in ref.~\cite{mallcm}. Instead of that, we will derive the Hamiltonian by Taylor 
series expansion with respect to the variables $\tau$, $a$ assuming that $\tau$, $a = c \tau$ have the same order of magnitude and $\lim\limits_{\tau \to 0} a = 0$ (as we are taking $c$ as a finite constant). 

\begin{align}\label{unitexpan}
 U(t, \tau) =  \lim_{\tau \to 0} U(t, \tau) + \tau \lim_{\tau \to 0} \frac{U(t, \tau) - U(t, 0) }{\tau} + \ldots 
 = \sigma_0 \otimes \sum_{x \in a \mathbb{Z}} \ket{x}\bra{x}  - i \frac{\tau}{\hbar}H_{\text{eff}}(t) 
 + \ldots ~~~~~~~~~~~~~~ 
\end{align} and hence the relation between system state
at time steps $t$ and $t + \tau$ can be written as follows
\begin{align}
 \ket{\psi(t + \tau)} = U(t, \tau)\ket{\psi(t)} 
 = \ket{\psi(t)} -  \frac{i\tau}{\hbar}H_{\text{eff}}(t) \ket{\psi(t)} + \mathcal{O}(\tau^2).
\end{align}
So, the effective Hamiltonian can be derived from the expansion upto the first order in $\tau$.  
But the zeroth order terms of the unitary operator:
\begin{align}\label{zerothoterm}
 U_{00}(t,0) =  \sum_x e^{i [\xi_1(x,t,0) + \xi_2(x,t,0)]} \big\{ F_2(x,t,0)F_1(x,t,0) 
-  G_2(x,t,0)G_1^*(x,t,0) \big\} \ket{x}\bra{x} , \nonumber\\ 
  U_{01}(t,0) = \sum_x e^{i [\xi_1(x,t,0) + \xi_2(x,t,0)]} \big\{ F_2(x,t,0) G_1(x,t,0) 
+  G_2(x,t,0)F_1^*(x,t,0) \big\} \ket{x}\bra{x},\nonumber\\ 
 U_{10}(t,0) =  - \sum_x e^{i [\xi_1(x,t,0) + \xi_2(x,t,0)]} \big\{ G_2^*(x,t,0) F_1(x,t,0)
 + F^*_2(x,t,0) G_1^*(x,t,0) \big\} \ket{x}\bra{x} ,\nonumber\\ 
   U_{11}(t,0) =  - \sum_x e^{i [\xi_1(x,t,0) + \xi_2(x,t,0)]} \big\{ G_2^*(x,t,0)G_1(x,t,0) 
- F^*_2(x,t,0)F^*_1(x,t,0) \big\} \ket{x}\bra{x}
\end{align}  will not simply be equal to the identity operator unless some constraints on the 
functions $\xi_j(x,t,0)$, $F_j(x,t,0)$ and \\ $G_j(x,t,0)$ have been imposed. The zeroth order term should 
be equal to the identity operator both in position and coin space in order to make the Hamiltonian, a bounded operator at $a \to 0$, $\tau \to 0$ for the validity of Taylor series expansion in eq.~(\ref{unitexpan}).

We are assuming that the $\xi_j(x,t,\tau),~ F_j(x,t,\tau), ~G_j(x,t,\tau)$ are 
analytic functions of $\tau$, so that we can do Taylor series expansion of them as well as the  
overall SS-DQW evolution operator. 
\begin{align}
 F_j(x,t, \tau) =  F_j(x,t,0) + \tau f_j(x,t) + \mathcal{O}(\tau^2),
 G_j(x,t, \tau) =   G_j(x,t,0) + \tau g_j(x,t) + \mathcal{O}(\tau^2),\nonumber\\
 \xi_j(x,t,\tau) =  \xi_j(x,t,0) + \tau \lambda_j(x,t) + \mathcal{O}(\tau^2).
 \end{align}
 Imposing the condition that $|F_j(x,t,\tau)|^2 + |G_j(x,t, \tau)|^2 = 1$ for all $x, t, \tau$; 
 as the coefficient of $\tau^n$ should be zero separately for each $n$, where $n \in \mathbb{N}$;
 we get \begin{align}
 \Re\big[F_j(x,t,0)f^*_j(x,t) + G_j(x,t,0)g^*_j(x,t)\big] = 0.
 \end{align}
 From the condition 
 \begin{align}
  |F_j(x + a,t,0)|^2 + |G_j(x + a,t, 0)|^2 =  |F_j(x - a,t,0)|^2 + |G_j(x - a,t, 0)|^2 
  =  |F_j(x,t,0)|^2 + |G_j(x,t, 0)|^2 = 1\nonumber\\
 \text{we have a difference equation} \hspace{13cm}\nonumber\\
 F_j(x + a,t,0) F^*_j(x + a,t,0) - F_j(x,t,0) F^*_j(x,t,0) + G_j(x + a,t,0) G^*_j(x + a,t,0) - G_j(x,t,0) G^*_j(x,t,0) = 0 \nonumber\\
 \text{which, after expansion upto the first order in}~a~\text{gives}~~ 
 \Re\big[F_j(x,t,0) \partial_x F^*_j(x,t,0) + G_j(x,t,0) \partial_x G_j^*(x,t,0)  \big] = 0, 
\end{align}where we have defined \begin{align*}
\partial_x F^*_j(x,t,0) \coloneqq \lim_{a \to 0} \frac{1}{a}\big[ F^*_j(x + a,t,0) - F^*_j(x,t,0) \big]
= \lim_{a \to 0} \frac{1}{a}\big[ F^*_j(x,t,0) - F^*_j(x-a,t,0) \big].~~~~~~~~~ \\
\text{The similar definition will be used for the functions}~F_j(x,t,0),~ G_j(x,t,0),~ G^*_j(x,t,0), ~\xi_j(x,t,0)~\forall~j = 1,2.\end{align*} 
We further assume that all the higher order difference equation in $a$ of all these functions are well defined, so that, 
at the limit $a \to 0$, we can neglect higher order terms compared to the first order term in $a$ in the Taylor series expansion 
with respect to the variable $a$.


\section{Modified Evolution Operator and Effective Hamiltonian}\label{modham}

Our conventional single-particle SS-DQW evolution operator does not directly satisfy the condition 
$\lim_{\tau \to 0} U(t, \tau) = \sigma_0 \otimes \sum_x \ket{x}\bra{x}$ in  eq.~(\ref{zerothoterm}),
unless we impose some extra conditions on the functions: $F_j(x,t,0),~ G_j(x,t,0), ~\xi_j(x,t,0)$.
But there may be a possibility that finding the valid conditions is not simple or may be the 
limit itself does not exist. Moreover, the condition will reduce the number of independent parameters. 
So, instead of using $U(t, \tau)$ as our evolution operator we will use 
$U^\dagger(t,0) \cdot U(t, \tau)$. Let's denote $\mathscr{U}(t, \tau) \coloneqq U^\dagger(t,0) \cdot U(t, \tau)$ where
$U^\dagger(t,0) =  C_1^\dagger(t,0) \cdot C_2^\dagger(t,0)$. We can see 
$\mathscr{U}(t, 0) = \sigma_0 \otimes \sum_x \ket{x}\bra{x}$.
Note that this modification won't affect the relation of Dirac cellular automaton (DCA) and SS-DQW established in ref.~\cite{mallcm}, 
because in that case $U^\dagger(t,0) = \sigma_0 \otimes \sum_x \ket{x}\bra{x}$.
We can write the matrix form of the modified evolution operator $\mathscr{U}(t, \tau)$ in the coin basis as:
 \begin{align}\label{unisnext}
  \mathscr{U}(t, \tau) =  \left( \begin{array}{cc} 1 & 0 \\ 0 & 0 \\ \end{array}\right)
  \otimes \mathscr{U}_{00}(t, \tau) 
 + \left( \begin{array}{cc} 0 & 1 \\ 0 & 0 \\ \end{array}\right) 
 \otimes \mathscr{U}_{01}(t, \tau)
 +  \left( \begin{array}{cc} 0 & 0 \\ 1 & 0 \\ \end{array}\right)
 \otimes \mathscr{U}_{10}(t, \tau) 
 + \left( \begin{array}{cc} 0 & 0 \\ 0 & 1 \\ \end{array}\right)\
 \otimes \mathscr{U}_{11}(t, \tau),
\end{align} where
\begin{align}\label{unisnext1}
\mathscr{U}_{00}(t, \tau) = U^\dagger_{00}(t,0) U_{00}(t,\tau) + U^\dagger_{10}(t,0) U_{10}(t,\tau),~~ 
\mathscr{U}_{01}(t, \tau) = U^\dagger_{00}(t,0) U_{01}(t,\tau) + U^\dagger_{10}(t,0) U_{11}(t,\tau) ,\nonumber\\
\mathscr{U}_{10}(t, \tau) =  U^\dagger_{01}(t,0) U_{00}(t,\tau) + U^\dagger_{11}(t,0) U_{10}(t,\tau),~~
\mathscr{U}_{11}(t, \tau) =  U^\dagger_{01}(t,0) U_{01}(t,\tau) + U^\dagger_{11}(t,0) U_{11}(t,\tau).
\end{align} The detailed forms of these operators are calculated in Appendix \ref{evoluform}.
 Expanding these operators upto first order in $\tau$ and $a$, we can calculate the effective Hamiltonian using the similar form of the eq.~(\ref{unitexpan})
 defined for the conventional SS-DQW evolution operator, i.e., now we use the definition:
 
 \begin{align}\label{effham}
  H_\text{eff}(t) \coloneqq i \hbar \lim_{\tau \to 0} \frac{1}{\tau}\big[ \mathscr{U}(t, \tau)
  - \mathscr{U}(t,0) \big].
 \end{align}

 For the detailed derivation of this Hamiltonian see Appendix \ref{dervham}.
 The derived effective Hamiltonian is of the form:
\begin{align}\label{derivedHam}
 H_{\text{eff}}(t) = \sum_{r=0}^3  \sigma_r \otimes \sum_x \Xi_r(x,t) \ket{x}\bra{x} + 
   c~\sum_{r=1}^3  \sigma_r \otimes \sum_x \Theta_r(x,t) \ket{x}\bra{x}~ \hat{p} .~~~~~~~~~~~
\end{align}  
The terms $\Xi_r(x,t)$, $\Theta_r(x,t)$ in Hamiltonian operator are explicit functions of 
$F_j(x,t,0),~ G_j(x,t,0), ~\xi_j(x,t,0)$, $f_j(x,t), ~g_j(x,t)$ and $\lambda_j(x,t)$ for $j = 1, 2$. 
The explicit expressions of these terms are given in Appendix \ref{dervham}.

\vspace{0.5cm}

Note that, in the standard Hamiltonian in eq.~(\ref{hamcurve}) in $(1+1)$ dimension, the total possible number of independent coefficients of the momentum operator $\hat{p}_1$ is 
two and they are $ c \alpha^{(0)} \otimes \frac{ e^1_{(0)} }{ e^0_{(0)} }, c \alpha^{(1)} \otimes \frac{ e^1_{(1)} }{ e^0_{(0)} }$. In the expression in (\ref{derivedHam}) 
three independent coefficients are possible, but they do not contain $\alpha^{(0)} = \sigma_0$ term. So, in order to match with the existing theory we need to do a careful choice of the 
coin parameters.  

\section{When The Coin Operation of modified SS-DQW is restricted  to span$\mathbf{\{\sigma_0, \sigma_1\}}$  only}\label{schoice}

The Hamiltonian, derived in the preceding section corresponds to the general $U(2)$ coin operation. 
Now we will consider a special case where the coin operations are only rotations about the spin-$x$-axis, such that
 $F_j(x,t,\tau) = \cos \theta_j(x, t , \tau)$,
$G_j(x,t, \tau ) = - i \sin \theta_j(x, t, \tau )$ and the overall phase $\xi_j(x,t,\tau)$ will be incorporated. 
 Further if $\theta_j(x,t, \tau) = \theta_j(x,t,0) + \tau \vartheta_j(x,t)$,
we have $f_j(x, t) = - \vartheta_j(x,t) \sin [\theta_j(x,t,0)]$
and $g_j(x, t) = - i \vartheta_j(x,t) \cos[\theta_j(x,t,0)]$. 
  \vspace{0.5cm}

In this case: \begin{align}\Theta_1  = 0, ~~ \Theta_2 = \cos[\theta_2(x,t,0)] \sin[2 \theta_1(x,t,0) + \theta_2(x,t,0)], 
 \Theta_3 = \frac{1}{2} \cos[2 \theta_1(x,t,0)] 
               + \frac{1}{2} \cos[2 \theta_1(x,t,0) + 2 \theta_2(x,t,0)],\end{align}
\begin{align}
\Xi_0 =  - \hbar [\lambda_1(x,t) + \lambda_2(x,t)]  + \frac{\hbar c}{2} \partial_x \xi_2(x,t,0),
~~
               \Xi_1 =  \hbar[\vartheta_1(x,t) + \vartheta_2(x,t)] - \frac{\hbar c}{2} \partial_x \theta_2(x,t,0),
       \end{align}
 \begin{align}               
 \Xi_3  = \frac{i \hbar c}{2}  \sin[2 \theta_1(x,t,0) + 2 \theta_2(x,t,0)]\partial_x \theta_2(x,t,0) 
 + i \hbar c \cos[\theta_2(x,t,0)] \sin[\theta_2(x,t,0) +  2 \theta_1(x,t,0)]\partial_x \theta_1(x,t,0)\nonumber\\
 + \frac{\hbar c}{2} \partial_x \xi_1  \Big[  \cos[2\theta_1(x,t,0)] + \cos[2\theta_1(x,t,0) + 2\theta_2(x,t,0)] \Big]
  + \frac{\hbar c}{2}\partial_x \xi_2 \cos[2\theta_1(x,t,0) + 2\theta_2(x,t,0)],~~~~~~~~ 
  \end{align}
and 
 \begin{align}
 \Xi_2 = - i \hbar c \cos[\theta_2(x,t,0)]  \cos[2 \theta_1(x,t,0) + \theta_2(x,t,0)] \partial_x \theta_1(x,t,0) 
 - \frac{i \hbar c}{2} \cos[2 \theta_1(x,t,0) + 2 \theta_2(x,t,0) ]\partial_x \theta_2(x,t,0) \nonumber\\
 + \hbar c  \partial_x \xi_1   \cos[\theta_2(x,t,0)] \sin[2 \theta_1(x,t,0) + \theta_2(x,t,0)] 
  + \frac{\hbar c}{2} \partial_x \xi_2(x,t,0)  \sin[2\theta_1(x,t,0) + 2\theta_2(x,t,0)].~~~~~~~~
\end{align}

Note that, for this choice $F_j(x,t,\tau) = \cos \theta_j(x, t , \tau)$,
$G_j(x,t, \tau ) = - i \sin \theta_j(x, t, \tau )$, the effective Hamiltonian in eq.~(\ref{derivedHam})
will reduce to the case of the flat space-time: $g^{\mu \nu} = \eta^{(a)(b)}$ when 
$\theta_2(x,t,0) = \theta_1(x,t,0) = 0$ for all $x, t$.

\subsection{Comparison with Curved Space-time Dirac Hamiltonian in ($\mathbf{1+1}$) dimensional}

In strictly $(1+1)$ dimension, the Dirac Hamiltonian in eq.~(\ref{hamcurve}) takes the form
 \begin{align}\label{11dimham}
 H =  - \hbar \sigma_0 \otimes  A_{0} 
 + c\alpha^{(a)} \otimes \Bigg[\frac{e^1_{(a)}}{e^0_{(0)}}\Bigg]
  \hat{p}_1   
- \frac{i \hbar c}{2}\alpha^{(a)}  \otimes \frac{\partial}{\partial x} \Bigg[ \frac{e^{1}_{(a)}}{e^{0}_{(0)}} \Bigg]
- \hbar \alpha^{(a)} \otimes \Bigg[\frac{e^1_{(a)}}{e^0_{(0)}}\Bigg]  A_{1}
+ c^2 \beta  \otimes \frac{m}{e^{0}_{(0)} }.
\end{align}
Here we have $\alpha^{(0)} = \sigma_0$; 
so, to compare this Hamiltonian with our derived Hamiltonian, given in  eq.~(\ref{derivedHam}), one possible choice is $ \theta_2(x,t,0) = - 2 \theta_1(x,t,0)$, 
 $e^1_{(0)} = 0$ and $\gamma^{(0)} = \sigma_1, \gamma^{(1)} = -i \sigma_2$ which implies $\alpha^{(1)}= \sigma_3$.
   
Then, 
   \begin{align}\Theta_1  = 0,~~
               \Theta_3 =  \cos[2 \theta_1(x,t,0)],~~
               \Theta_2 = 0 ,~~
               \Xi_0 =  - \hbar [\lambda_1(x,t) + \lambda_2(x,t)]  + \frac{\hbar c}{2} \partial_x \xi_2(x,t,0),~ \nonumber\\~ 
               \Xi_1 =  \hbar[\vartheta_1(x,t) + \vartheta_2(x,t)] +  \hbar c \partial_x \theta_1(x,t,0),~~ 
               \Xi_2 = - \frac{\hbar c}{2} \sin[2 \theta_1(x,t,0)] \partial_x \xi_2(x,t,0),\nonumber\\
 \Xi_3  =   i \hbar c  \sin[2 \theta_1(x,t,0)]\partial_x \theta_1(x,t,0) + \frac{\hbar c}{2} \cos[2 \theta_1(x,t,0)]
 \big[ 2 \partial_x \xi_1(x,t,0) + \partial_x \xi_2(x,t,0)\big].
\end{align}
   
After omitting all the zero-valued terms, the Hamiltonian in eq.~(\ref{11dimham}) becomes
\begin{align}
 H =  - \hbar \sigma_0 \otimes  A_{0} 
 + c\sigma_3 \otimes \Bigg[\frac{e^1_{(1)}}{e^0_{(0)}}\Bigg]
  \hat{p}_1 - \frac{i \hbar c}{2}\sigma_3  \otimes \frac{\partial}{\partial x} \Bigg[ \frac{e^{1}_{(1)}}{e^{0}_{(0)}} \Bigg]
- \hbar \sigma_3 \otimes \Bigg[\frac{e^1_{(1)}}{e^0_{(0)}}\Bigg]  A_{1} + c^2 \beta  \otimes \frac{m}{e^{0}_{(0)} },
\end{align} where we identify  $ \partial_x \xi_2(x,t,0) = 0$  and \begin{align}
   \Bigg[\frac{e^1_{(1)}}{e^0_{(0)}}\Bigg] = \cos[2 \theta_1(x,t,0)], 
   ~~\frac{m c^2}{e^0_{(0)}} = \hbar[\vartheta_1(x,t) + \vartheta_2(x,t)] + \hbar c \partial_x \theta_1(x,t,0),~ 
   A_{0} = \lambda_1(x,t) + \lambda_2(x,t),~~\nonumber\\ 
  A_{1} \Bigg[\frac{e^1_{(1)}}{e^0_{(0)}}\Bigg] = -  c ~\partial_x \xi_1(x,t,0)  \Rightarrow A_1 = - c \sec[2 \theta_1(x,t,0)] \partial_x \xi_1(x,t,0),
  ~~~~~~~~~~~~\nonumber\\  
    \text{Metric} ~ = \left( \begin{array}{cc}
                              g^{00} & g^{01} \\
                              g^{10} & g^{11} \\
                             \end{array} \right)
   = \left( \begin{array}{cc}
                             \Big[e^0_{(0)}\Big]^2 & 0 \\
                             0 & -  \Big[e^1_{(1)}\Big]^2 \\
                            \end{array}\right) = \Big[e^0_{(0)}\Big]^2\left( \begin{array}{cc}
                             1 & 0 \\
                             0 & -  \cos^2[2 \theta_1(x,t,0)] \\
                            \end{array}\right) .
 \end{align}
 In case we want to study the fundamental particle, the mass $m$ should be taken position-time independent, 
 we can choose $e^0_{(0)}$ = $ m c^2 \big( \hbar[\vartheta_1(x,t) + \vartheta_2(x,t)] + \hbar c \partial_x \theta_1(x,t,0) \big)^{-1}$. 
 In condensed matter, many kinds of emergent particles are possible whose masses may depend on both the time and position steps, so, we can  
 set $e^0_{(0)} = 1$ which implies $m c^2 = \hbar[\vartheta_1(x,t) + \vartheta_2(x,t)]+ \hbar c \partial_x \theta_1(x,t,0)$ . 
 As $\theta_1(x,t,0)$ can be an arbitrary function of $x, t$ but 
 $ - 1 \leq \cos[ 2\theta_1(x,t,0)] \leq 1 $, $g^{11}$ term of any metric can be captured 
by this through some constant value scaling.

  \subsection*{Numerical simulations} 
  The main purpose of this work is to unify all the possible background potential effects in the single particle massive Dirac Hamiltonian in its first quantized version and 
  simulate it in an operational form using quantum walks. For proper depiction one should do numerical analysis for all possible common mathematical 
  forms of the metric and gauge potentials. So that one can predict the mathematical forms of metric and gauge potentials corresponding 
  to the experimentally observed phenomena where the metric and gauge potential functions are unknown.
Here in the numerical section we have given examples of few common mathematical forms of metrics and external gauge potentials.
 Our numerical results are obtained by
 considering $\hbar = 1$ unit, $c = 1$ unit, $\tau \coloneqq \frac{1}{L}$ unit and $a \coloneqq \frac{1}{L}$ unit.
 Here, $L$ should not be 
 confused with the system size, we have used it merely to parameterize $\tau$ and $a$.
 We choose to work with the mass = $m = 0.04$ unit. 
 We have plotted the probability as a function of time (SS-DQW steps) and position for two different cases.
 This probability is irrespective of the coin state of the particle, i.e., we have traced over whole coin states at every time-step. 
 
\subsubsection{\normalsize {\bf A static metric}} 

For a static case we will run our simulation considering $L = 250$.

\begin{enumerate}

\item 
 \begin{figure}[h]
 \includegraphics[width = 12cm]{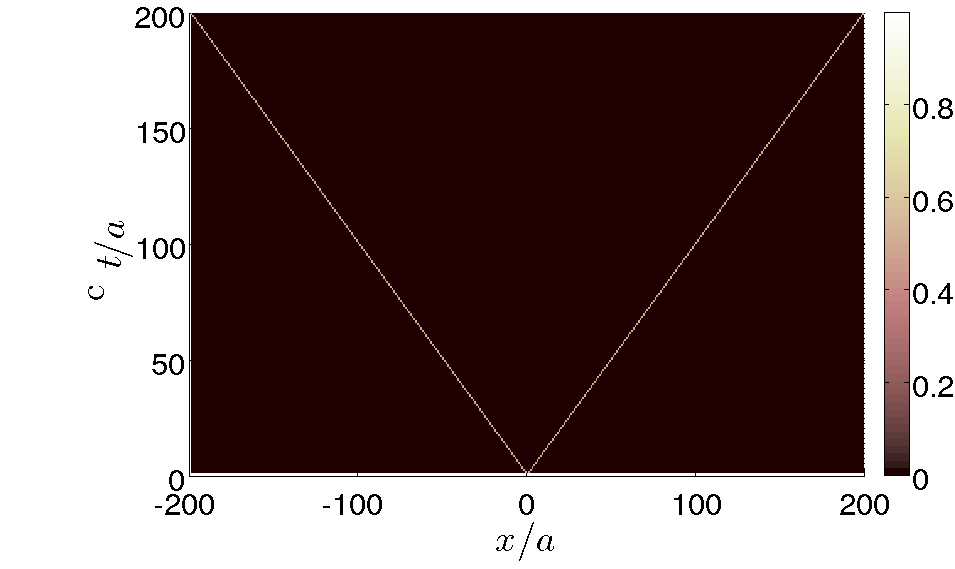} 
  \caption{(Color online) Probability as function of 200 time steps of the modified SS-DQW on a flat-lattice with 400 lattice points.
  The probability is for Minkowski metric system in absence of gauge potential with mass = 0.04 unit 
  and the initial state used for the evolution is $\frac{1}{\sqrt{2}}\big[ \ket{0} + i \ket{1} \big] \otimes \ket{x = 0}$.}
\label{fig3} 
\end{figure}

{\bf Fig.~\ref{fig3} is for flat space time without $U(1)$ potential:}
  
    $e^0_{(0)} = 1$, $e^1_{(1)} = 1$, the coin parameter functions are:
\begin{multline}
 ~ \xi_1(x,t,0) = 0,~\lambda_1(x,t) = 0,~\xi_2(x,t,0) = 0,~ \lambda_2(x,t) = 0,~~\\
 \theta_1(x,t,0) = 0 \Rightarrow \partial_x \theta_1(x,t,0) =  0,~ \vartheta_1(x,t) = 0, ~ \vartheta_2(x,t) = 0.04,~\\ 
 \Rightarrow~\text{our rotation angles are:}~~\theta_1(x,t,\tau) = 0,~~
 \theta_2(x,t,\tau) = \frac{0.04}{L},\\
 \text{our phases are:}~~\xi_1(x,t,\tau) = 0,~~
 \xi_2(x,t,\tau) = 0.
\end{multline}

\vspace{1cm}

\item     
 \begin{figure}[h]
 \includegraphics[width = 12cm]{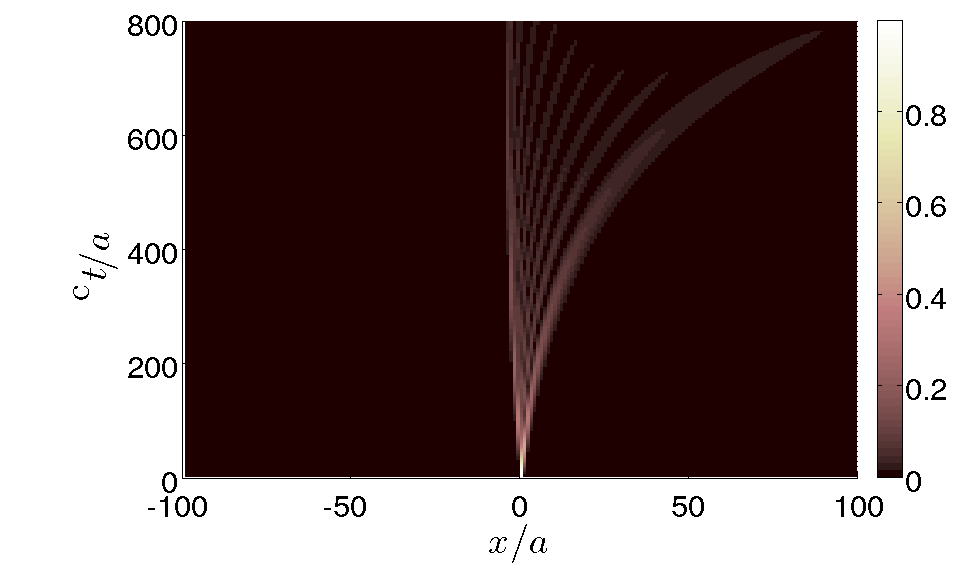}~~~~ 
 \caption{(Color online) Probability as a function of 800 time steps of the modified SS-DQW in a 
 flat-lattice with 200 lattice points. The probability  is for the metric system:~$g^{00} = 1, g^{01} = g^{10} = 0,
 g^{11} = - (x+5a)^2$ with mass = 0.04 unit
 and the initial state used for the evolution is $\frac{1}{\sqrt{2}}\big[ \ket{0} + i \ket{1} \big] \otimes \ket{x = 0}$.}
 \label{fig4}
\end{figure} 

{\bf Fig.~\ref{fig4} is for curved space-time without $U(1)$ potential:}

We choose to work with $e^0_{(0)} = 1,$ $e^1_{(1)} = x + 5 a.$

 The coin parameter functions are:\begin{multline}
 ~ \xi_1(x,t,0) = 0, ~ \lambda_1(x,t) = 0,~\xi_2(x,t,0) = 0,~  \lambda_2(x,t) = 0, \\
 \theta_1(x,t,0) = \frac{1}{2} \cos^{-1}[x + 5 a]
 \Rightarrow \partial_x \theta_1(x,t,0) =  - \frac{1}{2}  \big(1 - [x + 5a]^2 \big)^{- \frac{1}{2}},\\
 ~ \vartheta_1(x,t) =  \frac{1}{2} \big(1 -[x + 5 a]^2 \big)^{- \frac{1}{2}},~\vartheta_2(x,t) = 0.04,~
 \Rightarrow~\text{our rotation angles are:}~~\\
 \theta_1(x,t,\tau) = \frac{1}{2} \cos^{-1}[x + 5 a] 
 + \frac{\tau}{2} \big(1 -[x + 5 a]^2 \big)^{- \frac{1}{2}}, 
 \theta_2(x,t,\tau) = - \cos^{-1}[x + 5 a] + 0.04 \tau,\\
 \text{our phases are:}~~\xi_1(x,t,\tau) = 0,~~
 \xi_2(x,t,\tau) = 0.
\end{multline}
In  Fig.\,\ref{fig4}, the probability which spreads only to the right side of the origin is seen.

\vspace{1cm}

\item

\begin{figure}[h]
 \includegraphics[width = 12cm]{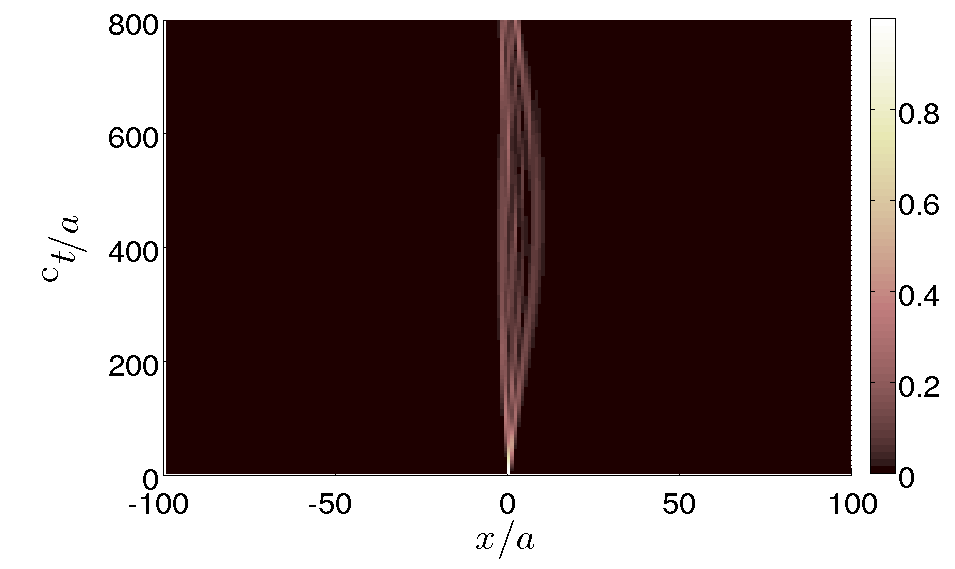}~~~~ 
 \caption{(Color online) Probability  as a function of 800 time steps of the modified SS-DQW in a 
 flat-lattice with 200 lattice points. The probability is for the metric system:~$g^{00} = 1, g^{01} = g^{10} = 0,
 g^{11} = - (x+5a)^2$  with mass = 0.04 unit
 and the initial state used for the evolution is 
 $\frac{1}{\sqrt{2}}\big[ \ket{0} + i \ket{1} \big] \otimes \ket{x = 0}$ in presence of gauge potential.}
 \label{fig5}
\end{figure}

{\bf Fig.~\ref{fig5} is for curved space-time with $U(1)$ potential:}

Here also the $e^0_{(0)} = 1,$ $e^1_{(1)} = x + 5 a.$ 
The gauge potential is captured by the parameters:
\begin{align}\xi_1(x,t,0) = 1000 x t, ~ \lambda_1(x,t) = 0.03 x, ~\xi_2(x,t,0) = 0,~\lambda_2(x,t) = 0. \nonumber \end{align}
The other coin parameter functions are:\begin{multline}
 \theta_1(x,t,0) = \frac{1}{2} \cos^{-1}[x + 5 a]
 \Rightarrow \partial_x \theta_1(x,t,0) =  - \frac{1}{2} \big(1 - [x + 5 a]^2 \big)^{- \frac{1}{2}}, 
 ~ \vartheta_1(x,t) =  \frac{1}{2}\big(1 - [x + 5 a]^2\big)^{-\frac{1}{2}},
 ~\vartheta_2(x,t) = 0.04,~\\
 \Rightarrow\text{our rotation angles are:}~
 \theta_1(x,t,\tau) = \frac{1}{2}\cos^{-1}[x+ 5 a] 
 + \frac{\tau}{2}\big(1 -[x+ 5 a]^2\big)^{-\frac{1}{2}}, 
 \theta_2(x,t,\tau) = -\cos^{-1}[x+ 5 a] + 0.04\tau,\\
 \text{our phases are:}~~\xi_1(x,t,\tau) = 1000 x t + \frac{0.03 x}{L},~~
 \xi_2(x,t,\tau) = 0.~~~~
\end{multline}

\end{enumerate}
\subsubsection{\normalsize{\bf A non-static metric}} 
 
 Here we will show the numerical simulation of a non-static case.  We will take $L = 150$.  

\begin{enumerate}

  \item   
 \begin{figure}[h]
 \includegraphics[width = 12cm]{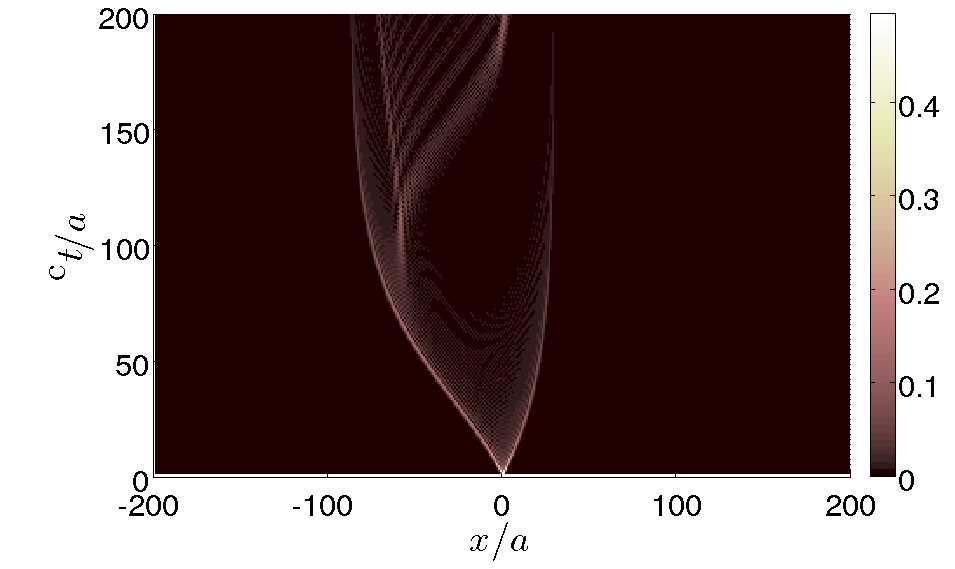} 
 \caption{(Color online) Probability as function of 200 time steps of the modified SS-DQW on a flat-lattice with 400 lattice points.
 The probability is for a non-static metric system:~$g^{00} = t^{-2}$, $g^{01} = g^{10} = 0$,
 $g^{11} = - \frac{t^{-2}}{2} \big[ \cos 4x + \sin 4x \big]^2$ in absence of gauge potential with mass = 0.04 unit.
 The initial state used for the evolution is $\frac{1}{\sqrt{2}}\big[ \ket{0} + i \ket{1} \big] \otimes \ket{x = 0}$.}
\label{fig2} 
\end{figure}  

   {\bf Fig.~\ref{fig2} is for curved space-time without $U(1)$ potential:}
    
    $e^0_{(0)} = \frac{1}{t}$, $e^1_{(1)} = \frac{1}{\sqrt{2} t} \big[ \cos 4x + \sin 4x \big]$, 
 the coin parameter functions are:
\begin{multline}
 ~ \xi_1(x,t,0) = 0, ~ \lambda_1(x,t) = 0, ~\xi_2(x,t,0) = 0,~\lambda_2(x,t) = 0,\\
 \theta_1(x,t,0) = \frac{\pi}{8} + 2 x \Rightarrow \partial_x \theta_1(x,t,0) =  2,~ \vartheta_1(x,t) = - 2, 
 ~ \vartheta_2(x,t) = 0.04 t,\\ 
 \Rightarrow~\text{our rotation angles are:}~~\theta_1(x,t,\tau) = \frac{\pi}{8} + 2 x - \frac{2}{L},~~
 \theta_2(x,t,\tau) = - \frac{\pi}{4} - 4 x + \frac{0.04 t}{L},\\
 \text{our phases are:}~~\xi_1(x,t,\tau) = 0,~~
 \xi_2(x,t,\tau) = 0.
\end{multline}
   
\vspace{1cm}    
   
   \item

\begin{figure}[h]
 \includegraphics[width = 12cm]{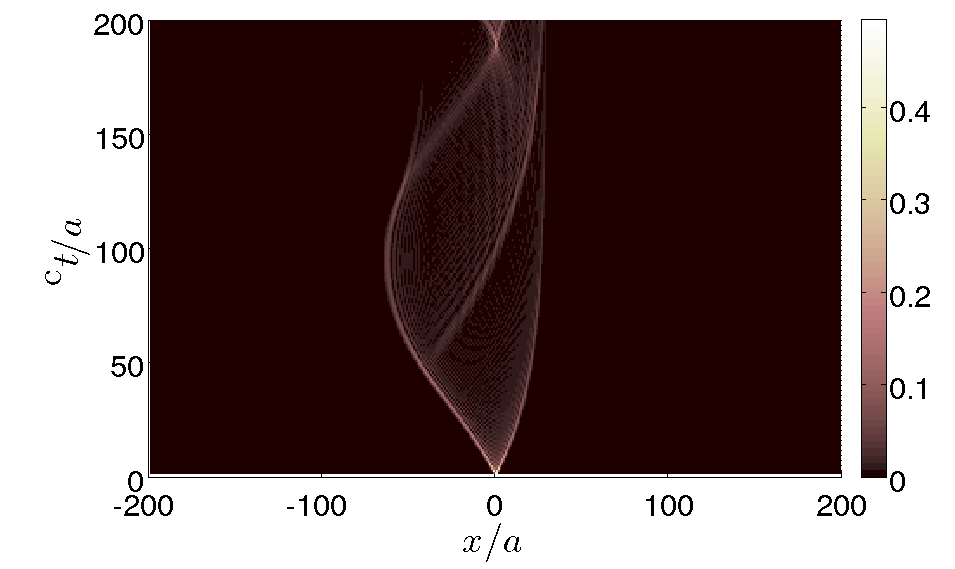} 
 \caption{(Color online) Probability as function of 200 time steps of the modified SS-DQW on a flat-lattice with 400 lattice points. 
 The probability is for a non-static metric system:~$g^{00} = t^{-2}$, $g^{01} = g^{10} = 0$, $g^{11}
 = - \frac{t^{-2}}{2} \big[ \cos 4x + \sin 4x \big]^2$ in presence of $U(1)$ gauge potential with mass = 0.04 unit. 
 The initial state used for the evolution is $\frac{1}{\sqrt{2}}\big[ \ket{0} + i \ket{1} \big] \otimes \ket{x = 0}$.}
 \label{fig1}
\end{figure}

   {\bf Fig.~\ref{fig1} is for curved space-time with $U(1)$ potential:} 
        
    $e^0_{(0)} = \frac{1}{t}$, $e^1_{(1)} = \frac{1}{\sqrt{2} t} \big[ \cos 4x + \sin 4x \big]$, 
 the coin parameter functions are:\begin{multline}
 ~ \xi_1(x,t,0) = 1000 x t, ~ \lambda_1(x,t) = 0.03 x,~ \xi_2(x,t,0) = 0,~ \lambda_2(x,t) = 0,\\
 \theta_1(x,t,0) = \frac{\pi}{8} + 2 x \Rightarrow \partial_x \theta_1(x,t,0) =  2,~ \vartheta_1(x,t) = - 2,
 ~ \vartheta_2(x,t) = 0.04 t,~\\
 \Rightarrow~\text{our rotation angles are:}~~\theta_1(x,t,\tau) = \frac{\pi}{8} + 2 x - \frac{2}{L},~~
 \theta_2(x,t,\tau) = - \frac{\pi}{4} - 4 x + \frac{0.04 t}{L}, \\
 \text{our phases are:}~~\xi_1(x,t,\tau) = 1000 x t + \frac{0.03 x}{L},~~
 \xi_2(x,t,\tau) = 0.
\end{multline}

 \end{enumerate}
 
\vspace{1cm}

In this work we should note that the initial state of the quantum walk system is taken to be a pure state $\in$ $\mathcal{H}_c \otimes \mathcal{H}_x$, 
and hence under the modified SS-DQW evolution which is also an unitary, the state will remain pure.
As we have dealt with a quantum walk particle which is always in a pure state $\ket{\psi(t)}\coloneqq \sum_x \big( \alpha_x\ket{0} + \beta_x \ket{1}\big) \otimes \ket{x}$, 
ensemble is the collection of the identically prepared quantum walk systems, all of them are in the state $\ket{\psi(t)}$ at time-step $t$.   
But during measurement of position irrespective of the coin state of the particle, we actually measure on the partial state of the system which is traced out over coin Hilbert space
= $\text{Tr}_c\big(\ket{\psi(t)}\bra{\psi(t)}\big)$. For example, the ensemble contains total $n_0 + n_1$ number of systems, at a particular time-step all are described by the state $\ket{\psi(t)}$,
$n_0$ among them are in the $\ket{0}$ coin state and other $n_1$ are in the $\ket{1}$ coin state (after coin measurement). 
Among the $n_0$ systems $r_0$ systems are in position $x$, among the $n_1$ systems $r_1$ systems are in position $x$. So the probability to be in the position $x$ is
 $\sum$ frequency of coin state $\times$ positional probability of that coin state = $\sum_{j=0}^1 \frac{n_j}{n_0 + n_1} \times \frac{r_j}{n_j}$.
 In that sense, the probabilities shown in the figs.~\ref{fig3}-\ref{fig1} are the averaged over the coin degrees of freedom
 and the corresponding probability reads
\begin{align}
 p(x,t) = \braket{x|\text{Tr}_c\big(\ket{\psi(t)}\bra{\psi(t)}\big) |x} = \sum_{j=0}^1 \frac{n_j}{n_0 + n_1} \times \frac{r_j}{n_j}~.
\end{align} This is the probability function which has been shown in all the figures as a function of position and time-step.
Different trajectories at a time-step correspond to the probabilities of obtaining the particle at different positions at that time-step.

For the static case the chosen vielbeins: $e^0_{(0)} = 1$ is constant and $e^1_{(1)} = x + 5 a$ is linear in position coordinate. 
In the non-static case we have chosen vielbeins: $e^0_{(0)}= \frac{1}{t}$ is inverse in time
and $e^1_{(1)} = \frac{1}{t} \sin \big[4x + \frac{\pi}{4}\big]$ is a combination of sinusoidal in position and inverse in time coordinate.
The choice of $U(1)$ gauge potential: $A_0 = 0.03x$ is linear in position coordinate and $A_1 = -1000 t (x+5a)^{-1}$ is linear in time and inverse in position coordinates.
The presence of the gauge potential increases localization of probability profiles in positions.
The flat space-time metric case: $e^0_{(0)}$ = $e^1_{(1)}$ = 1, has been shown to get a comparable idea about the other plots.

The parameters considered for the fig.~\ref{fig4} will give Hamiltonian $H =   \sigma_3 \otimes (x+5a)~\hat{p}_1 - \frac{i}{2}\sigma_3  \otimes \mathds{1}_x  + m \sigma_1  \otimes \mathds{1}_x$
at the continuum limit. This Hamiltonian is the same as the  
Rindler Hamiltonian: $H_R =  \sigma_3 \otimes (x+5a)~p_1 - \frac{i}{2} \sigma_3 \otimes \mathds{1}_x + \sigma_1 m \otimes (x+5a)$,  
except an additional potential term.  

In the fig.~\ref{fig4} from the probability profile it may seem that after long times the particle has a probability to exist outside the light-cone described by the fig.~\ref{fig3}. 
But in the fig.~\ref{fig4} case the light-cone should be described by equation:
\begin{align}\label{curvelight}
 ds^2 =  \Big[e_0^{(0)}\Big]^2 dt^2 - \Big[e_1^{(1)}\Big]^2 dx^2 = 0 \Rightarrow \frac{dx}{dt} = \pm (x+5a) \Rightarrow \ln(x+5a) - \ln(x_0 + 5a) = \pm t
\end{align} 
instead of the Minkowski light-cone described by the equation: $\frac{dx}{dt} = \pm 1$, where $ds$ is usually taken as the infinitesimal distance in world space-time 
and $x_0$ is the position of the particle at time $t = 0$. The trajectories should not cross the light-cone described by eq.~(\ref{curvelight}), 
as the coordinate system is not flat now. So it will not violate the causality principle even if it crosses the Minkowski light-cone.    
Although, in the unit system $c = 1$, $\hbar = 1$, $a = \frac{1}{L}$ that we have used while plotting the figures this light-cone in eq.~(\ref{curvelight}) will 
always remain within the Minkowski light-cone, and the particle trajectory never crosses the light-cone described by the eq.~(\ref{curvelight}). 
Because in the figures the axes labels are actually dimensionless quantities.

 
\subsection{Simulating $(2 + 1)$ space-time Hamiltonian by $(1+1)$ space-time dimensional SS-DQW}\label{2dim}
 
In $(2 + 1)$ space-time dimension when one of the spatial momentum components of the Dirac particle remains constant = $k_y$ unit
and all the operators in the Hamiltonian are simply function of the other spatial coordinate and time---the space-time become {\it effectively} $(1+1)$ dimensional. Under this consideration the effective Dirac Hamiltonian in $(2 + 1)$ space-time dimension given in eq.~(\ref{hamcurve}) can be written as
\begin{align}\label{2dham}
 H =  - \hbar \sigma_0 \otimes  A_{0} 
 + \Big\{ c\alpha^{(0)} \otimes q^1_{(0)} \hat{p}_1
 + c\alpha^{(0)} \otimes q^2_{(0)} k_y
 + c\alpha^{(1)} \otimes q^1_{(1)} \hat{p}_1
 + c\alpha^{(1)} \otimes q^2_{(1)} k_y 
 + c\alpha^{(2)} \otimes q^1_{(2)} \hat{p}_1
 + c\alpha^{(2)} \otimes q^2_{(2)} k_y \Big\}\nonumber\\
- \frac{i \hbar c}{2} \bigg\{ \alpha^{(0)}  \otimes \frac{\partial}{\partial x} q^1_{(0)}
+ \alpha^{(1)}  \otimes \frac{\partial}{\partial x} q^1_{(1)} 
+ \alpha^{(2)}  \otimes \frac{\partial}{\partial x}q^1_{(2)} \bigg\}
- \hbar \Big\{ \alpha^{(0)} \otimes q^1_{(0)}  A_{1}
+ \alpha^{(0)} \otimes q^2_{(0)}  A_{2} 
+ \alpha^{(1)} \otimes q^1_{(1)}  A_{1}\nonumber\\
+ \alpha^{(1)} \otimes q^2_{(1)}  A_{2} 
+ \alpha^{(2)} \otimes q^1_{(2)}  A_{1}
+ \alpha^{(2)} \otimes q^2_{(2)}  A_{2} \Big\}
+ c^2 \beta  \otimes \frac{m}{e^{0}_{(0)}}, ~~~~~~~~~~~~ \text{where}~~q^\mu_{(j)} \coloneqq \Bigg[\frac{e^\mu_{(j)}}{e^0_{(0)}}\Bigg]~~~~~~~~~~~
\end{align}
and we have taken all the operators in the Hamiltonian as functions of $x$, $t$ only.
We had $\alpha^{(0)} = \sigma_0$, now if we consider $\gamma^{(0)} = \sigma_1$,~ $\gamma^{(1)} = - i \sigma_2$,~ $\gamma^{(2)} = i \sigma_3$ it implies 
 $\alpha^{(1)} = \sigma_3,~ \alpha^{(2)} =  \sigma_2$. In order to compare the Hamiltonian in eq.~(\ref{2dham}) with our Hamiltonian in eq.~(\ref{derivedHam}) derived from the modified SS-DQW, 
 we have to make $e^1_{(0)} = 0$ which reduces the Hamiltonian in eq.~(\ref{2dham})
 to the form, \begin{align}\label{2dhamA}
 H =  - \hbar \sigma_0 \otimes  A_{0} 
 + \Big\{ c\sigma_0 \otimes q^2_{(0)} k_y
 + c\sigma_3 \otimes q^1_{(1)} \hat{p}_1
 + c\sigma_3 \otimes q^2_{(1)} k_y 
 + c\sigma_2 \otimes q^1_{(2)} \hat{p}_1
 + c\sigma_2 \otimes q^2_{(2)} k_y \Big\}\nonumber\\ 
- \frac{i \hbar c}{2} \bigg\{ \sigma_3  \otimes \frac{\partial}{\partial x} q^1_{(1)}
+ \sigma_2  \otimes \frac{\partial}{\partial x} q^1_{(2)} \bigg\}
- \hbar \Big\{ \sigma_0 \otimes q^2_{(0)}  A_{2} 
+ \sigma_3 \otimes q^1_{(1)}  A_{1}
+ \sigma_3 \otimes q^2_{(1)}  A_{2}\nonumber\\
+ \sigma_2 \otimes q^1_{(2)}  A_{1}
+ \sigma_2 \otimes q^2_{(2)}  A_{2}\Big\}
+ c^2 \sigma_1  \otimes \frac{m}{e^{0}_{(0)} }.
\end{align}
 
In this case: \begin{align}\label{2dim1}
q^1_{(2)} = \Theta_2(x,t) = \frac{1}{2} \sin[2 \theta_1(x,t,0)] 
               + \frac{1}{2} \sin[2 \theta_1(x,t,0) + 2 \theta_2(x,t,0)] ,
\end{align} \begin{align}\label{2dim2}
 q^1_{(1)} = \Theta_3(x,t) = \frac{1}{2} \cos[2 \theta_1(x,t,0)] 
               + \frac{1}{2} \cos[2 \theta_1(x,t,0) + 2 \theta_2(x,t,0)],\end{align}
\begin{align}\label{2dim3}
 - \hbar A_0 +  q^2_{(0)} ( k_y c - \hbar A_2 ) 
 = \Xi_0(x,t) =  - \hbar [\lambda_1(x,t) + \lambda_2(x,t)]  + \frac{\hbar c}{2} \partial_x \xi_2(x,t,0),
 \end{align}
 \begin{align} \label{2dim4}
  \frac{m  c^2 }{e^{0}_{(0)} }  =   \Xi_1(x,t) 
 =  \hbar[\vartheta_1(x,t) + \vartheta_2(x,t)] - \frac{\hbar c}{2} \partial_x \theta_2(x,t,0) , 
       \end{align}
 \begin{align} \label{2dim5}
  q^2_{(1)} ( k_y c - \hbar  A_2) - \hbar q^1_{(1)} A_1  = 
  \hbar c \partial_x \xi_1(x,t,0) \Theta_3(x,t)  + \frac{\hbar c}{2}\partial_x \xi_2(x,t,0) \cos[2\theta_1(x,t,0) + 2\theta_2(x,t,0)], 
  \end{align}
 \begin{align}\label{2dim6}
 q^2_{(2)} ( k_y c - \hbar  A_2)
 - \hbar q^1_{(2)} A_1  =  
 \hbar c  \partial_x \xi_1(x,t,0) \Theta_2(x,t) 
 + \frac{\hbar c}{2} \partial_x \xi_2(x,t,0) \sin[2 \theta_2(x,t,0) + 2 \theta_1(x,t,0)] .
\end{align}

The total number of variables $\big\{A_0, A_1, A_2, m, e^0_{(0)}, e^1_{(1)}, e^1_{(2)}, e^2_{(0)}, e^2_{(1)}, e^2_{(2)} \big\}$ 
for the set of the eqs.~(\ref{2dim1})-(\ref{2dim6}) are larger than the total 
number of the equations. So, unique solution is not possible. One possible solution is 
\begin{align}
 A_0 = \lambda_1(x,t) + \lambda_2(x,t),
 ~ A_1 = -  c \partial_x \xi_1(x,t,0),~ A_2 = -  c \partial_x \xi_2(x,t,0) + \frac{k_y c}{\hbar}, \nonumber\\
 ~ \Bigg[\frac{e^2_{(0)}}{e^0_{(0)}}\Bigg] = \frac{1}{2},~
\Bigg[\frac{e^1_{(2)}}{e^0_{(0)}}\Bigg] =  \frac{1}{2} \sin[2 \theta_1(x,t,0)] 
               + \frac{1}{2} \sin[2 \theta_1(x,t,0) + 2 \theta_2(x,t,0)] ,\nonumber\\
 \Bigg[\frac{e^1_{(1)}}{e^0_{(0)}}\Bigg] =  \frac{1}{2} \cos[2 \theta_1(x,t,0)] 
               + \frac{1}{2} \cos[2 \theta_1(x,t,0) + 2 \theta_2(x,t,0)], 
       \Bigg[\frac{e^2_{(1)}}{e^0_{(0)}}\Bigg] = \frac{1}{2} \cos[2 \theta_1(x,t,0) + 2 \theta_2(x,t,0)],       \nonumber\\
        \Bigg[\frac{e^2_{(2)}}{e^0_{(0)}}\Bigg] = \frac{1}{2} \sin[2 \theta_1(x,t,0) + 2 \theta_2(x,t,0)],~
  \frac{m c^2}{e^{0}_{(0)} }  =  \hbar[\vartheta_1(x,t) + \vartheta_2(x,t)] - \frac{\hbar c}{2} \partial_x \theta_2(x,t,0).              
\end{align}
Therefore, the metric
 \begin{align}
                         =  \left( \begin{array}{ccc}
                                   g^{00} & g^{01} & g^{02} \\
                                   g^{10} & g^{11} & g^{12} \\
                                   g^{20} & g^{21} & g^{22} \\
                                   \end{array} \right)
                                   = \left( \begin{array}{ccc}
                                   \Big[e^{0}_{(0)}\Big]^2 & 0 & e^{0}_{(0)} e^{2}_{(0)} \\ \\
                                   0 & - \Big[e^{1}_{(1)}\Big]^2 -  \Big[e^{1}_{(2)}\Big]^2 & - e^{1}_{(1)} e^{2}_{(1)} - e^{1}_{(2)} e^{2}_{(2)} \\ \\
                                  e^{0}_{(0)} e^{2}_{(0)} & - e^{1}_{(1)} e^{2}_{(1)} - e^{1}_{(2)} e^{2}_{(2)} 
                                  &  \Big[e^{2}_{(0)}\Big]^2 -  \Big[e^{2}_{(1)}\Big]^2 -  \Big[e^{2}_{(2)}\Big]^2 \\
                                   \end{array} \right) \nonumber\\
                                   = \Big[e^{0}_{(0)}\Big]^2\left(\begin{array}{ccc}
                                            1  &  0   & \frac{1}{2}\\  \\
                                            0  & - \frac{1}{4} - \frac{1}{2} \cos^2 [\theta_2(x,t,0)] &  - \frac{1}{2} \cos^2 [\theta_2(x,t,0)] \\  \\
                                            \frac{1}{2} & - \frac{1}{2} \cos^2 [\theta_2(x,t,0)] & 0 \\
                                           \end{array}\right),
                        \end{align} 
where we have used the definition:  $g^{\mu \nu} = e^\mu_{(0)} e^\nu_{(0)} - e^\mu_{(1)} e^\nu_{(1)} - e^\mu_{(2)} e^\nu_{(2)}$ with the sign convention: 
$\eta^{(0)(0)} = 1$, $\eta^{(1)(1)} = - 1$, $\eta^{(2)(2)} = -1$.
We should note here that this kind of choice implies that the effect of the momentum $k_y$ of the hidden coordinate express itself as a part of the gauge potential $A_2$. 
Other choices are possible which may give rise to different metrics.


\section{Implementation of our scheme in Qubit-system}\label{qubitsim}
The shift operations $S_\pm$ in eq.~(\ref{shiftssqw}),
and the coin operations $C_j(t, \tau)$ in eq.~(\ref{coinssqw}) are kinds of {\it controlled-unitary} 
operations. The shift operations $S_\pm$ change the 
position distribution while the coin state acts as the controller, and the coin operations $C_j(t,\tau)$ change the coin state while positions act as controllers. Coin state is 
already represented by a qubit (a 2-dimensional quantum-state), 
but the position space is $\mathcal{N}$ dimensional if the total number of lattice sites is $\mathcal{N}$, so, in general it can be any dimensional. Here, we will represent the position states by $n$-qubit system such that the total number of position will now be $2^n$ and each position 
is indexed by the decimal value---of the corresponding binary bits expression. 
Although $\mathcal{N} = 2^n$ represents only a particular kind of numbers, any general number of lattice sites can be constructed by neglecting some extra degrees of freedom. Below we demonstrate this scheme by a simple example.

Suppose our working system is a {\it periodic} lattice with 4 lattice sites, i.e.~lattice system is 
$\{\ket{x}~\text{such that}~x~\in~\mathbb{Z}_4\}$. We can build it
by 2-qubit only---representing each qubit in the computational basis 
$\{\ket{0} \equiv (1 ~ 0)^T,~ \ket{1} \equiv (0 ~ 1)^T ~\text{which are the eigenbasis of the conventional Pauli matrix}~\sigma_3\}$
we can write the basis of the 2-qubit system as $\{\ket{00}, \ket{01}, \ket{10}, \ket{11}\}$.
We will use the definition: position state $\ket{0} \coloneqq \ket{00}$, position state $\ket{a} \coloneqq \ket{01}$, 
position state $\ket{2a} \coloneqq \ket{10}$, position state $\ket{3a} \coloneqq \ket{11}$. 
So, in this representation \begin{align}\label{shifmat+}
              \sum_x \ket{x+a}\bra{x} = \ket{01}\bra{00} + \ket{10}\bra{01} + \ket{11}\bra{10} + \ket{00}\bra{11}  
              = \left( \begin{array}{cccc}
                        0 & 0 & 0 & 1 \\
                        1 & 0 & 0 & 0 \\
                        0 & 1 & 0 & 0 \\
                        0 & 0 & 1 & 0 \\
                       \end{array}\right) \nonumber \\
    =    \frac{1}{4} \Big[ (\sigma_0 + \sigma_3) \otimes (\sigma_1 - i \sigma_2) 
    + (\sigma_1 - i \sigma_2) \otimes (\sigma_1 + i \sigma_2) + (\sigma_0 - \sigma_3) \otimes (\sigma_1 - i \sigma_2) 
    + (\sigma_1 + i \sigma_2) \otimes (\sigma_1 + i \sigma_2) \Big]   \nonumber\\
    = \frac{1}{2} \Big[ \sigma_0 \otimes (\sigma_1 - i \sigma_2) 
    + \sigma_1  \otimes (\sigma_1 + i \sigma_2) \Big]. 
                \end{align}
 Similarly, \begin{align}\label{shifmat-}
              \sum_x \ket{x-a}\bra{x} = \ket{00}\bra{01} + \ket{01}\bra{10} + \ket{10}\bra{11} + \ket{11}\bra{00}  
              = \left( \begin{array}{cccc}
                        0 & 1 & 0 & 0 \\
                        0 & 0 & 1 & 0 \\
                        0 & 0 & 0 & 1 \\
                        1 & 0 & 0 & 0 \\
                       \end{array}\right) \\
    = \frac{1}{2} \Big[ \sigma_0 \otimes (\sigma_1 + i \sigma_2) 
    + \sigma_1  \otimes (\sigma_1 - i \sigma_2) \Big]. \nonumber
                \end{align} 
  \begin{align}\label{shifmati}
              \sum_x \ket{x}\bra{x} = \ket{00}\bra{00} + \ket{01}\bra{01} + \ket{10}\bra{10} + \ket{11}\bra{11}  
              = \left( \begin{array}{cccc}
                        1 & 0 & 0 & 0 \\
                        0 & 1 & 0 & 0 \\
                        0 & 0 & 1 & 0 \\
                        0 & 0 & 0 & 1 \\
                       \end{array}\right) 
    =  \sigma_0 \otimes \sigma_0~.
                \end{align}

For the periodic lattice case with total $\mathcal{N}$  number of lattice sites we use the relation 
 \begin{align}\label{fourier}
                     \ket{k} = \frac{1}{\sqrt{\mathcal{N}}} \sum_{x=0}^{\mathcal{N}-1} e^{\frac{i k x}{\hbar}} \ket{x}~\Rightarrow~\ket{x} = \sum_k \ket{k} \braket{k|x} = 
                     \frac{1}{\sqrt{\mathcal{N}}} \sum_k e^{-\frac{i k x}{\hbar}} \ket{k}.  \nonumber\\
                     \ket{x + \mathcal{N}a} = \ket{x} \Rightarrow  \frac{k \mathcal{N} a}{\hbar} = 2 \pi n_k \Rightarrow k =  \frac{2 \pi n_k \hbar}{\mathcal{N} a}
                    \end{align} 
                    where $n_k \in \mathbb{Z}$. Therefore the minimum possible gap between two $k$'s = $\Delta k$ = $\frac{2 \pi \hbar}{\mathcal{N} a}$.
                    Here $\ket{k}$ is the eigenvectors of the generator of the positional translation operator of the quantum walk: $\sum_x \ket{x \pm a}\bra{x}$. 
                    The normalization condition $\braket{x|x} = 1$ implies that $n_k$ can take only $\mathcal{N}$ distinct values.

{Possible choices of $k$:} For odd $\mathcal{N}$, we can choose $k \in \Big\{ - \frac{2 \pi \hbar (\mathcal{N} - 1)}{2\mathcal{N} a},
- \frac{2 \pi \hbar (\mathcal{N} - 3)}{2\mathcal{N} a}, \ldots,- \frac{2 \pi \hbar}{\mathcal{N} a}, 0, \frac{2 \pi \hbar}{\mathcal{N} a}, \ldots,
\frac{2 \pi \hbar (\mathcal{N} - 3)}{2\mathcal{N} a}, \frac{2 \pi \hbar (\mathcal{N} - 1)}{2\mathcal{N} a} \Big\}$.
For even $\mathcal{N}$, we can choose  $k \in \Big\{- \frac{2 \pi \hbar (\mathcal{N} - 2)}{2\mathcal{N} a}, \ldots,- \frac{2 \pi \hbar}{\mathcal{N} a}, 0, \frac{2 \pi \hbar}{\mathcal{N} a},
\ldots, \frac{2 \pi \hbar (\mathcal{N} - 2)}{2\mathcal{N} a}, \frac{\pi\hbar}{a} \Big\}$. So for both cases $-\frac{\pi \hbar}{a} < k \leq \frac{\pi \hbar}{a}$, this domain 
describes the first Brillouin zone in condensed matter physics.

The UV cutoff momentum = $\frac{\pi \hbar}{a}$ as this is the largest possible values in our case --- for $a \to 0$ limit this approaches to $\infty$. The IR divergence in momentum does not arise as minimum modulus value of momentum can be zero. 

One may question as here we are only showing scheme for $\mathcal{N} = 4$ case, but we believe that extension to large $\mathcal{N}$ can be simply developed based on our scheme. 
Just for the reader friendly demonstration we are dealing with $\mathcal{N} = 4$ here. 

Now we are going to use another qubit for the coin space having basis---$\{\ket{0}_c, \ket{1}_c\}$. 
In this case the shift operators take the forms: 
\begin{align}
     S_+ = \ket{0}_c \bra{0} \otimes \sum_x \ket{x+a}\bra{x}
                                  + \ket{1}_c \bra{1}\otimes \sum_x \ket{x}\bra{x} 
      =  \left(\begin{array}{cc}
                1_c & 0_c\\
                0_c & 0_c \\
               \end{array}\right) \otimes 
 \left( \begin{array}{cccc}
                        0 & 0 & 0 & 1 \\
                        1 & 0 & 0 & 0 \\
                        0 & 1 & 0 & 0 \\
                        0 & 0 & 1 & 0 \\
                       \end{array}\right) + \left(\begin{array}{cc}
                0_c & 0_c\\
                0_c & 1_c \\
               \end{array}\right) \otimes \left( \begin{array}{cccc}
                        1 & 0 & 0 & 0 \\
                        0 & 1 & 0 & 0 \\
                        0 & 0 & 1 & 0 \\
                        0 & 0 & 0 & 1 \\
                       \end{array}\right)  \nonumber\\
                                  = \frac{1}{4}(\sigma_{0c} + \sigma_{3c} )
                                  \otimes  \big[ \sigma_0 \otimes (\sigma_1 - i \sigma_2) 
    + \sigma_1  \otimes (\sigma_1 + i \sigma_2) \big] + \frac{1}{2}(\sigma_{0c} - \sigma_{3c} )
                                  \otimes \sigma_0 \otimes \sigma_0,  
    \end{align}
    
    \begin{align}
        S_- =  \ket{0}_c \bra{0} \otimes \sum_x \ket{x}\bra{x}
                                  + \ket{1}_c \bra{1} \otimes \sum_x \ket{x-a}\bra{x} 
  =  \left(\begin{array}{cc}
                1_c & 0_c\\
                0_c & 0_c \\
               \end{array}\right) \otimes \left( \begin{array}{cccc}
                        1 & 0 & 0 & 0 \\
                        0 & 1 & 0 & 0 \\
                        0 & 0 & 1 & 0 \\
                        0 & 0 & 0 & 1 \\
    \end{array}\right)  + \left(\begin{array}{cc}
                0_c & 0_c\\
                0_c & 1_c \\
               \end{array}\right) \otimes \left( \begin{array}{cccc}
                        0 & 1 & 0 & 0 \\
                        0 & 0 & 1 & 0 \\
                        0 & 0 & 0 & 1 \\
                        1 & 0 & 0 & 0 \\
                       \end{array}\right)  \nonumber\\
               =   \frac{1}{2}(\sigma_{0c} + \sigma_{3c})\otimes \sigma_0 \otimes \sigma_0 
          + \frac{1}{4}(\sigma_{0c} - \sigma_{3c}) \otimes  \big[ \sigma_0 \otimes (\sigma_1 + i \sigma_2) 
    + \sigma_1  \otimes (\sigma_1 - i \sigma_2) \big].
    \end{align}
The two coin operations for $j = 1, 2 $  are defined as
                   \begin{align}
                         C_j(t,\tau) =  \Big[e^{- i \xi_j(x = 0,t,\tau)} e^{- i \theta_j(x = 0,t,\tau) \sigma_{1c}} \otimes \ket{00}\bra{00} + 
                        e^{- i \xi_j(x = a,t,\tau)} e^{- i \theta_j(x = a,t,\tau) \sigma_{1c}} \otimes \ket{01}\bra{01}~~~~ \nonumber\\
                         + e^{- i \xi_j(x = 2a,t,\tau)} e^{- i \theta_j(x = 2 a,t,\tau) \sigma_{1c}} \otimes \ket{10}\bra{10} +
                         e^{- i \xi_j(x = 3a,t,\tau)} e^{- i \theta_j(x = 3a,t,\tau) \sigma_{1c}} \otimes \ket{11}\bra{11} \Big]~~~~ \nonumber\\
         = \frac{1}{4}\Big[ e^{- i \xi_j(x = 0,t,\tau)} e^{- i \theta_j(x = 0,t,\tau) \sigma_{1c}}
         \otimes (\sigma_0 + \sigma_3) \otimes (\sigma_0 + \sigma_3)
         + e^{- i \xi_j(x = a,t,\tau)} e^{- i \theta_j(x = a,t,\tau) \sigma_{1c}}
         \otimes (\sigma_0 + \sigma_3) \otimes (\sigma_0 - \sigma_3) \nonumber\\
         + e^{- i \xi_j(x = 2a,t,\tau)} e^{- i \theta_j(x = 2a,t,\tau) \sigma_{1c}}
         \otimes (\sigma_0 - \sigma_3) \otimes (\sigma_0 + \sigma_3)
         +e^{- i \xi_j(x = 3a,t,\tau)} e^{- i \theta_j(x = 3a,t,\tau) \sigma_{1c}}
         \otimes (\sigma_0 - \sigma_3) \otimes (\sigma_0 - \sigma_3)\Big].
 \end{align}

Thus the whole evolution operator 
$\mathscr{U}(t, \tau) = C_1^\dagger(t,0) \cdot C_2^\dagger(t,0) \cdot \big[ S_+ \cdot C_2(t, \tau) \cdot S_- \cdot C_1(t, \tau) \big]$
is implementable by a simple qubit system.

There is opinion of unnecessity of a quantum simulator for the simulation of the dynamics of a single particle quantum system---properly chosen
classical simulator can do the whole job. But the following two aspects can be used to counter this opinion.
(i) A single quantum particle can be in a superposition of wave and particle state according to the ref.~\cite{rab}. 
But classical particle and wave are two independent entities and they never mix.
(ii) Entanglement between two different degrees of freedom (coin and position in our case) in a single quantum particle have contextual origin,
but classical physics shows non-contextual behavior. For detailed discussion please look into the
ref.~\cite{markiewicz}.
So unless one explicitly proves that, in case of quantum simulation these two aspects are not important 
or can be captured by classical means after some kinds of encoding, 
it is better to work with quantum simulators.  
  
 
\section{Inclusion of $U(N)$ potential in our SS-DQW scheme} \label{uN}
 
 In the above cases we are able to include the effect of the $U(1)$ gauge potential. But we can define the coin operations in such a way that 
 influence of general $U(N)$ potential on single Dirac particle in $(1+1)$ dimension can also be derived. Here we will follow a similar kind of procedure 
 as in the ref.~\cite{arnault}. 
 
 In this case the coin Hilbert space $\mathcal{H}_c = \text{span}\{(1~ 0 ~ 0~\ldots~0~0)^T, (0~ 1 ~ 0~\ldots~0~0)^T, \ldots, (0~ 0 ~ 0~\ldots~0~1)^T\}$
 is a $2N$ dimensional vector space instead of only two dimensional. The position Hilbert space will be the same as it was. 
 
 The shift operators are now defined as 
 \begin{align}\label{shiftssqwN}
     S_+ = \sum_{x} \left(\begin{array}{cc}
                                   1 &  0 \\
                                   0 &  0 \\
                                  \end{array} \right) \otimes \mathds{1}_N \otimes \ket{x+a}\bra{x}
                                  + \left(\begin{array}{cc}
                                   0 &  0 \\
                                   0 &  1 \\
                                  \end{array} \right) \otimes \mathds{1}_N  \otimes \ket{x}\bra{x}, \nonumber\\
        S_- = \sum_{x} \left(\begin{array}{cc}
                                   1 &  0 \\
                                   0 &  0 \\
                                  \end{array} \right) \otimes \mathds{1}_N  \otimes \ket{x}\bra{x}
                                  + \left(\begin{array}{cc}
                                   0 &  0 \\
                                   0 &  1 \\
                                  \end{array} \right) \otimes \mathds{1}_N  \otimes \ket{x-a}\bra{x}.                        
    \end{align}where $\mathds{1}_N $ is the $N \times N$ identity matrix defined on the coin Hilbert space. The coin operations 
    are now defined as    
\begin{align}\label{coinssqwN}
  C_j(t,\tau) = \sum_{x} e^{i \xi_j(x,t, \tau)}\ovast[ \left(\begin{array}{cc}
                                    F_j(x,t, \tau) & G_j(x,t, \tau) \\\\
                                   -G^*_j(x,t, \tau) &  F^*_j(x,t, \tau) \\
                                  \end{array} \right) \otimes \mathds{1}_N \ovast] \cdot 
                                  \ovast[ \left(\begin{array}{cc}
                                                 e^{- i \tau \sum_{q=0}^{N^2 - 1} \omega^q_j(x,t) \Lambda_q}  &  0 \\
                                                 0  &  e^{- i \tau \sum_{q=0}^{N^2 - 1} \Omega^q_j(x,t) \Lambda_q} \\ 
                                                \end{array} \right) \ovast] \otimes \ket{x}\bra{x}
 \end{align} where the matrices $\Lambda_q$ are the generators of $U(N)$ group. We will define our evolution operator as 
 $\mathscr{U}_N(t, \tau) = C_1^\dagger(t,0) \cdot C_2^\dagger(t,0) \cdot \big[ S_+ \cdot C_2(t, \tau) \cdot S_- \cdot C_1(t, \tau) \big]$ 
 which is similar to the case having $U(1)$ potential only. Using the definition of the effective Hamiltonian as in eq.~(\ref{effham}), we get 
\begin{align}
 H_{\text{eff},N} = \sum_{r=0}^3 (\sigma_r \otimes \mathds{1}_N ) \otimes \sum_x \Xi_r(x,t) \ket{x}\bra{x} + 
   c~\sum_{r=1}^3  (\sigma_r \otimes \mathds{1}_N) \otimes \sum_x \Theta_r(x,t) \ket{x}\bra{x}~ \hat{p} \nonumber\\
   + \sum_{r=0}^3 \sigma_r \otimes \sum_x \sum_{q = 0}^{N^2 - 1} \Lambda_q \chi^q_r(x,t) \otimes \ket{x}\bra{x}
\end{align} where 
\begin{align}
\chi^q_0(x,t) = \frac{\hbar}{2}\Big[ \omega^q_1(x,t) + \Omega^q_1(x,t) + \omega^q_2(x,t) + \Omega^q_2(x,t)\Big] , \nonumber\\
\chi^q_3(x,t) = \frac{\hbar}{2}\Big[ \omega^q_1(x,t) - \Omega^q_1(x,t) + \{\omega^q_2(x,t) - \Omega^q_2(x,t)\}\{|F_1(x,t,0)|^2 - |G_1(x,t,0)|^2\} \Big] , \nonumber\\
\chi^q_1(x,t) = \hbar \Re[G_1(x,t,0) F_1^*(x,t,0)]\big[\omega^q_2(x,t) - \Omega_2^q(x,t)\big],~~ 
\chi^q_2(x,t) = - \hbar \Im[G_1(x,t,0) F_1^*(x,t,0)]\big[\omega^q_2(x,t) - \Omega_2^q(x,t)\big].
\end{align}
 
The term  $\sum_{r=0}^3 \sigma_r \otimes \sum_x \sum_{q = 1}^{N^2 - 1} \Lambda_q \chi^q_r(x,t) \otimes \ket{x}\bra{x}$ 
describes the effect of nonabelian potentials, where we have taken $\Lambda_0 = \mathds{1}_N$. 
In our $(1+1)$ dimensional case we can work with the choice: $\chi^q_2(x,t) = \chi^q_1(x,t) = 0$ for all $q$. 
 For a proper choice of $N$ the $U(N)$ can be the composition of all possible abelian and non-abelian gauge potential effects, 
 and hence, the derived Hamiltonian can capture all the possible fundamental force effects on a single Dirac particle. 
 For example we can include $SU(3)$ and $SU(2)$ interactions by choosing $N = 2 \times 3 = 6$. 
One important point is that the dynamical characters of these potentials have not been considered, they act as background potentials on the single Dirac particle.

Sometimes the fermion doubling problem \cite{nielsen1, nielsen2} appears, when the fermion particle dynamics is discussed in lattice position framework. 
The corresponding no-go theorem---Nielsen-Ninomiya theorem describes the impossibility of lattice simulation of local
fermion field theory consistently without avoiding the fermion doubling problem. 
In refs.~\cite{birula, quinn}, it is discussed that the Nielsen-Ninomiya theorem may not be applicable for discrete time evolution. 

In our case all the evolution operators are defined on discrete time and discrete position. In the homogeneous SS-DQW case when we allow only rotation about the spin-$x$ axis in the coin operations, according to the ref.~\cite{mallcm} the positive energy eigenvalue 
\begin{align}\label{dispers}
 E(k) = \frac{\hbar}{\tau} \cos^{-1}\bigg[ \cos \theta_1 \cos \theta_2  \cos\bigg(\frac{ka}{\hbar}\bigg) - \sin \theta_1 \sin \theta_2 \bigg]~
\end{align} 
is a monotonic function of the modulus of momentum: $|k| \in \big[0, \frac{\pi \hbar}{a} \big]$.
For a massless case: $\theta_1$ = $\theta_2$ = 0 the relation in eq.~(\ref{dispers}) reduces to $E(k) = k c$ which is very consistent with the Weyl fermion case.  
Because of the monotonicity there does not exist two different $|k|$ for which the positive energy eigenvalues are the same.
This implies no fermion doubling. This is independent of the cutoff scale $a$.

In case of position, time-step dependent coin parameters, the overall effect can be thought as a introduction of space-time dependent 
potential effects on the homogeneous SS-DQW case. It is expected that for the scalar potential, i.e.~while the potential does not depend on the chirality of the particle,
it does not change the monotonic nature of the energy as a function of the modulus of momentum.  So, in those cases, fermion doubling does not occur.
But for chirality dependent potentials, it is not so obvious that the fermion doubling problem does not appear, so these cases need further investigations.

 
\section{Extending our ($\mathbf{1+1}$) dimensional SS-DQW scheme to two-particle case } \label{twopar}
 
Here we will apply our SS-DQW framework into a two-particle system. 
In order to extend to two-particle case we will use entangled coin operations and the separable shift operations.
We extend the conventional evolution operator that evolves a two-particle state at time $t$ to a state at time $t + \tau$, 
 \begin{align}\label{unissqwtwo}
  U^\text{two}(t, \tau) = \big[ S^1_+ \otimes S^2_+ \big] \cdot C_2(t, \tau) \cdot \big[ S^1_- \otimes S^2_- \big] \cdot C_1(t, \tau)~~~~
\end{align} such that the modified or actual evolution operator will now be  
\begin{align}
 \mathscr{U}^\text{two}(t, \tau) =  C^\dagger_1(t, 0) \cdot C^\dagger_2(t, 0) \cdot
 \big[ S^1_+ \otimes S^2_+ \big] \cdot C_2(t, \tau) \cdot \big[ S^1_- \otimes S^2_- \big] \cdot C_1(t, \tau)
\end{align}
acting on the Hilbert space $\mathcal{H}_{c_1} \otimes \mathcal{H}_{c_2}
\otimes \mathcal{H}_{x_1} \otimes \mathcal{H}_{x_2}$ $\equiv$ $\mathcal{H}^{\otimes 2}_c \otimes \mathcal{H}^{\otimes 2}_x$.
 $\mathcal{H}_{x_1} = \text{span}\{\ket{x_1} : x_1 \in a \mathbb{Z}\}$,
$\mathcal{H}_{x_2} = \text{span}\{\ket{x_2} : x_2 \in a \mathbb{Z}\}$ correspond to the 
position Hilbert spaces of the first and second particles, respectively. 
$\mathcal{H}_{c_1} = \text{span}\{(1~ 0)^T, (0 ~ 1)^T\}$, $\mathcal{H}_{c_2} = \text{span}\{(1~ 0)^T, (0 ~ 1)^T\}$
correspond to the coin Hilbert spaces of the first and second particles, respectively.
Note that, we have synchronized the time-steps of both the particles to the time-step $t$ --- same for both, which is a special case.   
The shift operators for the individual particle are now defined as
\begin{align}\label{shiftssqwtwo}
     S^r_+ = \sum_{x_r} \left(\begin{array}{cc}
                                   1 &  0 \\
                                   0 &  0 \\
                                  \end{array} \right) \otimes \ket{x_r + a}\bra{x_r}
                                  + \left(\begin{array}{cc}
                                   0 &  0 \\
                                   0 &  1 \\
                                  \end{array} \right) \otimes \ket{x_r}\bra{x_r},~~ 
        S^r_- = \sum_{x_r} \left(\begin{array}{cc}
                                   1 &  0 \\
                                   0 &  0 \\
                                  \end{array} \right) \otimes \ket{x_r}\bra{x_r}
                                  + \left(\begin{array}{cc}
                                   0 &  0 \\
                                   0 &  1 \\
                                  \end{array} \right) \otimes \ket{x_r-a}\bra{x_r},                        
    \end{align} where $r = 1$ and $r = 2$ are for the first and the second particles, respectively.
    Therefore, \begin{align}\label{twofsh}
  S^1_+ \otimes S^2_+ = \sum_{x_1, x_2} \left(\begin{array}{cc}
                                   1 &  0 \\
                                   0 &  0 \\
                                  \end{array} \right) \otimes \left(\begin{array}{cc}
                                   1 &  0 \\
                                   0 &  0 \\
                                  \end{array} \right) \otimes \ket{x_1 + a, x_2 + a}\bra{x_1, x_2}
                                  + \left(\begin{array}{cc}
                                   1 &  0 \\
                                   0 &  0 \\
                                  \end{array} \right) \otimes \left(\begin{array}{cc}
                                   0 &  0 \\
                                   0 &  1 \\
                                  \end{array} \right) \otimes \ket{x_1 + a, x_2}\bra{x_1, x_2} \nonumber\\
                                + \left(\begin{array}{cc}
                                   0 &  0 \\
                                   0 &  1 \\
                                  \end{array} \right) 
                                  \otimes \left(\begin{array}{cc}
                                   1 &  0 \\
                                   0 &  0 \\
                                  \end{array} \right)\otimes \ket{x_1, x_2 + a}\bra{x_1, x_2}
                                + \left(\begin{array}{cc}
                                   0 &  0 \\
                                   0 &  1 \\
                                  \end{array} \right) \otimes \left(\begin{array}{cc}
                                   0 &  0 \\
                                   0 &  1 \\
                                  \end{array} \right)  \otimes \ket{x_1, x_2}\bra{x_1, x_2}
 \end{align}
 and, \begin{align}\label{twobsh}
      S^1_- \otimes S^2_- = \sum_{x_1, x_2}\left(\begin{array}{cc}
                                   1 &  0 \\
                                   0 &  0 \\
                                  \end{array} \right) \otimes \left(\begin{array}{cc}
                                   1 &  0 \\
                                   0 &  0 \\
                                  \end{array} \right) \otimes \ket{x_1, x_2}\bra{x_1, x_2}
                                  +  \left(\begin{array}{cc}
                                   1 &  0 \\
                                   0 &  0 \\
                                  \end{array} \right) \otimes \left(\begin{array}{cc}
                                   0 &  0 \\
                                   0 &  1 \\
                                  \end{array} \right) \otimes \ket{x_1, x_2 - a}\bra{x_1, x_2} \nonumber\\
                                + \left(\begin{array}{cc}
                                   0 &  0 \\
                                   0 &  1 \\
                                  \end{array} \right) 
                                  \otimes \left(\begin{array}{cc}
                                   1 &  0 \\
                                   0 &  0 \\
                                  \end{array} \right) \otimes \ket{x_1-a, x_2}\bra{x_1, x_2}
                                + \left(\begin{array}{cc}
                                   0 &  0 \\
                                   0 &  1 \\
                                  \end{array} \right) \otimes \left(\begin{array}{cc}
                                   0 &  0 \\
                                   0 &  1 \\
                                  \end{array} \right) \otimes \ket{x_1-a, x_2-a}\bra{x_1, x_2}.
     \end{align}   The coin operators are now defined as
 \begin{align}\label{twocoin}
  C_j(t, \tau) = \sum_{x_1, x_2} \exp\Bigg(- i \sum_{\alpha, \beta = 0}^3 
  \mathscr{C}_j^{\alpha \beta}(x_1, x_2, t, \tau)~ \sigma_\alpha \otimes \sigma_\beta \Bigg) \otimes 
 \ket{x_1}\bra{x_1} \otimes \ket{x_2}\bra{x_2}~~\text{for}~~j = 1, 2 ~~~~~~~\nonumber\\
\text{where}~\mathscr{C}_j^{\alpha \beta}(x_1, x_2, t, \tau)~\text{has to be real 
 for all}~j, \alpha, \beta, x_1, x_2, t, \tau~\text{in order to make the coin operations unitary}. 
  \end{align} 
 The shift operators in eqs.~(\ref{twofsh})-(\ref{twobsh}) are symmetric in joint exchange of coin and position indices of the two particles. 
 The coin operator in (\ref{twocoin})
 is not in general symmetric in joint exchange of coin and position indices of the two particles unless symmetrization imposed on the functions
 $\mathscr{C}_j^{\alpha \beta}(x_1, x_2, t, \tau)$. Hence, this kind of evolution operator 
 can capture distinguishable as well as indistinguishable two-particle evolution depending on the functional 
 form of $\mathscr{C}_j^{\alpha \beta}(x_1, x_2, t, \tau)$.   
 
 In the separable coin operation case while there is no interaction among the particles we must have 
 $\mathscr{C}_j^{\alpha \beta}(x_1, x_2, t, \tau) = 0$ for all $\alpha, \beta \in \{ 1, 2, 3\}$
  and any $x,t,\tau$. Thus for the nontrivial case --- when these coefficients are nonzero, the particles can be in general entangled in their coin space 
  by the whole SS-DQW evolution. Here our main purpose is to study the emergence of the curvature effects from the  
  coin-coin entanglement, so, we will choose to work in a special entangled coin operations:
  
   $$\mathscr{C}_j^{11}(x_1, x_2, t, \tau) = \theta_j(x_1, x_2, t, \tau)~\text{and all other}~\mathscr{C}_j^{\alpha \beta}(x_1, x_2, t, \tau) = 0.$$ 
  
Therefore,
\begin{align}
  C_j(t, \tau) = \mathlarger{\mathlarger{\sum}}_{x_1, x_2}\vast[ \cos[\theta_j(x_1, x_2, t, \tau)] \left( \begin{array}{cccc}
                                                1 & 0 & 0 & 0 \\
                                                0 & 1 & 0 & 0 \\
                                                0 & 0 & 1 & 0 \\
                                                0 & 0 & 0 & 1\\
                                               \end{array}\right)
                                               - i \sin[\theta_j(x_1, x_2, t, \tau)] \left( \begin{array}{cccc}
                                                0 & 0 & 0 & 1 \\
                                                0 & 0 & 1 & 0 \\
                                                0 & 1 & 0 & 0 \\
                                                1 & 0 & 0 & 0\\
                                               \end{array}\right)\vast] \otimes 
 \ket{x_1}\bra{x_1} \otimes \ket{x_2}\bra{x_2}.
\end{align}
Our choice of the whole coin operators is already symmetric in two-particle coin states, so it may describe
indistinguishable particles if $\theta_j(x_1, x_2, t, \tau) = \theta_j(|x_1-x_2|, t, \tau)$.

We will consider the case when $\theta_j(x_1, x_2, t, \tau)$ are analytic in all of their arguments for all $j= 1, 2$.
So, we can consider the Taylor series expansion in variable $\tau$ as
\begin{align}
 \theta_j(x_1, x_2, t, \tau) = \theta_j(x_1, x_2, t, 0) + \tau \vartheta_j(x_1, x_2, t),
\end{align}  
where the higher order terms in $\tau$ are chosen to be zero.
 
\vspace{0.6 cm} 
 
Following the same procedure as in the case of the single particle, we get the effective two-particle Hamiltonian: 
\begin{align}\label{twoeffham}
  H^\text{two}_\text{eff}  =  (\sigma_{0} \otimes \sigma_{3}) \otimes \Xi_{03}(\hat{x}_1, \hat{x}_2, t)
   +  c (\sigma_{0} \otimes \sigma_{3}) \otimes \Theta^2_{03}(\hat{x}_1, \hat{x}_2, t)  \mathbb{I}_1 \otimes \hat{p}_2 \nonumber\\
   + (\sigma_{3} \otimes \sigma_{0}) \otimes  \Xi_{30}(\hat{x}_1, \hat{x}_2, t)
  + c(\sigma_{3} \otimes \sigma_{0}) \otimes  \Theta^1_{30}(\hat{x}_1, \hat{x}_2, t) \hat{p}_1\otimes \mathbb{I}_2 \nonumber\\
  + (\sigma_{1} \otimes \sigma_{2}) \otimes  \Xi_{12}(\hat{x}_1, \hat{x}_2, t) 
   +  c (\sigma_{1} \otimes \sigma_{2}) \otimes \Theta^2_{12}(\hat{x}_1, \hat{x}_2, t)  \mathbb{I}_1 \otimes \hat{p}_2 \nonumber\\
  + (\sigma_{2} \otimes \sigma_{1}) \otimes  \Xi_{21}(\hat{x}_1, \hat{x}_2, t)
  + c(\sigma_{2} \otimes \sigma_{1}) \otimes \Theta^1_{21}(\hat{x}_1, \hat{x}_2, t) \hat{p}_1\otimes \mathbb{I}_2 \nonumber\\
  + (\sigma_{1} \otimes \sigma_{1}) \otimes  \Xi_{11}(\hat{x}_1, \hat{x}_2, t).\hspace{5cm}~~~~
\end{align}

where $\Xi_{\mu \nu}(\hat{x}_1, \hat{x}_2, t) \coloneqq \sum\limits_{x_1, x_2} \Xi_{\mu \nu}(x_1, x_2, t) \ket{x_1, x_2} \bra{x_1, x_2}$ 
and $\Theta_{\mu \nu}(\hat{x}_1, \hat{x}_2, t) \coloneqq \sum\limits_{x_1, x_2} \Theta_{\mu \nu}(x_1, x_2, t) \ket{x_1, x_2} \bra{x_1, x_2}$.

 For detailed derivation and the explicit expression of the coefficient functions of the two-particle Hamiltonian
 see Appendix \ref{twoparham}. 
 
The two-particle effective Hamiltonian can be split into three parts as
  \begin{align}
  H^\text{two}_\text{eff} = H^1_\text{eff} \otimes  \sigma_0 + 
  \sigma_0 \otimes H^2_\text{eff} + H^\text{inter}_\text{eff}.
  \end{align}
 Here
  \begin{align}
        H^1_\text{eff} = \sigma_{3}  \otimes \big[ \Xi_{30}(\hat{x}_1, \hat{x}_2, t)
  + c \Theta^1_{30}(\hat{x}_1, \hat{x}_2, t) \hat{p}_1\otimes \mathbb{I}_2 \big]
       \end{align} looks like local Hamiltonian part for the first particle whose effective mass = 0 and 
      the curved nature of space-time which is influenced by the presence of the second 
       particle, is captured by the term $\Theta^1_{30}(x_1, x_2, t)$, and 
           \begin{align}
         H^2_\text{eff} = \sigma_{3}  \otimes  \big[\Xi_{03}(\hat{x}_1, \hat{x}_2, t)
  + c \Theta^2_{03}(\hat{x}_1, \hat{x}_2, t) \mathbb{I}_1 \otimes \hat{p}_2 \big]
       \end{align} looks like local Hamiltonian part for the second particle whose effective mass = 0 and 
       the curved nature of space-time which is influenced by the presence of the first
       particle, is captured by the term $\Theta^2_{03}(x_1, x_2, t)$. 
       
The part $H^\text{inter}_\text{eff}$ of the Hamiltonian has no proper local analogy.  This appears as a purely two-particle interaction term originated from the entangled coin operations. 

\vspace{1cm}

{\bf Note:} The coin operation is global, which can entangle two separable particles, and this entanglement has in general nonlocal features. 
Thus implementation by local operation is in general impossible. But the coefficients (or strength) 
of the interaction term controlled by $\mathscr{C}_j^{\alpha \beta}(x_1, x_2, t, \tau)$
are functions of positions of both the particles and time. 
If one consider the functions $\mathscr{C}_j^{\alpha \beta}(x_1, x_2, t, \tau)$ = $0$ outside the light-cone, the interaction can be made local. 
In case of quantum simulation these particles are usually very near to each other, i.e., the distance between the particles is hardly space-like.
Almost of all the simulation cases they remain within time-like distance, so that information transfer from one to another is possible during any bipartite local operation.
Once local and two-particle controlled local simulators implement $\sigma_\alpha \otimes \sigma_\beta$ operations, an entanglement between these particles can be created. 
After that it is possible that they possess nonlocal nature in Bell inequality violation sense, when get separated beyond light-like distance.
At the current stage we do not have any explanation of this nonlocal interaction in terms of any gauge boson exchange.
This is very interesting point, and need further investigation.

 \subsubsection{\normalsize {\bf For a special functional forms of the coin parameters:}} 
%
  For an example, we will deal with a case when $\theta_2(x_1, x_2, t, 0) = - 2 \theta_1(x_1, x_2, t, 0) = \cos^{-1}\big[ (x_1 - x_2)^2 \big]$. 
  In this case the terms of the two-particle effective Hamiltonian in eq.~(\ref{twoeffham}) become
  \begin{align}
    \Xi_{03}(x_1, x_2, t) = \frac{i \hbar c}{2} \sin[\theta_2(x_1, x_2, t, 0)] \partial_{x_2}\theta_2(x_1, x_2, t, 0) = i \hbar c (x_1 - x_2) ,~~ \nonumber\\
     \Theta^2_{03}(x_1, x_2, t) = \cos[\theta_2(x_1, x_2, t, 0)] = (x_1 - x_2)^2, \nonumber\\ 
      \Xi_{30}(x_1, x_2, t) = \frac{i \hbar c}{2} \sin[\theta_2(x_1, x_2, t, 0)] \partial_{x_1}\theta_2(x_1, x_2, t, 0) = - i \hbar c (x_1 - x_2),~~\nonumber\\ 
    \Theta^1_{30}(x_1, x_2, t) = \cos[\theta_2(x_1, x_2, t, 0)] = (x_1 - x_2)^2,\nonumber\\
    \Xi_{11}(x_1, x_2, t) = - \frac{\hbar c}{2}\big[\partial_{x_2}\theta_2(x_1, x_2, t, 0) + \partial_{x_1}\theta_2(x_1, x_2, t, 0)\big]
    + \hbar \big[\vartheta_1(x_1, x_2, t) + \vartheta_2(x_1, x_2, t) \big]\nonumber\\ = \hbar \big[\vartheta_1(x_1, x_2, t) + \vartheta_2(x_1, x_2, t) \big], \nonumber\\
       \Xi_{12}(x_1, x_2, t) = 0,~~ 
     \Theta^2_{12}(x_1, x_2, t) = 0,~~
     \Xi_{21}(x_1, x_2, t) = 0,~~ 
    \Theta^1_{21}(x_1, x_2, t) = 0.~~~~
    \end{align}
  
  Therefore the Hamiltonian takes the form 
  \begin{align}
   H^\text{two}_\text{eff}  = i \hbar c (\sigma_{0} \otimes \sigma_{3}) \otimes ( \hat{x}_1 \otimes \mathbb{I}_2 - \mathbb{I}_1 \otimes \hat{x}_2 )
    +  c (\sigma_{0} \otimes \sigma_{3}) \otimes ( \hat{x}_1 \otimes \mathbb{I}_2 - \mathbb{I}_1 \otimes \hat{x}_2 )^2  \mathbb{I}_1 \otimes \hat{p}_2 \nonumber\\
    - i \hbar c (\sigma_{3} \otimes \sigma_{0}) \otimes ( \hat{x}_1 \otimes \mathbb{I}_2 - \mathbb{I}_1 \otimes \hat{x}_2 )
   + c(\sigma_{3} \otimes \sigma_{0}) \otimes  ( \hat{x}_1 \otimes \mathbb{I}_2 - \mathbb{I}_1 \otimes \hat{x}_2 )^2 \hat{p}_1\otimes \mathbb{I}_2 \nonumber\\
    + \hbar (\sigma_{1} \otimes \sigma_{1}) \otimes \big[\vartheta_1(\hat{x}_1, \hat{x}_2, t) + \vartheta_2(\hat{x}_1, \hat{x}_2, t) \big].
 \end{align}

\section{Conclusion}\label{conclu}
 
 In this work we are able to show that single-step SS-DQW with slight modification, can simulate massive Dirac particle dynamics under the influence of external abelian gauge potential and curved space-time.  The modification of evolution operator is just an extra coin operation after applying the conventional SS-DQW. We have shown that the same Hamiltonian can capture pseudo $(1+1)$ dimensional
 or $(2+1)$ dimensional Dirac particle dynamics when the momentum along the hidden dimension remains fixed.
 We provided an implementation scheme by qubit systems which is realizable in current experimental set-up.
  By increasing the dimension of the coin-space, 
 the influence of general $U(N)$ gauge potential has been included in our scheme which paves a way towards simulation of 
 four fundamental force effects on a single Dirac particle.
 We extended our study to the case of two-particle SS-DQW where the interaction of the particles is solely 
 comes from the entangled coin operations and showed that the parameters of this entanglement can be included in the curvature effect. 
 Our study shows a way to investigate non-classical properties as well as the curvature effects which are difficult to observe 
 in real situation.

 \section*{ACKNOWLEDGMENT} Authors like to thank Avijit Nath, Sagnik Chakraborty for useful mathematical discussion.
 CMC would like to thank SERB, Department of Science and Technology, Government of India for the Ramanujan Fellowship grant No.:SB/S2/RJN-192/2014.

 \vspace{0.5cm} 
 

                              $~~~~~~~~~~~\hspace{5cm}~~~~~~~~~~$\underline{\Large{\bf Appendix}}
 
{\footnotesize

\appendix

      \section{Derivation of Schrödinger like equation form curved space-time Dirac equation}\label{schroder}
Flat space-time Dirac equation is given by
\begin{align*}
 \left(i \hbar \gamma^{(a)}\partial_{(a)}- m c^2 \right )\psi=0,
\end{align*} where $\partial_{(a)}~\text{or later used}~\partial_\mu \in \left\{ \partial_{t},  c~ \partial_{x^i} ~\text{such that}~i=1, 2, 3.\right\}$.
Generalization to the curved space-time is given by
\begin{align}
 \left(i \hbar e^{\mu}_{(a)}\gamma^{(a)}\nabla_{\mu} - m c^2 \right)\psi=0,
 \label{Dirac Curved Eq appendix}
\end{align}
where $\nabla_\mu = \partial_\mu + \Gamma_\mu - i A_\mu$, $\Gamma_{\mu}=-\frac{i}{4}S_{(c)(d)}e^{(c)\nu}\left(\frac{\partial e^{(d)}_{\nu}}{\partial x^{\mu}}-\Gamma^{\lambda}_{\mu\nu}e^{(d)}_{\lambda}\right)$,
$\Gamma^{\sigma}_{\lambda\mu}=\frac{1}{2}g^{\nu\sigma}\left(\partial_{\lambda}g_{\mu\nu}+\partial_{\mu}g_{\lambda\nu}-\partial_{\nu}g_{\mu\lambda}\right)$,
and $S_{(c)(d)}$ are the flat spinor matrices: $S_{(c)(d)}=\frac{i}{2}[\gamma_{(c)},\gamma_{(d)}]$, $A_\mu$ is the $U(1)$ potential.
Now in view of the following relations,
\begin{align*}
\gamma_{(a)}S_{(b)(c)}=\frac{1}{2}[\gamma_{(a)},S_{(b)(c)}]+\frac{1}{2}\{\gamma_{(a)},S_{(b)(c)}\},~~
[\gamma_{(a)},S_{(b)(c)}]=2i\left(\eta_{(a)(b)}\gamma_{(c)}-\eta_{(a)(c)}\gamma_{(b)}\right),\\
\{\gamma_{(a)},S_{(b)(c)}\}=-2i\epsilon_{(a)(b)(c)(d)}\gamma^{(d)}\gamma_{5};\,\,\gamma_{5}=\gamma_{(0)}\gamma_{(1)}\gamma_{(2)}\gamma_{(3)},
~~~~~~~~~~~~~~~~~~~~~~~~~~~~~~~~\end{align*}
it is possible to write eq.~(\ref{Dirac Curved Eq appendix}) as,
\begin{align}
 \frac{i \hbar}{2}\gamma^{(a)} \bigg[ \bigg\{e^{\mu}_{(a)},\bigg(\frac{\partial}{\partial x^{\mu}} - i A_{\mu} \bigg) \bigg\} 
 + e^{\rho}_{(a)}\Gamma^{\mu}_{\mu\rho} \bigg]\psi
 +\frac{i \hbar}{2}\gamma^{(a)}\gamma_{5}\mathcal{B}_{(a)}\psi = m c^2 \psi,
 \label{dirac equation appendix}
\end{align}
where $\mathcal{B}_{(a)}=\frac{1}{2}\epsilon_{(a)(b)(c)(d)}e^{(b)\mu}e^{(c)\nu}\frac{\partial e^{(d)}_{\nu}}{\partial x^{\mu}}$; For $(1+1)$ and $(2+1)$ dimensions
$\epsilon_{(a)(b)(c)(d)}$ is always zero, so $\mathcal{B}_{(a)}=0$.

To derive the current density we need to derive also the dual equation satisfied by $\bar{\psi}=\psi^{\dagger}\beta$, where $\beta=\gamma^{(0)}$
and it is given by the following equation, with the assumption that all the vielbeins are real,
\begin{align}
 \frac{i\hbar}{2} \bigg[ \bigg\{ e^{\mu}_{(a)},\bigg(\frac{\partial}{\partial x^{\mu}} + i A_{\mu} \bigg) \bigg\} 
 + e^{\rho}_{(a)}\Gamma^{\mu}_{\mu\rho} \bigg] \bar{\psi}\gamma^{(a)}
 -\frac{i \hbar }{2}\gamma^{(a)}\gamma_{5}\mathcal{B}_{(a)}\bar{\psi} = - m c^2 \bar{\psi},
 \label{dual dirac Curved Eq appendix}
\end{align}
From eq.~(\ref{dirac equation appendix}) and eq.~(\ref{dual dirac Curved Eq appendix}) it is possible to derive the four vector current $j^{\mu}$, and they are 
given as 
\begin{align}
 j^{\mu}=\sqrt{-g}e^{\mu}_{(a)}\bar{\psi}\gamma^{(a)}\psi ~\Rightarrow~ j^{0}=\sqrt{-g}e^{0}_{(0)}\psi^{\dagger}\psi+\sqrt{-g}e^{0}_{(i)}\bar{\psi}\gamma^{(i)}\psi,
\label{current}
\end{align}
where $g = \det(g_{\mu\nu})$ and the current is conserved, i.e.~ $\frac{\partial j^{\mu}}{\partial x^{\mu}} = 0$.
We want to write the curved space-time Dirac equation in the following Schrödinger equation like form
\begin{align}
 i \hbar \frac{\partial\chi}{\partial t} = H \chi,
 \label{Schrodinger equation}
\end{align}
where $H$ is the Hermitian Hamiltonian operator. So the probability density is given by, $j^{0}=\chi^{\dagger}\chi.$
After we multiply eq.~(\ref{dirac equation appendix}) by $\beta$, we get a similar equation like eq.~(\ref{Schrodinger equation})
\begin{align}
 \frac{i \hbar}{2}\alpha^{(a)}\left[\left\{e^{\mu}_{(a)},\bigg( \frac{\partial}{\partial x^{\mu}}-i A_{\mu} \bigg)\right\}
 + e^{\rho}_{(a)}\Gamma^{\mu}_{\mu\rho}\right]\psi = m c^2 \beta\psi, \nonumber
  \end{align}
 
 \begin{align}
 \frac{i \hbar}{2}\left\{e^{0}_{(0)},\left(\frac{\partial}{\partial t}-iA_{0}\right)\right\}\psi
 = - \frac{i \hbar }{2}\alpha^{(a)}\left\{e^{i}_{(a)}, c \frac{\partial}{\partial x^{i}}-iA_{i}\right\}
 \psi - \frac{i \hbar }{2}\alpha^{(a)}e^{\rho}_{(a)}\Gamma^{\mu}_{\mu\rho}\psi + m c^2 \beta\psi
 \label{dirac equation after beta multiplication}
\end{align}
where $\alpha^{(a)}=\beta\gamma^{(a)}$. However this Hamiltonian is not
hermitian and the current is also not same as eq.~(\ref{current}). In this case current is given by,
\begin{equation}
j^{0}=\sqrt{-g}e^{0}_{(0)}\psi^{\dagger}\psi.
\label{second current}
\end{equation}
Comparisons of eq.~(\ref{current}) and eq.~(\ref{second current}) suggests that
we must make nonunitary transformation (with the assumption $e^{0}_{(i)}=0$),
\begin{align}
 \chi=(-g)^{\frac{1}{4}}\Big[e^{0}_{(0)}\Big]^{\frac{1}{2}}\psi.
\end{align}
Now we will use this transformation in eq.~(\ref{dirac equation after beta multiplication}) to write $\psi$ in terms of $\chi$.
\begin{align}
& \left\{e^{0}_{(0)},\left(\frac{\partial}{\partial t}-iA_{0}\right)\right\}\psi
= 2e^{0}_{(0)}\frac{\partial\psi}{\partial t}-2ie^{0}_{(0)}A_{0}\psi+\frac{\partial e^{0}_{(0)}}{\partial t}
 \psi\nonumber\\
&=(-g)^{-\frac{1}{4}}\left[-\Big[e^{0}_{(0)}\Big]^{-\frac{1}{2}}\frac{\partial e^{0}_{(0)}}{\partial t}\chi+2\Big[e^{0}_{(0)}\Big]^{\frac{1}{2}}\frac{\partial\chi}{\partial t}
+\frac{\partial e^{0}_{(0)}}{\partial t}\Big[e^{0}_{(0)}\Big]^{-\frac{1}{2}}\chi\right]
+ 2\Big[e^{0}_{(0)}\Big]^{\frac{1}{2}}\frac{\partial(-g)^{-\frac{1}{4}}}{\partial t}\chi - 2i\Big[e^{0}_{(0)}\Big]^{\frac{1}{2}}A_{0}(-g)^{-\frac{1}{4}}\chi\nonumber\\
& = (-g)^{-\frac{1}{4}}2\Big[e^{0}_{(0)}\Big]^{\frac{1}{2}}\frac{\partial\chi}{\partial t}
+ 2\Big[e^{0}_{(0)}\Big]^{\frac{1}{2}}\frac{\partial(-g)^{-\frac{1}{4}}}{\partial t}\chi  -  2i\Big[e^{0}_{(0)}\Big]^{\frac{1}{2}}A_{0}(-g)^{-\frac{1}{4}}\chi ~~.
\end{align}
Similarly,
\begin{align}
 &\left\{e^{i}_{(a)},\frac{\partial}{\partial x^{i}}-iA_{i}\right\}\psi=2e^{i}_{(a)}\frac{\partial\psi}{\partial x^{i}}+\frac{\partial e^{i}_{(a)}}{\partial x^{i}}\psi
 -2ie^{i}_{(a)}A_{i}\psi\nonumber\\
 &=2e^{i}_{(a)}\left[\Big[e^{0}_{(0)}\Big]^{-\frac{1}{2}}\frac{\partial(-g)^{-\frac{1}{4}}}{\partial x^{i}}
 \chi+(-g)^{-\frac{1}{4}}\Big[e^{0}_{(0)}\Big]^{-\frac{1}{2}}\frac{\partial\chi}
 {\partial x^{i}}+(-g)^{-\frac{1}{4}}\frac{\partial\Big[e^{0}_{(0)}\Big]^{-\frac{1}{2}}}{\partial x^{i}}\chi\right]\nonumber\\
 &+\frac{\partial e^{i}_{(a)}}{\partial x^{i}}(-g)^{-\frac{1}{4}}\Big[e^{0}_{(0)}\Big]^{-\frac{1}{2}}\chi-2ie^{i}_{(a)}
 A_{i}(-g)^{-\frac{1}{4}}\Big[e^{0}_{(0)}\Big]^{-\frac{1}{2}}\chi
\end{align}
and,
\begin{align}\label{conn1}
 \Gamma^{\mu}_{\mu\rho}=\frac{1}{2}g^{\mu\lambda}\left\{\frac{\partial g_{\lambda\mu}}{\partial x^{\rho}}+\frac{\partial g_{\lambda\rho}}{\partial x^{\mu}}
 -\frac{\partial g_{\mu\rho}}{\partial x^{\lambda}}\right\}
=\frac{1}{2}\left\{g^{\mu\lambda}\frac{\partial g_{\lambda\mu}}{\partial x^{\rho}}+\frac{\partial g_{\lambda\rho}}{\partial x^{\lambda}}
 -\frac{\partial g_{\mu\rho}}{\partial x^{\rho}}\right\}=\frac{1}{2}g^{\mu\lambda}\frac{\partial g_{\lambda\mu}}{\partial x^{\rho}} .
\end{align}
We can evaluate this easily by using the following relation for any arbitrary matrix M,
\begin{align}\label{conn2}
 \text{Tr}\left\{M^{-1}(x)\frac{\partial}{\partial x^{\lambda}}M(x)\right\}=\frac{\partial}{\partial x^{\lambda}}\ln [\det M(x)]
\end{align}
So, $\Gamma^{\mu}_{\mu\rho}=\frac{1}{2}\frac{\partial}{\partial x^{\rho}}\text{ln}\,g=\frac{1}{\sqrt{g}}\frac{\partial}{\partial x^{\rho}}\sqrt{g}$.
Finally using all the relations described above, we can write,
\begin{align}
&i \hbar \frac{\partial\chi}{\partial t} = \Big[e^{0}_{(0)}\Big]^{-1} \vast(
- \hbar\Big[e^{0}_{(0)}\Big] A_{0} + \frac{i \hbar}{4} \Big[e^{0}_{(0)}\Big] \frac{ \partial \ln (-g)}{\partial t}
- i\hbar \alpha^{(a)}e^{i}_{(a)}\Bigg[ - \frac{c}{4}\frac{\partial \ln (-g) }{\partial x^{i}}\nonumber\\
&+  c \frac{\partial}{\partial x^{i}}
- \frac{c}{2}  \frac{\partial \ln e^{0}_{(0)}}{\partial x^{i}}\Bigg]
-\frac{i\hbar}{2}\alpha^{(a)} c \frac{\partial e^{i}_{(a)}}{\partial x^{i}}
- \hbar \alpha^{(a)}e^{i}_{(a)}A_{i} - \frac{i \hbar}{2}\alpha^{(a)}e^{\rho}_{(a)}\Gamma^{\mu}_{\mu\rho} + m c^2 \beta \vast)\chi
\label{Schrodinger equation 2}
\end{align}
Now using $e^{0}_{(i)}=0$ (which will not make any lose of generalization as the number of independent vielbeins in the metric is 
less than the total number of vielbeins---see ref.~\cite{olddir} for details) and the properties in  eqs.~(\ref{conn1}), (\ref{conn2})
we can show that second, third, and eighth terms of the above equation
will cancel with each other. Finally we can write,
\begin{align}\label{schrohamcur}
i \hbar \frac{\partial\chi}{\partial t} = \Big[e^{0}_{(0)}\Big]^{-1} \vast(
- \hbar\Big[e^{0}_{(0)}\Big] A_{0} - i\hbar c ~ \alpha^{(a)}e^{i}_{(a)}\Bigg[
   \frac{\partial}{\partial x^{i}}
- \frac{1}{2}  \frac{\partial \ln e^{0}_{(0)}}{\partial x^{i}}\Bigg]
-\frac{i\hbar}{2}\alpha^{(a)} c \frac{\partial e^{i}_{(a)}}{\partial x^{i}}
- \hbar \alpha^{(a)}e^{i}_{(a)}A_{i} + m c^2 \beta \vast)\chi \nonumber\\
\Rightarrow i \hbar \frac{\partial\chi}{\partial t} =  
- \hbar A_{0}  \chi - i\hbar c ~ \alpha^{(a)}  \frac{e^{i}_{(a)}}{e^0_{(0)}} \frac{\partial  \chi}{\partial x^{i}}
- \frac{i \hbar c}{2} \alpha^{(a)}  \frac{\partial}{\partial x^{i}} \Bigg[\frac{e^i_{(a)}}{e^0_{(0)}}\Bigg] \chi
- \hbar \alpha^{(a)}\Bigg[\frac{e^i_{(a)}}{e^0_{(0)}}\Bigg] A_{i} \chi + \beta \frac{m c^2 }{e^0_{(0)}} \chi
\end{align}

So, in operator terms the above  eq.~(\ref{schrohamcur}) can be expressed as:

\begin{align}
H =  - \hbar A_{0} + c ~ \alpha^{(a)}\Bigg[\frac{e^i_{(a)}}{e^0_{(0)}}\Bigg] \hat{p}_i
- \frac{i \hbar c}{2} \alpha^{(a)}  \frac{\partial}{\partial x^{i}} \Bigg[\frac{e^i_{(a)}}{e^0_{(0)}}\Bigg]
- \hbar \alpha^{(a)}\Bigg[\frac{e^i_{(a)}}{e^0_{(0)}}\Bigg] A_{i} + \beta \frac{m c^2 }{e^0_{(0)}}.
\end{align}

          \section{Expansion of the single particle conventional SS-DQW Evolution operator}\label{evoluform}

 The conventional single particle SS-DQW unitary evolution operator  
 \begin{align}\label{convenevol}
  U(t, \tau) = \sum_x \left( \begin{array}{cc}
\ket{x+a}\bra{x} & 0 \\
0 & \ket{x}\bra{x}
\end{array}\right) \boldsymbol{\cdot} \sum_x \left( \begin{array}{cc}
e^{i \xi_2(x,t,\tau)} F_2(x,t, \tau)\ket{x}\bra{x} & e^{i \xi_2(x,t,\tau)} G_2(x,t, \tau)\ket{x}\bra{x}\\ \\
-e^{i \xi_2(x,t,\tau)} G^*_2(x,t, \tau)\ket{x}\bra{x} & e^{i \xi_2(x,t,\tau)} F^*_2(x,t, \tau)\ket{x}\bra{x} \\
\end{array}\right) \nonumber\\
\boldsymbol{\cdot}    \sum_x \left( \begin{array}{cc}
\ket{x}\bra{x} & 0 \\
0 & \ket{x-a}\bra{x}
\end{array}\right) \boldsymbol{\cdot} \sum_x \left( \begin{array}{cc}
e^{i \xi_1(x,t,\tau)} F_1(x, t, \tau)\ket{x}\bra{x} & e^{i \xi_1(x, t,\tau)} G_1(x, t, \tau)\ket{x}\bra{x}\\ \\
-e^{i \xi_1(x,t,\tau)} G^*_1(x, t, \tau)\ket{x}\bra{x} & e^{i \xi_1(x, t,\tau)} F^*_1(x, t, \tau)\ket{x}\bra{x} \\
\end{array}\right)  \nonumber\\
=  \sum_x \left( \begin{array}{cc}
e^{i \xi_2(x,t,\tau)} F_2(x,t, \tau)\ket{x+a}\bra{x} & e^{i \xi_2(x,t,\tau)} 
G_2(x,t, \tau)\ket{x+a}\bra{x} \\ \\
-e^{i \xi_2(x,t,\tau)} G^*_2(x,t, \tau)\ket{x}\bra{x} & e^{i \xi_2(x,t,\tau)} 
F^*_2(x,t, \tau)\ket{x}\bra{x} \\
\end{array}\right) \nonumber\\
\boldsymbol{\cdot} \sum_x \left( \begin{array}{cc}
e^{i \xi_1(x,t,\tau)} F_1(x,t, \tau) \ket{x}\bra{x} & e^{i \xi_1(x,t,\tau)}
G_1(x,t, \tau) \ket{x}\bra{x}\\ \\
-e^{i \xi_1(x,t,\tau)} G^*_1(x,t, \tau)\ket{x-a}\bra{x} & e^{i \xi_1(x,t,\tau)}
F^*_1(x,t, \tau) \ket{x-a}\bra{x}\\
\end{array}\right)  \nonumber\\                                                              
 \end{align}

We now expand the unitary evolution operators in  eq.~(\ref{convenevol}) upto first order in variables $\tau$ and $a$. 
We use here the definition of generator of translation as: 
$$ e^{\mp\frac{i \hat{p} a}{\hbar}} = \sum_x \ket{x \pm a}\bra{x}
\Rightarrow \ket{x \pm a} = e^{\mp\frac{i \hat{p} a}{\hbar}} \ket{x} 
= \ket{x} \mp \frac{ia}{\hbar} \hat{p} \ket{x} + \mathcal{O}(a^2) \ket{x}.$$ 

\vspace{0.1cm}
 {\bf From the calculation in eq.~(\ref{convenevol}) we get the matrix elements of $U(t, \tau)$ in coin basis as follows}
\vspace{0.1cm}

\begin{itemize}
 \item The first-row first-column term of SS-DQW Evolution operator in coin-basis
\begin{align}
 U_{00}(t,\tau) =  \sum_x e^{i [\xi_1(x,t,\tau) + \xi_2(x,t,\tau)]}  F_2(x,t,\tau)F_1(x,t,\tau) \ket{x+a}\bra{x}  
- e^{i [\xi_1(x,t,\tau) + \xi_2(x-a,t,\tau)]} G_2(x-a,t,\tau)G_1^*(x,t,\tau) \ket{x}\bra{x} \nonumber\\
= \sum_x e^{i [\xi_1(x-a,t,\tau) + \xi_2(x-a,t,\tau)]} F_2(x-a,t,\tau)
F_1(x-a,t,\tau) \ket{x} \bra{x} e^{-\frac{i \hat{p} a}{\hbar}} 
-  e^{i [\xi_1(x,t,\tau) + \xi_2(x-a,t,\tau)]} G_2(x-a,t,\tau)G_1^*(x,t,\tau) \ket{x}\bra{x}.
\end{align}
\item The first-row second-column term of SS-DQW Evolution operator in coin-basis
\begin{align}
U_{01}(t,\tau) =  \sum_x  e^{i [\xi_1(x,t,\tau) + \xi_2(x,t,\tau)]} F_2(x,t,\tau) G_1(x,t,\tau) \ket{x+a}\bra{x} 
+  e^{i [\xi_1(x,t,\tau) + \xi_2(x-a,t,\tau)]} G_2(x-a,t,\tau)F_1^*(x,t,\tau) \ket{x}\bra{x} \nonumber\\
= \sum_x  e^{i [\xi_1(x-a,t,\tau) + \xi_2(x-a,t,\tau)]} F_2(x-a,t,\tau) G_1(x-a,t,\tau) 
\ket{x}\bra{x} e^{\frac{-i\hat{p}a}{\hbar}} 
+  e^{i [\xi_1(x,t,\tau) + \xi_2(x-a,t,\tau)]} G_2(x-a,t,\tau)F_1^*(x,t,\tau) \ket{x}\bra{x}.
\end{align}
\item The second-row first-column term of SS-DQW Evolution operator in coin-basis
\begin{align}
U_{10}(t,\tau) =  \sum_x -  e^{i [\xi_1(x,t,\tau) + \xi_2(x,t,\tau)]} G_2^*(x,t,\tau) F_1(x,t,\tau) \ket{x}\bra{x}
-  e^{i [\xi_1(x,t,\tau) + \xi_2(x-a,t,\tau)]} F^*_2(x-a,t,\tau) G_1^*(x,t,\tau)\ket{x-a}\bra{x} \nonumber\\
= \sum_x -  e^{i [\xi_1(x,t,\tau) + \xi_2(x,t,\tau)]} G_2^*(x,t,\tau) F_1(x,t,\tau) \ket{x}\bra{x}
-  e^{i [\xi_1(x+a,t,\tau) + \xi_2(x,t,\tau)]} 
F^*_2(x,t,\tau) G_1^*(x+a,t,\tau)\ket{x}\bra{x} e^{\frac{i \hat{p} a}{\hbar}}. 
\end{align}
\item The second-row second-column term of SS-DQW Evolution operator in coin-basis
\begin{align}
U_{11}(t,\tau) =  \sum_x -  e^{i [\xi_1(x,t,\tau) + \xi_2(x,t,\tau)]} G_2^*(x,t,\tau)G_1(x,t,\tau) \ket{x}\bra{x}
+  e^{i [\xi_1(x,t,\tau) + \xi_2(x-a,t,\tau)]} F^*_2(x-a,t,\tau)F^*_1(x,t,\tau)\ket{x-a}\bra{x}\nonumber\\
= \sum_x -  e^{i [\xi_1(x,t,\tau) + \xi_2(x,t,\tau)]} G_2^*(x,t,\tau)G_1(x,t,\tau) \ket{x}\bra{x} 
+  e^{i [\xi_1(x+a,t,\tau) + \xi_2(x,t,\tau)]} F^*_2(x,t,\tau)F^*_1(x+a,t,\tau)\ket{x}\bra{x}e^{\frac{i \hat{p} a}{\hbar}}.
\end{align}
\end{itemize}

{\bf Therefore,}

\subsection{the first-row first-column term of our modified Evolution operator in coin-basis}\label{evoluform00}

\begin{align}
 \mathscr{U}_{00}(t,\tau) = U_{00}^\dagger(t,0) U_{00}(t,\tau) + U_{10}^\dagger(t,0) U_{10}(t,\tau)
 = \sum_x e^{- i [\xi_1(x,t,0) + \xi_2(x,t,0)]} \Big[  F^*_2(x,t,0) F^*_1(x,t,0) -  G^*_2(x,t,0)G_1(x,t,0) \Big] \nonumber\\
\times \Big[ e^{i [\xi_1(x-a,t,\tau) + \xi_2(x-a,t,\tau)]} F_2(x-a,t,\tau)F_1(x-a,t,\tau) \ket{x} \bra{x} e^{-\frac{i \hat{p} a}{\hbar}}
- e^{i [\xi_1(x,t,\tau) + \xi_2(x-a,t,\tau)]} G_2(x-a,t,\tau)G_1^*(x,t,\tau) \ket{x}\bra{x} \Big]\nonumber\\
+ \sum_x -  e^{-i [\xi_1(x,t,0) + \xi_2(x,t,0)]} \Big[ G_2(x,t,0) F^*_1(x,t,0) + F_2(x,t,0) G_1(x,t,0) \Big] \nonumber\\
 \times \Big[-  e^{i [\xi_1(x,t,\tau) + \xi_2(x,t,\tau)]} G_2^*(x,t,\tau) F_1(x,t,\tau) \ket{x}\bra{x}
 - e^{i [\xi_1(x+a,t,\tau) + \xi_2(x,t,\tau)]} F^*_2(x,t,\tau) G_1^*(x+a,t,\tau)\ket{x}\bra{x} e^{\frac{i \hat{p} a}{\hbar}} \Big]
\end{align}

\begin{align}
\Rightarrow \mathscr{U}_{00}(t,\tau) - \sum_x \ket{x}\bra{x} = ~~~~~~~~~~~~~~~~~~~~~~~~~~~~~~~~~~ \nonumber\\
 - \frac{ i a}{\hbar}\sum_x \Big[ |F_2(x,t,0)|^2 |F_1(x,t,0)|^2 - |F_2(x,t,0)|^2 |G_1(x,t,0)|^2 
 - 2 \Re\{G_2^*(x,t,0) F_1(x,t,0) F_2(x,t,0) G_1(x,t,0)\} \Big] \ket{x}\bra{x}~\hat{p} \nonumber\\
 +  \sum_x  \bigg\{ \tau  \Big[ F_1^*(x,t,0) f_1(x,t,0) + g_1^*(x,t,0) G_1(x,t,0) + G_2^*(x,t,0) g_2(x,t,0) + F_2(x,t,0) f_2^*(x,t,0) \Big] \nonumber\\
 + 2 i  \tau \Im\Big[ f_2(x,t,0) F_2^*(x,t,0) |F_1(x,t,0)|^2 - f_2(x,t,0) G_1(x,t,0) F_1(x,t,0) G_2^*(x,t,0)
 + g_2^*(x,t,0) G_2(x,t,0) |F_1(x,t,0)|^2  \nonumber\\
 + g_2^*(x,t,0) F_2(x,t,0) F_1(x,t,0) G_1(x,t,0)\Big] 
 + a \partial_x F_2(x,t,0) \Big[F_1(x,t,0) G_1(x,t,0) G_2^*(x,t,0) - |F_1(x,t,0)|^2 F_2^*(x,t,0) \Big]\nonumber\\ 
 + a \partial_x F_1(x,t,0) \Big[F_2(x,t,0) G_1(x,t,0) G_2^*(x,t,0)
 - |F_2(x,t,0)|^2 F_1^*(x,t,0) \Big] \nonumber\\
 + a \partial_x G_2(x,t,0) \Big[ F_2^*(x,t,0) F_1^*(x,t,0) G_1^*(x,t,0) - |G_1(x,t,0)|^2 G_2^*(x,t,0) \Big] \nonumber\\
 + a \partial_x G_1^*(x,t,0) \Big[ G_2(x,t,0) F_1^*(x,t,0) F_2^*(x,t,0) + G_1(x,t,0) |F_2(x,t,0)|^2 \Big]
 + i \tau [\lambda_1(x,t,0) + \lambda_2(x,t,0)]  \nonumber\\
 - i a \partial_x \xi_1(x,t,0) \Big( |F_2(x,t,0)|^2 |F_1(x,t,0)|^2 - |F_2(x,t,0)|^2|G_1(x,t,0)|^2 
 - 2 \Re[F_2(x,t,0) F_1(x,t,0) G_2^*(x,t,0) G_1(x,t,0)] \Big) \nonumber\\
 - i a \partial_x \xi_2(x,t,0) \Big( |F_2(x,t,0)|^2 |F_1(x,t,0)|^2 + |G_2(x,t,0)|^2 |G_1(x,t,0)|^2 
 - 2 \Re[F_2(x,t,0) F_1(x,t,0) G_2^*(x,t,0) G_1(x,t,0)]\Big) \bigg\} \ket{x}\bra{x} + \mathcal{O}(\tau^2),
\end{align}

where we have used the definition \begin{align}
 F_j(x,t, \tau) =  F_j(x,t,0) + \tau f_j(x,t) + \mathcal{O}(\tau^2),
 G_j(x,t, \tau) =   G_j(x,t,0) + \tau g_j(x,t) + \mathcal{O}(\tau^2),
 \xi_j(x,t,\tau) =  \xi_j(x,t,0) + \tau \lambda_j(x,t) + \mathcal{O}(\tau^2),
 \end{align}

\subsection{the first-row second-column term of our modified Evolution operator in coin-basis}\label{evoluform01}

\begin{align}
 \mathscr{U}_{01}(t,\tau) = U_{00}^\dagger(t,0) U_{01}(t,\tau) + U_{10}^\dagger(t,0) U_{11}(t,\tau) \nonumber\\
 = \sum_x  e^{-i [\xi_1(x,t,0) + \xi_2(x,t,0)]} \bigg[ F_2^*(x,t,0) F_1^*(x,t,0)  -  G^*_2(x,t,0)G_1(x,t,0)\bigg] \nonumber\\
 \bigg[ e^{i [\xi_1(x-a,t,\tau) + \xi_2(x-a,t,\tau)]} F_2(x-a,t,\tau) G_1(x-a,t,\tau) \ket{x}\bra{x} e^{\frac{-i\hat{p}a}{\hbar}} 
+  e^{i [\xi_1(x,t,\tau) + \xi_2(x-a,t,\tau)]} G_2(x-a,t,\tau)F_1^*(x,t,\tau) \ket{x}\bra{x} \bigg] \nonumber\\
+ \sum_x e^{-i [\xi_1(x,t,0) + \xi_2(x,t,0)]} \bigg[- G_2(x,t,0) F^*_1(x,t,0) - F_2(x,t,0) G_1(x,t,0)\bigg] \nonumber\\
\bigg[ -  e^{i [\xi_1(x,t,\tau) + \xi_2(x,t,\tau)]} G_2^*(x,t,\tau)G_1(x,t,\tau) \ket{x}\bra{x}
+ e^{i [\xi_1(x+a,t,\tau) + \xi_2(x,t,\tau)]} F^*_2(x,t,\tau)F^*_1(x+a,t,\tau)\ket{x}\bra{x}e^{\frac{i \hat{p} a}{\hbar}} \bigg]
\end{align}

$\Rightarrow$

\begin{align}
 \mathscr{U}_{01}(t, \tau) = \sum_x - \frac{i a}{\hbar}
 \big[ 2 |F_2(x,t,0)|^2 G_1(x,t,0) F_1^*(x,t,0) - F_2(x,t,0) G_2^*(x,t,0) [G_1(x,t,0)]^2
 + F_2^*(x,t,0) [ F_1^*(x,t,0)]^2 G_2(x,t,0)  \big] \ket{x}\bra{x}~\hat{p} \nonumber\\
 + \sum_x \bigg\{ - a \partial_x G_1(x,t,0) \Big[|F_2(x,t,0)|^2 F_1^*(x,t,0) - G_2^*(x,t,0) G_1(x,t,0) F_2(x,t,0) \Big]\nonumber\\
 -  a \partial_x F_2(x,t,0) \Big[ F_2^*(x,t,0) F_1^*(x,t,0) G_1(x,t,0) - G_2^*(x,t,0) [G_1(x,t,0)]^2  \Big] \nonumber\\
 - a \partial_x G_2(x,t,0) \Big[ F_2^*(x,t,0) [F_1^*(x,t,0)]^2 - F_1^*(x,t,0) G_1(x,t,0) G_2^*(x,t,0) \Big] \nonumber\\
 - a \partial_x F_1^*(x,t,0) \Big[G_2(x,t,0) F_1^*(x,t,0) F_2^*(x,t,0) + G_1(x,t,0) |F_2(x,t,0)|^2 \Big] \nonumber\\
 + \tau \Big[ g_1(x,t,0) F_1^*(x,t,0)  - f_1^*(x,t,0) G_1(x,t,0)  \Big] 
 + \tau g_2^*(x,t,0) \Big[ F_2(x,t,0) [G_1(x,t,0)]^2 + F_1^*(x,t,0) G_1(x,t,0) G_2(x,t,0) \Big]\nonumber\\
 - \tau f_2^*(x,t,0) [G_2(x,t,0) [F_1^*(x,t,0)]^2 + G_1(x,t,0) F_2(x,t,0) F_1^*(x,t,0) ] \nonumber\\
 +\tau g_2(x,t,0) \Big[[F_1^*(x,t,0)]^2 F_2^*(x,t,0) - F_1^*(x,t,0) G_1(x,t,0) G_2^*(x,t,0) \Big] \nonumber\\
 + \tau f_2(x,t,0) \Big[ G_1(x,t,0) F_1^*(x,t,0) F_2^*(x,t,0) - [G_1(x,t,0)]^2 G_2^*(x,t,0) \Big] \nonumber\\
  - i a \partial_x \xi_1(x,t,0) \Big[ 2 |F_2(x,t,0)|^2 G_1(x,t,0) F_1^*(x,t,0) 
  + G_2(x,t,0) F_2^*(x,t,0) [F_1^*(x,t,0)]^2 - G_2^*(x,t,0) F_2(x,t,0) [G_1(x,t,0)]^2 \Big] \nonumber\\
 - i a \partial_x \xi_2(x,t,0) \Big[ G_1(x,t,0) F_1^*(x,t,0)\big[ |F_2(x,t,0)|^2 - |G_2(x,t,0)|^2 \big]  \nonumber\\
  - F_2(x,t,0) G_2^*(x,t,0) [G_1(x,t,0)]^2 + G_2(x,t,0) F_2^*(x,t,0) [F_1^*(x,t,0)]^2 \Big] \bigg\} \ket{x}\bra{x} + \mathcal{O}(\tau^2),
\end{align}

where we have used the definition \begin{align}
 F_j(x,t, \tau) =  F_j(x,t,0) + \tau f_j(x,t) + \mathcal{O}(\tau^2),
 G_j(x,t, \tau) =   G_j(x,t,0) + \tau g_j(x,t) + \mathcal{O}(\tau^2),
 \xi_j(x,t,\tau) =  \xi_j(x,t,0) + \tau \lambda_j(x,t) + \mathcal{O}(\tau^2),
 \end{align}

\subsection{the second-row first-column term of our modified Evolution operator in coin-basis}\label{evoluform10}

\begin{align}
 \mathscr{U}_{10}(t, \tau) = U_{01}^\dagger(t, 0) U_{00}(t, \tau) + U_{11}^\dagger(t,0) U_{10}(t, \tau) \nonumber\\
 = \sum_x e^{-i [\xi_1(x,t,0) + \xi_2(x,t,0)]} \bigg[ F^*_2(x,t,0) G^*_1(x,t,0) +  G^*_2(x,t,0)F_1(x,t,0) \bigg] \nonumber\\
 \bigg[ e^{i [\xi_1(x-a,t,\tau) + \xi_2(x-a,t,\tau)]} F_2(x-a,t,\tau) F_1(x-a,t,\tau) \ket{x} \bra{x} e^{-\frac{i \hat{p} a}{\hbar}}
- e^{i [\xi_1(x,t,\tau) + \xi_2(x-a,t,\tau)]} G_2(x-a,t,\tau)G_1^*(x,t,\tau) \ket{x}\bra{x} \bigg] \nonumber\\
+ \sum_x e^{-i [\xi_1(x,t,0) + \xi_2(x,t,0)]} \bigg[- G_2(x,t,0)G^*_1(x,t,0) +  F_2(x,t,0) F_1(x,t,0) \bigg] \nonumber\\
\bigg[ - e^{i [\xi_1(x,t,\tau) + \xi_2(x,t,\tau)]} G_2^*(x,t,\tau) F_1(x,t,\tau) \ket{x}\bra{x}
- e^{i [\xi_1(x+a,t,\tau) + \xi_2(x,t,\tau)]} F^*_2(x,t,\tau) G_1^*(x+a,t,\tau)\ket{x}\bra{x} e^{\frac{i \hat{p} a}{\hbar}}\bigg]
\end{align}

$\Rightarrow$

\begin{align}
 \mathscr{U}_{10}(t, \tau) = \sum_x - \frac{i a}{\hbar}
 \big[ 2 |F_2(x,t,0)|^2 G^*_1(x,t,0) F_1(x,t,0) - F^*_2(x,t,0) G_2(x,t,0) [G_1^*(x,t,0)]^2 
 + F_2(x,t,0) [ F_1(x,t,0)]^2 G^*_2(x,t,0)  \big] \ket{x}\bra{x}~\hat{p} \nonumber\\
 + \sum_x \bigg\{ -  a \partial_x F_2(x,t,0) \Big[ F_1(x,t,0) F_2^*(x,t,0) G_1^*(x,t,0) + [F_1(x,t,0)]^2 G_2^*(x,t,0)  \Big] \nonumber\\
  - a \partial_x F_1(x,t,0) \Big[ |F_2(x,t,0)|^2 G_1^*(x,t,0) + F_2(x,t,0) F_1(x,t,0) G_2^*(x,t,0)  \Big] \nonumber\\
  + a \partial_x G_2(x,t,0) \Big[ F_2^*(x,t,0) [G_1^*(x,t,0)]^2 + G_1^*(x,t,0) G_2^*(x,t,0) F_1(x,t,0) \Big] \nonumber\\
   - a \partial_x G_1^*(x,t,0) \Big[ |F_2(x,t,0)|^2 F_1(x,t,0) - F_2^*(x,t,0) G_2(x,t,0) G_1^*(x,t,0) \Big] \nonumber\\
   + \tau \Big[f_1(x,t,0) G_1^*(x,t,0) - g_1^*(x,t,0) F_1(x,t,0) \Big] 
   + \tau f_2(x,t,0) \Big[F_1(x,t,0) F_2^*(x,t,0) G_1^*(x,t,0) + [F_1(x,t,0)]^2 G_2^*(x,t,0)  \Big] \nonumber\\
   - \tau g_2(x,t,0) \Big[ F_2^*(x,t,0) [G_1^*(x,t,0)]^2 + F_1(x,t,0) G_1^*(x,t,0) G_2^*(x,t,0) \Big] \nonumber\\
  - \tau g_2^*(x,t,0) \Big[ F_2(x,t,0) [F_1(x,t,0)]^2 - F_1(x,t,0) G_2(x,t,0) G_1^*(x,t,0) \Big] \nonumber\\
  - \tau f_2^*(x,t,0) \Big[ G_1^*(x,t,0) F_1(x,t,0) F_2(x,t,0) - G_2(x,t,0) [G_1^*(x,t,0)]^2  \Big]\nonumber\\
   - i a \partial_x \xi_1(x,t,0) \Big[ 2 |F_2(x,t,0)|^2 F_1(x,t,0) G_1^*(x,t,0) + [F_1(x,t,0)]^2 F_2(x,t,0) G_2^*(x,t,0)
   - F_2^*(x,t,0) [G_1^*(x,t,0)]^2 G_2(x,t,0) \Big] \nonumber\\
   - i a \partial_x \xi_2(x,t,0) \Big [ |F_2(x,t,0)|^2 F_1(x,t,0) G_1^*(x,t,0) 
   - |G_2(x,t,0)|^2 F_1(x,t,0) G_1^*(x,t,0) + [F_1(x,t,0)]^2 F_2(x,t,0) G_2^*(x,t,0)\nonumber\\
   - G_2(x,t,0) [G_1^*(x,t,0)]^2 F_2^*(x,t,0) \Big] \bigg\} \ket{x}\bra{x} + \mathcal{O}(\tau^2),
\end{align}

where we have used the definition \begin{align}
 F_j(x,t, \tau) =  F_j(x,t,0) + \tau f_j(x,t) + \mathcal{O}(\tau^2),
 G_j(x,t, \tau) =   G_j(x,t,0) + \tau g_j(x,t) + \mathcal{O}(\tau^2),
 \xi_j(x,t,\tau) =  \xi_j(x,t,0) + \tau \lambda_j(x,t) + \mathcal{O}(\tau^2),
 \end{align}

\subsection{the second-row second-column term of our modified Evolution operator in coin-basis}\label{evoluform11}

\begin{align}
 \mathscr{U}_{11}(t, \tau) = U_{01}^\dagger(t,0) U_{01}(t, \tau) + U_{11}^\dagger(t, 0) U_{11}(t, \tau) \nonumber\\
 = \sum_x  e^{-i [\xi_1(x,t,0) + \xi_2(x,t,0)]} \bigg[ F^*_2(x,t,0) G^*_1(x,t,0) +  G^*_2(x,t,0)F_1(x,t,0)\bigg] \nonumber\\
\bigg[ e^{i [\xi_1(x-a,t,\tau) + \xi_2(x-a,t,\tau)]} F_2(x-a,t,\tau) G_1(x-a,t,\tau) 
\ket{x}\bra{x} e^{\frac{-i\hat{p}a}{\hbar}} 
+ e^{i [\xi_1(x,t,\tau) + \xi_2(x-a,t,\tau)]} G_2(x-a,t,\tau)F_1^*(x,t,\tau) \ket{x}\bra{x}\bigg] \nonumber\\
+ \sum_x  e^{-i [\xi_1(x,t,0) + \xi_2(x,t,0)]} \bigg[ - G_2(x,t,0)G^*_1(x,t,0) +  F_2(x,t,0)F_1(x,t,0) \bigg] \nonumber\\
\bigg[ - e^{i [\xi_1(x,t,\tau) + \xi_2(x,t,\tau)]} G_2^*(x,t,\tau)G_1(x,t,\tau) \ket{x}\bra{x}
+ e^{i [\xi_1(x+a,t,\tau) + \xi_2(x,t,\tau)]} F^*_2(x,t,\tau)F^*_1(x+a,t,\tau)\ket{x}\bra{x}e^{\frac{i \hat{p} a}{\hbar}} \bigg]
\end{align}

$\Rightarrow$

\begin{align}
 \mathscr{U}_{11}(t, \tau) - \sum_x \ket{x}\bra{x}~~~~~~~~~~~~~~~~~~~~~~~~~~~~~~~~\nonumber\\
 = \sum_x \frac{-i a}{\hbar} \big[ |F_2(x,t,0)|^2 |G_1(x,t,0)|^2 - |F_2(x,t,0)|^2 |F_1(x,t,0)|^2 
 + 2 \Re\{F_1(x,t,0) F_2(x,t,0) G_1(x,t,0) G^*_2(x,t,0) \} \big] \ket{x}\bra{x}~\hat{p} \nonumber\\
+ \sum_x \bigg\{ - a \partial_x G_2(x,t,0) \Big[ F_1^*(x,t,0) F_2^*(x,t,0) G_1^*(x,t,0) 
+ |F_1(x,t,0)|^2 G_2^*(x,t,0) \Big] \nonumber\\
 - a \partial_x F_2(x,t,0) \Big[ F_2^*(x,t,0) |G_1(x,t,0)|^2 
 + G_1(x,t,0) F_1(x,t,0) G_2^*(x,t,0) \Big] \nonumber\\
 - a \partial_x G_1(x,t,0) \Big[ |F_2(x,t,0)|^2 G_1^*(x,t,0) 
 + F_1(x,t,0) F_2(x,t,0) G_2^*(x,t,0) \Big] \nonumber\\
 + a \partial_x F_1^*(x,t,0) \Big[ F_1(x,t,0) |F_2(x,t,0)|^2 - F_2^*(x,t,0) G_2(x,t,0) G_1^*(x,t,0) \Big]\nonumber\\
 + \tau \Big[ g_1(x,t,0) G_1^*(x,t,0) + F_1(x,t,0) f_1^*(x,t,0) 
 + g^*_2(x,t,0) G_2(x,t,0) + f_2(x,t,0) F_2^*(x,t,0) \Big] \nonumber\\
 + 2 i \tau \Im \Big[ g_2(x,t,0) F_1^*(x,t,0) F_2^*(x,t,0) G_1^*(x,t,0) 
 + g_2(x,t,0) G_2^*(x,t,0) |F_1(x,t,0)|^2 + f_2(x,t,0) G_1(x,t,0) F_1(x,t,0) G_2^*(x,t,0)\nonumber\\
 - f_2(x,t,0) F_2^*(x,t,0) |F_1(x,t,0)|^2 \Big]
 + i \tau [\lambda_1(x,t,0) + \lambda_2(x,t,0)] \nonumber\\
 - i a \partial_x \xi_1(x,t,0) \Big[ |F_2(x,t,0)|^2 |G_1(x,t,0)|^2 - |F_2(x,t,0)|^2 |F_1(x,t,0)|^2 
 + 2 \Re[F_1(x,t,0) F_2(x,t,0) G_1(x,t,0) G_2^*(x,t,0)] \Big]  \nonumber\\
 - i a \partial_x \xi_2(x,t,0) \Big[|F_2(x,t,0)|^2 |G_1(x,t,0)|^2 + |G_2(x,t,0)|^2 |F_1(x,t,0)|^2 
 +   2 \Re[F_1(x,t,0) F_2(x,t,0) G_1(x,t,0) G_2^*(x,t,0)]  \Big] \bigg\} \ket{x}\bra{x} + \mathcal{O}(\tau^2), 
\end{align}

where we have used the definition \begin{align}
 F_j(x,t, \tau) =  F_j(x,t,0) + \tau f_j(x,t) + \mathcal{O}(\tau^2),
 G_j(x,t, \tau) =   G_j(x,t,0) + \tau g_j(x,t) + \mathcal{O}(\tau^2),
 \xi_j(x,t,\tau) =  \xi_j(x,t,0) + \tau \lambda_j(x,t) + \mathcal{O}(\tau^2).
 \end{align}

\section{Calculating the operator terms of the effective Hamiltonian for the single particle}

From the previous Sections \ref{evoluform00}-\ref{evoluform11} we get 

\begin{align}
 \mathscr{U}_{00}(t, \tau) + \mathscr{U}_{11}(t, \tau) - 2\sum_x \ket{x}\bra{x} = \hspace{5cm} \nonumber\\
 \sum_x \bigg\{ 2 i \tau [\lambda_1(x,t,0) + \lambda_2(x,t,0)] - a i \Im\Big[F_2^*(x,t,0)\partial_x F_2(x,t,0) +  G_2^*(x,t,0)\partial_x G_2(x,t,0) \Big] \nonumber\\
 + 2 i a |F_2(x,t,0)|^2  \Im \Big[F_1(x,t,0) \partial_x F_1^*(x,t,0) + G_1(x,t,0) \partial_x G_1^*(x,t,0) \Big]
 + 2 i a \Im \Big[ F_1^*(x,t,0) F_2^*(x,t,0) G_2(x,t,0) \partial_x G_1^*(x,t,0) \nonumber\\
 + F_2(x,t,0) G_1(x,t,0) G_2^*(x,t,0) \partial_x F_1(x,t,0)  \Big] -  i a \partial_x \xi_2(x,t,0) \bigg\} \ket{x}\bra{x} + \mathcal{O}(\tau^2).
\end{align}

\begin{align}
 \mathscr{U}_{00}(t, \tau) - \mathscr{U}_{11}(t, \tau)  = \hspace{7cm} \nonumber\\
 - \frac{ 2 i a}{\hbar}\sum_x \Big[ |F_2(x,t,0)|^2 |F_1(x,t,0)|^2 - |F_2(x,t,0)|^2 |G_1(x,t,0)|^2 
 - 2 \Re\{G_2^*(x,t,0) F_1(x,t,0) F_2(x,t,0) G_1(x,t,0)\} \Big] \ket{x}\bra{x}~\hat{p} \nonumber\\
 + \sum_x  \bigg\{ 2 i \tau \Im \Big[ F_1^*(x,t,0) f_1(x,t,0) + g_1^*(x,t,0) G_1(x,t,0) 
 + G_2^*(x,t,0) g_2(x,t,0) + F_2(x,t,0) f_2^*(x,t,0) \Big] \nonumber\\
 + 4 i \tau \Im\Big[ f_2(x,t,0) F_2^*(x,t,0) |F_1(x,t,0)|^2 - f_2(x,t,0) G_1(x,t,0) F_1(x,t,0) G_2^*(x,t,0) \nonumber\\
 + g_2^*(x,t,0) G_2(x,t,0) |F_1(x,t,0)|^2 + g_2^*(x,t,0) F_2(x,t,0) F_1(x,t,0) G_1(x,t,0)\Big] \nonumber\\
 + a \partial_x F_2(x,t,0) \Big[ 2 F_1(x,t,0) G_1(x,t,0) G_2^*(x,t,0) + F_2^*(x,t,0) |G_1(x,t,0)|^2 
 - |F_1(x,t,0)|^2 F_2^*(x,t,0) \Big]\nonumber\\
 + 2 a |F_2(x,t,0)|^2 \Re \Big[ G_1(x,t,0) \partial_x G_1^*(x,t,0) - F_1(x,t,0) \partial_x F_1^*(x,t,0) \Big] \nonumber\\
 + a \partial_x G_2(x,t,0) \Big[2 F_2^*(x,t,0) F_1^*(x,t,0) G_1^*(x,t,0)  + |F_1(x,t,0)|^2 G_2^*(x,t,0)
 - |G_1(x,t,0)|^2 G_2^*(x,t,0)\Big] \nonumber\\
 + 2 a \Re \Big[ F_2(x,t,0) G_1(x,t,0) G_2^*(x,t,0) \partial_x F_1(x,t,0) 
 + F_1(x,t,0) F_2(x,t,0) G_2^*(x,t,0) \partial_x G_1(x,t,0) \Big]\nonumber\\
  -  i a \partial_x \xi_1(x,t,0)  \Big[ 2 |F_2(x,t,0)|^2 \big( |F_1(x,t,0)|^2 - |G_1(x,t,0)|^2 \big) 
  -  4 \Re[F_2(x,t,0) F_1(x,t,0) G_1(x,t,0) G_2^*(x,t,0)]\Big]\nonumber\\
   - i a \partial_x \xi_2(x,t,0) \Big[ \big(|G_2(x,t,0)|^2 - |F_2(x,t,0)|^2 \big) \big(|G_1(x,t,0)|^2  
   - |F_1(x,t,0)|^2\big) \nonumber\\
   - 4 \Re [F_2(x,t,0) F_1(x,t,0) G_1(x,t,0) G_2^*(x,t,0)] \Big] \bigg\} \ket{x}\bra{x} + \mathcal{O}(\tau^2).
\end{align}

\begin{align}
 \mathscr{U}_{01}(t, \tau) + \mathscr{U}_{10}(t, \tau) = \hspace{7cm} \nonumber\\
 \sum_x - \frac{2 i a}{\hbar}
\Re \Big[ 2 |F_2(x,t,0)|^2 G_1(x,t,0) F_1^*(x,t,0) 
- F_2(x,t,0) G_2^*(x,t,0) [G_1(x,t,0)]^2 + F_2^*(x,t,0) [ F_1^*(x,t,0)]^2 G_2(x,t,0)  \Big] \ket{x}\bra{x}~\hat{p} \nonumber\\
 + \sum_x \bigg\{ 2 i \tau \Im \Big[ g_1(x,t,0) F_1^*(x,t,0) - f_1^*(x,t,0) G_1(x,t,0)  \Big]  \nonumber\\
 +  2 i \tau \Im  \Big[g_2^*(x,t,0) F_2(x,t,0) [G_1(x,t,0)]^2 + g_2^*(x,t,0) F_1^*(x,t,0) G_1(x,t,0) G_2(x,t,0)\nonumber\\
 -  f_2^*(x,t,0) G_2(x,t,0) [F_1^*(x,t,0)]^2 -  f_2^*(x,t,0) G_1(x,t,0) F_2(x,t,0) F_1^*(x,t,0) \nonumber\\
 +  g_2(x,t,0) [F_1^*(x,t,0)]^2 F_2^*(x,t,0) - g_2(x,t,0) F_1^*(x,t,0) G_1(x,t,0) G_2^*(x,t,0) \nonumber\\
 +  f_2(x,t,0) G_1(x,t,0) F_1^*(x,t,0) F_2^*(x,t,0) 
 - f_2(x,t,0) [G_1(x,t,0)]^2 G_2^*(x,t,0) \Big]  \nonumber\\
 - 2 a \Re \Big[ F_1^*(x,t,0) |F_2(x,t,0)|^2 \partial_x G_1(x,t,0) - G_2^*(x,t,0) G_1(x,t,0) F_2(x,t,0) \partial_x G_1(x,t,0) 
 + F_1^*(x,t,0) F_2^*(x,t,0) G_2(x,t,0) \partial_x F_1^*(x,t,0) \nonumber\\
 + G_1(x,t,0) |F_2(x,t,0)|^2 \partial_x F_1^*(x,t,0) \Big] - 2 a  \Re \Big[ F_1^*(x,t,0) G_1(x,t,0)\Big] 
 \Big[ F_2^*(x,t,0) \partial_x F_2(x,t,0) - G_2^*(x,t,0) \partial_x G_2(x,t,0) \Big] \nonumber\\
  - a G_2^*(x,t,0) \partial_x F_2(x,t,0) \big( [F_1(x,t,0)]^2 - [G_1(x,t,0)]^2 \big)
  -  a F_2^*(x,t,0) \partial_x G_2(x,t,0) \big( [F_1^*(x,t,0)]^2 - [G_1^*(x,t,0)]^2 \big) \nonumber\\
  - i a \partial_x \xi_1(x,t,0) \Big[ 4 |F_2(x,t,0)|^2 \Re\big(G_1(x,t,0) F_1^*(x,t,0)\big)
  - 2 \Re\big(G_2^*(x,t,0) F_2(x,t,0) [G_1(x,t,0)]^2 \big) 
  + 2 \Re\big( G_2 F_2^* [F_1^*]^2 \big) \Big] \nonumber\\
  - i a \partial_x \xi_2(x,t,0) \Big[ 2 |F_2(x,t,0)|^2 \Re\big( G_1(x,t,0) F_1^*(x,t,0)\big) - 2 |G_2(x,t,0)|^2 \Re\big( G_1(x,t,0) F_1^*(x,t,0)\big) \nonumber\\
  + 2 \Re \big( [F_1(x,t,0)]^2 F_2(x,t,0) G_2^*(x,t,0) \big) - 2 \Re\big( F_2(x,t,0) G_2^*(x,t,0) [G_1(x,t,0)]^2 \big) \Big] \bigg\} \ket{x}\bra{x} + \mathcal{O}(\tau^2).
\end{align}

\begin{align}
 \mathscr{U}_{01}(t, \tau) - \mathscr{U}_{10}(t, \tau) = \hspace{7cm} \nonumber\\
 \frac{2 a}{\hbar} \sum_x 
\Im \Big[ 2 |F_2(x,t,0)|^2 G_1(x,t,0) F_1^*(x,t,0) - F_2(x,t,0) G_2^*(x,t,0) [G_1(x,t,0)]^2
+ F_2^*(x,t,0) [ F_1^*(x,t,0)]^2 G_2(x,t,0)  \Big] \ket{x}\bra{x}~\hat{p} \nonumber\\
 + \sum_x \bigg\{ - 2 i a \Im\Big[ \partial_x G_1(x,t,0) \big(|F_2(x,t,0)|^2 F_1^*(x,t,0)
 - G_2^*(x,t,0) G_1(x,t,0) F_2(x,t,0) \big)\Big] \nonumber\\
- 2 i a \Im \Big[ \partial_x F_1^*(x,t,0) \big( G_2(x,t,0) F_1^*(x,t,0) F_2^*(x,t,0) + G_1(x,t,0) |F_2(x,t,0)|^2 \big) \Big]
 - a \partial_x F_2(x,t,0) \Big[2 i F_2^*(x,t,0) \Im\big(F_1^*(x,t,0) G_1(x,t,0) \big) \nonumber\\
 - G_2^*(x,t,0)  [F_1(x,t,0)]^2 -  G_2^*(x,t,0) [G_1(x,t,0)]^2  \Big]  
  - a \partial_x G_2(x,t,0) \Big[ - 2 i G_2^*(x,t,0) \Im \big(F_1^*(x,t,0) G_1(x,t,0) \big) \nonumber\\
  + F_2^*(x,t,0) \big( [F_1^*(x,t,0)]^2 + [G_1^*(x,t,0)]^2 \big) \Big]
 + 2 \tau \Re \Big[ g_1(x,t,0) F_1^*(x,t,0) - f_1^*(x,t,0) G_1(x,t,0)  \nonumber\\
 + g_2^*(x,t,0) F_2(x,t,0) [G_1(x,t,0)]^2 + g_2^*(x,t,0) F_1^*(x,t,0) G_1(x,t,0) G_2(x,t,0)
 - f_2^*(x,t,0) G_2(x,t,0) [F_1^*(x,t,0)]^2 \nonumber\\
 - f_2^*(x,t,0) G_1(x,t,0) F_2(x,t,0) F_1^*(x,t,0) 
 + g_2(x,t,0) F_2^*(x,t,0) [F_1^*(x,t,0)]^2  \nonumber\\
 -  g_2(x,t,0) F_1^*(x,t,0) G_1(x,t,0) G_2^*(x,t,0) 
 + f_2(x,t,0) G_1(x,t,0) F_1^*(x,t,0) F_2^*(x,t,0) 
 - f_2(x,t,0) [G_1(x,t,0)]^2 G_2^*(x,t,0) \Big] \nonumber\\
 + 2 a  \partial_x \xi_1(x,t,0) \Big[2 |F_2(x,t,0)|^2 \Im \big(G_1(x,t,0) F_1^*(x,t,0)\big)
 + \Im \big( G_2(x,t,0) F_2^*(x,t,0) [F^*_1(x,t,0)]^2 - G_2^*(x,t,0) F_2(x,t,0) [G_1(x,t,0)]^2 \big) \Big] \nonumber\\
 + 2 a \partial_x \xi_2(x,t,0) \Big[ \big( |F_2(x,t,0)|^2 - |G_2(x,t,0)|^2 \big) \Im \big(G_1(x,t,0) F_1^*(x,t,0) \big)  
 + \Im \big( G_2(x,t,0) F_2^*(x,t,0) [F_1^*(x,t,0)]^2 \nonumber\\
 - F_2(x,t,0) G_2^*(x,t,0) [G_1(x,t,0)]^2 \big) \Big] \bigg\} \ket{x}\bra{x} + \mathcal{O}(\tau^2).
\end{align}

\subsection{Effective Hamiltonian}\label{dervham}

Using the definition
\begin{align}
\mathscr{U}(t, \tau)  =  \exp \bigg( - i \frac{H_{\text{eff}}(t)\tau}{\hbar} \bigg)
\end{align}
we can write 

\begin{align}
 H_{\text{eff}} = i \hbar \lim_{\tau \to 0} \frac{1}{\tau} \left( \begin{array}{cc}
                                      \mathscr{U}_{00}(t, \tau) - \sum_x \ket{x}\bra{x} & \mathscr{U}_{01}(t, \tau) \\ \\
                                       \mathscr{U}_{10}(t, \tau)  & \mathscr{U}_{11}(t, \tau) - \sum_x \ket{x}\bra{x} \\
                                                                  \end{array}\right) 
   = \sigma_0 \otimes i \hbar \lim_{\tau \to 0} \frac{1}{2\tau} \Big(  \mathscr{U}_{00}(t, \tau) + \mathscr{U}_{11} (t, \tau)
   - 2\sum_x \ket{x}\bra{x}\Big)\nonumber\\
   + \sigma_3 \otimes i \hbar \lim_{\tau \to 0} \frac{1}{2\tau} \Big(  \mathscr{U}_{00}(t, \tau) - \mathscr{U}_{11}(t, \tau) \Big) 
   + \sigma_1 \otimes i \hbar \lim_{\tau \to 0} \frac{1}{2\tau} \Big(  \mathscr{U}_{01}(t, \tau) + \mathscr{U}_{10}(t, \tau) \Big)
   - \sigma_2 \otimes  \hbar \lim_{\tau \to 0} \frac{1}{2\tau} \Big(  \mathscr{U}_{01}(t, \tau) - \mathscr{U}_{10}(t, \tau) \Big) \nonumber\\
   \coloneqq \sum_{r=0}^3  \sigma_r \otimes \sum_x \Xi_r(x,t) \ket{x}\bra{x} + 
   c~\sum_{r=1}^3  \sigma_r \otimes \sum_x \Theta_r(x,t) \ket{x}\bra{x}~ \hat{p} .~~~~~~~~~~~~~~~~~~~~~~~
                                                               \end{align}

Here for notational convenience we will omit the arguement $(x,t,0)$ from all the functions $F_j(x,t,0)$, $G_j(x,t,0)$, $\xi_j(x,t,0)$, and 
$(x,t)$ from $\lambda_j(x,t)$, $f_j(x,t)$, $g_j(x,t)$ and will be represented as $F_j$, $G_j$, $\xi_j$, $\lambda_j$, $f_j$, $g_j$ respectively. 
Then the operator terms of this effective Hamiltonian can be written as follows

\begin{align}
 \Xi_0(x,t) =  \hspace{10cm} \nonumber\\ 
 - \hbar [\lambda_1 + \lambda_2] + \frac{\hbar c}{2} \Im\Big[F_2^*\partial_x F_2 +  G_2^*\partial_x G_2 \Big] 
 - \hbar c |F_2|^2  \Im \Big[F_1 \partial_x F_1^* + G_1 \partial_x G_1^* \Big]
 - \hbar c \Im \Big[ F_1^* F_2^* G_2 \partial_x G_1^* + F_2 G_1 G_2^* \partial_x F_1  \Big] + \frac{\hbar c}{2} \partial_x \xi_2,
\end{align}
\begin{align}
 \Xi_3(x,t)  = \hspace{10cm} \nonumber\\
 - \hbar \Im \Big[ F_1^* f_1 + g_1^* G_1 + G_2^* g_2 + F_2 f_2^* \Big] 
 - 2 \hbar \Im\Big[ f_2 F_2^* |F_1|^2 - f_2 G_1 F_1 G_2^* + g_2^* G_2 |F_1|^2 + g_2^* F_2 F_1 G_1\Big]\nonumber\\
 + \frac{i \hbar c}{2} \partial_x F_2 \Big[ 2 F_1 G_1 G_2^* + F_2^* |G_1|^2 - |F_1|^2 F_2^* \Big]
 + i \hbar c |F_2|^2 \Re \Big[ G_1 \partial_x G_1^* - F_1 \partial_x F_1^* \Big] \nonumber\\
 + \frac{i \hbar c}{2} \partial_x G_2 \Big[2 F_2^* F_1^* G_1^*  + |F_1|^2 G_2^* - |G_1|^2 G_2^*\Big]
 + i \hbar c \Re \Big[ F_2 G_1 G_2^* \partial_x F_1 + F_1 F_2 G_2^* \partial_x G_1 \Big]\nonumber\\
  + \frac{\hbar c}{2} \partial_x \xi_1  \Big[ 2 |F_2|^2 \big( |F_1|^2 - |G_1|^2 \big) -  4 \Re[F_2 F_1 G_1 G_2^*]\Big]
  + \frac{\hbar c}{2}\partial_x \xi_2 \Big[ \big(|G_2|^2 - |F_2|^2 \big) \big(|G_1|^2 - |F_1|^2\big) - 4 \Re [F_2 F_1 G_1 G_2^*] \Big],
\end{align}

\begin{align}
 \Xi_1(x,t) =  \hspace{10cm} \nonumber\\
 - \hbar \Im  \Big[g_2^* F_2 [G_1]^2 + g_2^* F_1^* G_1 G_2 -  f_2^*G_2 [F_1^*]^2 -  f_2^* G_1 F_2 F_1^* 
 +  g_2[F_1^*]^2 F_2^* - g_2 F_1^* G_1 G_2^* +  f_2 G_1 F_1^* F_2^* - f_2 [G_1]^2 G_2^* \Big]  \nonumber\\
 - i \hbar c \Re \Big[ F_1^* |F_2|^2 \partial_x G_1 - G_2^* G_1 F_2 \partial_x G_1 + F_1^* F_2^* G_2 \partial_x F_1^*
 + G_1 |F_2|^2 \partial_x F_1^* \Big] - i \hbar c  \Re \Big[ F_1^* G_1\Big] \Big[ F_2^*\partial_x F_2 - G_2^* \partial_x G_2 \Big] \nonumber\\
  - \frac{i \hbar c}{2} G_2^* \partial_x F_2 \big( [F_1]^2 - [G_1]^2 \big) -  \frac{i \hbar c}{2} F_2^* \partial_x G_2 \big( [F_1^*]^2 - [G_1^*]^2 \big)
  + \frac{\hbar c}{2} \partial_x \xi_1 \Big[ 4 |F_2|^2 \Re\big(G_1 F_1^*\big) - 2 \Re\big(G_2^* F_2 [G_1]^2 \big) 
  + 2 \Re\big( G_2 F_2^* [F_1^*]^2 \big) \Big] \nonumber\\
  + \frac{\hbar c}{2} \partial_x \xi_2 \Big[ 2 |F_2|^2 \Re\big( G_1 F_1^*\big) - 2 |G_2|^2 \Re\big( G_1 F_1^*\big)
  + 2 \Re \big( [F_1]^2 F_2 G_2^* \big) - 2 \Re\big( F_2 G_2^* [G_1]^2 \big) \Big]- \hbar \Im \Big[ g_1 F_1^* - f_1^* G_1  \Big], 
\end{align}

\begin{align}
 \Xi_2(x,t) =  \hspace{10cm} \nonumber\\
 i \hbar c \Im\Big[ \partial_x G_1 \big(|F_2|^2 F_1^* - G_2^* G_1 F_2 \big)\Big] 
+ i \hbar c \Im \Big[ \partial_x F_1^* \big( G_2 F_1^* F_2^* + G_1 |F_2|^2 \big) \Big]
 + \frac{\hbar c}{2} \partial_x F_2 \Big[2 i F_2^* \Im\big(F_1^* G_1 \big) - G_2^*  [F_1]^2 -  G_2^* [G_1]^2  \Big] \nonumber\\
 + \frac{\hbar c}{2} \partial_x G_2 \Big[ - 2 i G_2^* \Im \big(F_1^* G_1 \big) + F_2^* \big( [F_1^*]^2 + [G_1^*]^2 \big) \Big]
 - \hbar  \Re \Big[ g_1 F_1^* - f_1^* G_1 + g_2^* F_2 [G_1]^2 + g_2^* F_1^* G_1 G_2 - f_2^* G_2 [F_1^*]^2 \nonumber\\
 - f_2^* G_1 F_2 F_1^* + g_2 F_2^* [F_1^*]^2  -  g_2 F_1^* G_1 G_2^* + f_2 G_1 F_1^* F_2^* - f_2 [G_1]^2 G_2^* \Big] 
 - \hbar c  \partial_x \xi_1 \Big[2 |F_2|^2 \Im \big(G_1 F_1^*\big) + \Im \big( G_2 F_2^* [F_1^*]^2 - G_2^* F_2 [G_1]^2 \big) \Big] \nonumber\\
 - \hbar c \partial_x \xi_2 \Big[ \big( |F_2|^2 - |G_2|^2 \big) \Im \big(G_1 F_1^* \big) + \Im \big( G_2 F_2^* [F_1^*]^2 - F_2 G_2^* [G_1]^2 \big) \Big],
\end{align}

\begin{align}
 \Theta_3(x,t) = -   \Big[ |F_2|^2 |G_1|^2 - |F_2|^2 |F_1|^2 
 + 2 \Re\{F_1 F_2 G_1 G^*_2 \} \Big], \nonumber\\
 \Theta_1(x,t) =   \Re\Big[ 2 |F_2|^2 G^*_1 F_1 - F^*_2 G_2 [G_1^*]^2 + F_2 [ F_1]^2 G^*_2  \Big], \nonumber\\
 \Theta_2(x,t) =   \Im \Big[ 2 |F_2|^2 G^*_1 F_1 - F^*_2 G_2 [G_1^*]^2 + F_2 [ F_1]^2 G^*_2  \Big] .
\end{align}

\section{Calculating the modified Evolution Operator for the two-particle case}

Here we will calculate the modified evolution operator $\mathscr{U}^\text{two}(t, \tau)$ in section \ref{twopar} parts by parts.

\subsection{The matrix elements of the operator 
$\big[S^1_+ \otimes S^2_+ \big] \cdot C_2(t, \tau)$ in coin-basis}

In coin basis the matrix elements of the operator $\big[S^1_+ \otimes S^2_+ \big] \cdot C_j(t, \tau)$
(the suffixes ``$i, j$'' used in the next calculations to denote the row, column numbers respectively of matrix in coin-basis )---

\begin{align}
  \bigg( \big[S^1_+ \otimes S^2_+ \big] \cdot C_2(t, \tau)\bigg)_{i = 1, j = 1} =  
  \sum_{x_1, x_2}  \cos[\theta_2(x_1, x_2, t, \tau)]  \ket{x_1 + a, x_2 + a}\bra{x_1, x_2} ,\nonumber\\
  \bigg( \big[S^1_+ \otimes S^2_+ \big] \cdot C_2(t, \tau)\bigg)_{i = 1, j = 4} =  
  - i \sum_{x_1, x_2}  \sin[\theta_2(x_1, x_2, t, \tau)]  \ket{x_1 + a, x_2 + a}\bra{x_1, x_2}, \nonumber\\
   \bigg( \big[S^1_+ \otimes S^2_+ \big] \cdot C_2(t, \tau)\bigg)_{i = 2, j = 2} =
   \sum_{x_1, x_2}  \cos[\theta_2(x_1, x_2, t, \tau)] \ket{x_1 + a, x_2}\bra{x_1, x_2} , \nonumber\\
   \bigg( \big[S^1_+ \otimes S^2_+ \big] \cdot C_2(t, \tau)\bigg)_{i = 2, j = 3} = 
 - i \sum_{x_1, x_2}  \sin[\theta_2(x_1, x_2, t, \tau)] \ket{x_1 + a, x_2}\bra{x_1, x_2} , \nonumber\\
    \bigg( \big[S^1_+ \otimes S^2_+ \big] \cdot C_2(t, \tau)\bigg)_{i = 3, j = 2} =
 - i  \sum_{x_1, x_2}  \sin[\theta_2(x_1, x_2, t, \tau)] \ket{x_1, x_2 + a}\bra{x_1, x_2} , \nonumber\\
  \bigg( \big[S^1_+ \otimes S^2_+ \big] \cdot C_2(t, \tau)\bigg)_{i = 3, j = 3} =
   \sum_{x_1, x_2}   \cos[\theta_2(x_1, x_2, t, \tau)] \ket{x_1, x_2 + a}\bra{x_1, x_2}, \nonumber\\
  \bigg( \big[S^1_+ \otimes S^2_+ \big] \cdot C_2(t, \tau)\bigg)_{i = 4, j = 1} =  
  - i \sum_{x_1, x_2}  \sin[\theta_2(x_1, x_2, t, \tau)]  \ket{x_1, x_2}\bra{x_1, x_2}, \nonumber\\
  \bigg( \big[S^1_+ \otimes S^2_+ \big] \cdot C_2(t, \tau)\bigg)_{i = 4, j = 4} =
  \sum_{x_1, x_2}   \cos[\theta_2(x_1, x_2, t, \tau)] \ket{x_1, x_2}\bra{x_1, x_2}.
 \end{align}

All the other matrix elements of the operator $\big[S^1_+ \otimes S^2_+ \big] \cdot C_2(t, \tau)$ zeros.

\subsection{The matrix elements of the operator 
$\big[S^1_- \otimes S^2_- \big] \cdot C_1(t, \tau)$ in coin-basis}

\begin{align}
  \bigg( \big[S^1_- \otimes S^2_- \big] \cdot C_1(t, \tau)\bigg)_{i = 1, j = 1} =  
 \sum_{x_1, x_2}  \cos[\theta_1(x_1, x_2, t, \tau)] \ket{x_1, x_2}\bra{x_1, x_2}, \nonumber\\
  \bigg( \big[S^1_- \otimes S^2_- \big] \cdot C_1(t, \tau)\bigg)_{i = 1, j = 4} =  
 - i \sum_{x_1, x_2}  \sin[\theta_1(x_1, x_2, t, \tau)] \ket{x_1, x_2}\bra{x_1, x_2}, \nonumber\\ 
    \bigg( \big[S^1_- \otimes S^2_- \big] \cdot C_1(t, \tau)\bigg)_{i = 2, j = 2} =
\sum_{x_1, x_2}   \cos[\theta_1(x_1, x_2, t, \tau)] \ket{x_1, x_2-a}\bra{x_1, x_2}, \nonumber\\ 
   \bigg( \big[S^1_- \otimes S^2_- \big] \cdot C_1(t, \tau)\bigg)_{i = 2, j = 3} = 
  - i \sum_{x_1, x_2} \sin[\theta_1(x_1, x_2, t, \tau)] \ket{x_1, x_2-a}\bra{x_1, x_2} , \nonumber\\
    \bigg( \big[S^1_- \otimes S^2_- \big] \cdot C_1(t, \tau)\bigg)_{i = 3, j = 2} =
  - i \sum_{x_1, x_2} \sin[\theta_1(x_1, x_2, t, \tau)] \ket{x_1-a, x_2}\bra{x_1, x_2},  \nonumber\\
  \bigg( \big[S^1_- \otimes S^2_- \big] \cdot C_1(t, \tau)\bigg)_{i = 3, j = 3} =
   \sum_{x_1, x_2}  \cos[\theta_1(x_1, x_2, t, \tau)] \ket{x_1-a, x_2}\bra{x_1, x_2}, \nonumber\\
   \bigg( \big[S^1_- \otimes S^2_- \big] \cdot C_1(t, \tau)\bigg)_{i = 4, j = 1} =
 - i \sum_{x_1, x_2} \sin[\theta_1(x_1, x_2, t, \tau)] \ket{x_1-a, x_2-a}\bra{x_1, x_2}, \nonumber\\
  \bigg( \big[S^1_- \otimes S^2_- \big] \cdot C_1(t, \tau)\bigg)_{i = 4, j = 4} =
 \sum_{x_1, x_2} \cos[\theta_1(x_1, x_2, t, \tau)] \ket{x_1-a, x_2-a}\bra{x_1, x_2}.
 \end{align}

All the other matrix elements of the operator $\big[S^1_- \otimes S^2_- \big] \cdot C_1(t, \tau)$ are zeros.

\subsection{The matrix elements of the operator 
$\big[S^1_+ \otimes S^2_+ \big] \cdot C_2(t, \tau) \cdot \big[S^1_- \otimes S^2_- \big] \cdot C_1(t, \tau)$ in coin-basis}

 \subsubsection{\bf First row first column term}
 
\begin{align}
 \bigg( \big[S^1_+ \otimes S^2_+ \big] \cdot C_2(t, \tau) \cdot \big[S^1_- \otimes S^2_- \big] \cdot C_1(t, \tau)\bigg)_{i=1,j=1} 
 = \sum_{x_1, x_2}  \cos[\theta_2(x_1, x_2, t, \tau)] \cos[\theta_1(x_1, x_2, t, \tau)] \ket{x_1 + a, x_2 + a}\bra{x_1, x_2} \nonumber\\
 - \sin[\theta_2(x_1-a, x_2-a, t, \tau)] \sin[\theta_1(x_1, x_2, t, \tau)] \ket{x_1, x_2}\bra{x_1, x_2} \nonumber\\
  = \sum_{x_1, x_2}  \cos[\theta_2(x_1-a, x_2-a, t, \tau)] \cos[\theta_1(x_1-a, x_2-a, t, \tau)] \ket{x_1, x_2}\bra{x_1, x_2}
  e^{-\frac{i (\hat{p}_1 \otimes \mathbb{I}_2 +  \mathbb{I}_1 \otimes \hat{p}_2)a}{\hbar}}\nonumber\\
 - \sin[\theta_2(x_1-a, x_2-a, t, \tau)] \sin[\theta_1(x_1, x_2, t, \tau)] \ket{x_1, x_2}\bra{x_1, x_2} 
  \end{align}

  \begin{align}
  = - \frac{i a}{\hbar} \sum_{x_1, x_2}  \cos[\theta_2(x_1, x_2, t, 0)] \cos[\theta_1(x_1, x_2, t, 0)] 
  \ket{x_1, x_2}\bra{x_1, x_2}  (\hat{p}_1 \otimes \mathbb{I}_2 +  \mathbb{I}_1 \otimes \hat{p}_2) \nonumber\\
  +  \sum_{x_1, x_2} \Big\{ \cos[\theta_2(x_1, x_2, t, 0)] - \sin[\theta_2(x_1, x_2, t, 0)] \Big(- a \partial_{x_1} \theta_2(x_1, x_2, t, 0)
  - a \partial_{x_2} \theta_2(x_1, x_2, t, 0) + \tau \vartheta_2(x_1, x_2, t, 0)\Big) \Big\} \nonumber\\
  \Big\{ \cos[\theta_1(x_1, x_2, t, 0)] - \sin[\theta_1(x_1, x_2, t, 0)] \Big(-a \partial_{x_1} \theta_1(x_1, x_2, t, 0)
  - a \partial_{x_2} \theta_1(x_1, x_2, t, 0) + \tau \vartheta_1(x_1, x_2, t, 0)\Big) \Big\}
  \ket{x_1, x_2}\bra{x_1, x_2} \nonumber\\
  - \Big\{ \sin[\theta_2(x_1, x_2, t, 0)] + \cos[\theta_2(x_1, x_2, t, 0)] \Big( - a \partial_{x_1} \theta_2(x_1, x_2, t, 0)
  - a \partial_{x_2} \theta_2(x_1, x_2, t, 0) + \tau \vartheta_2(x_1, x_2, t, 0)\Big) \Big\}  \nonumber\\
  \Big\{ \sin[\theta_1(x_1, x_2, t, 0)] 
  + \tau \vartheta_1(x_1, x_2, t, 0)\cos[\theta_1(x_1, x_2, t, 0)]\Big\}\ket{x_1, x_2}\bra{x_1, x_2} + \mathcal{O}(\tau^2).
\end{align}
 
 \subsubsection{\bf First row fourth column term}
 
\begin{align}
 \bigg( \big[S^1_+ \otimes S^2_+ \big] \cdot C_2(t, \tau) \cdot \big[S^1_- \otimes S^2_- \big] \cdot C_1(t, \tau)\bigg)_{i=1,j=4} 
 = - i \sum_{x_1, x_2}  \cos[\theta_2(x_1, x_2, t, \tau)] \sin[\theta_1(x_1, x_2, t, \tau)] \ket{x_1 + a, x_2 + a}\bra{x_1, x_2} \nonumber\\
 + \sin[\theta_2(x_1-a, x_2-a, t, \tau)] \cos[\theta_1(x_1, x_2, t, \tau)] \ket{x_1, x_2}\bra{x_1, x_2} \nonumber\\
  = - i \sum_{x_1, x_2}  \cos[\theta_2(x_1-a, x_2-a, t, \tau)] \sin[\theta_1(x_1-a, x_2-a, t, \tau)] \ket{x_1, x_2}\bra{x_1, x_2}
  e^{-\frac{i (\hat{p}_1 \otimes \mathbb{I}_2 +  \mathbb{I}_1 \otimes \hat{p}_2)a}{\hbar}}\nonumber\\
 + \sin[\theta_2(x_1-a, x_2-a, t, \tau)] \cos[\theta_1(x_1, x_2, t, \tau)] \ket{x_1, x_2}\bra{x_1, x_2} 
  \end{align}
 
  \begin{align}
   = - \frac{a}{\hbar} \sum_{x_1, x_2} \cos[\theta_2(x_1, x_2, t, 0)] \sin[\theta_1(x_1, x_2, t, 0)] \ket{x_1, x_2}\bra{x_1, x_2}
   (\hat{p}_1 \otimes \mathbb{I}_2 +  \mathbb{I}_1 \otimes \hat{p}_2) \nonumber\\
   - i \sum_{x_1, x_2}  \Big\{ \cos[\theta_2(x_1, x_2, t, 0)] - \sin[\theta_2(x_1, x_2, t, 0)]
   \Big( - a \partial_{x_1}\theta_2(x_1, x_2, t, 0) 
   - a \partial_{x_1}\theta_2(x_1, x_2, t, 0) + \tau \vartheta_2(x_1, x_2, t, 0)\Big)\Big\}  \nonumber\\
  \Big\{ \sin[\theta_1(x_1, x_2, t, 0)] +\cos[\theta_1(x_1, x_2, t, 0)]
   \Big( - a \partial_{x_1}\theta_1(x_1, x_2, t, 0) 
   - a \partial_{x_1}\theta_1(x_1, x_2, t, 0) + \tau \vartheta_1(x_1, x_2, t, 0)\Big) \Big\}\ket{x_1, x_2}\bra{x_1, x_2} \nonumber\\
   + \Big\{ \sin[\theta_2(x_1, x_2, t, 0)] + \cos[\theta_2(x_1, x_2, t, 0)]
   \Big( - a \partial_{x_1}\theta_2(x_1, x_2, t, 0) 
   - a \partial_{x_1}\theta_2(x_1, x_2, t, 0) + \tau \vartheta_2(x_1, x_2, t, 0)\Big)\Big\} \nonumber\\
   \Big\{  \cos[\theta_1(x_1, x_2, t, 0)] - \tau \vartheta_1(x_1, x_2, t, 0) \sin[\theta_1(x_1, x_2, t, 0)] \Big\}\ket{x_1, x_2}\bra{x_1, x_2}
   + \mathcal{O}(\tau^2).
  \end{align}

 \subsubsection{\bf Second row second column term}
 
 \begin{align}
  \bigg( \big[S^1_+ \otimes S^2_+ \big] \cdot C_2(t, \tau) \cdot \big[S^1_- \otimes S^2_- \big] \cdot C_1(t, \tau)\bigg)_{i=2,j=2} \nonumber\\
 =  \sum_{x_1, x_2}  \cos[\theta_2(x_1, x_2 -a, t , \tau)] \cos[\theta_1(x_1, x_2, t , \tau)] \ket{x_1 + a, x_2-a}\bra{x_1, x_2} 
- \sin[\theta_2(x_1-a, x_2, t , \tau)] \sin[\theta_1(x_1, x_2, t , \tau)] \ket{x_1, x_2}\bra{x_1, x_2}\nonumber\\
  =  \sum_{x_1, x_2} \cos[\theta_2(x_1-a, x_2, t , \tau)] \cos[\theta_1(x_1-a, x_2+a, t , \tau)]
 \ket{x_1, x_2}\bra{x_1, x_2}e^{-\frac{i (\hat{p}_1 \otimes \mathbb{I}_2 -  \mathbb{I}_1 \otimes \hat{p}_2)a}{\hbar}}  \nonumber\\
 - \sin[\theta_2(x_1-a, x_2, t , \tau)] \sin[\theta_1(x_1, x_2, t , \tau)] \ket{x_1, x_2}\bra{x_1, x_2}
  \end{align}
 \begin{align}
 = -\frac{i a }{\hbar} \sum_{x_1, x_2}  \cos[\theta_2(x_1, x_2, t , 0)] \cos[\theta_1(x_1, x_2, t , 0)]
 \ket{x_1, x_2}\bra{x_1, x_2}(\hat{p}_1 \otimes \mathbb{I}_2 -  \mathbb{I}_1 \otimes \hat{p}_2)  \nonumber\\
 + \sum_{x_1, x_2} \Big\{\cos[\theta_2(x_1, x_2, t , 0)] - \sin[\theta_2(x_1, x_2, t , 0)] 
 \Big(- a  \partial_{x_1} \theta_2(x_1, x_2, t , 0) + \tau \vartheta_2(x_1, x_2, t, 0) \Big)  \Big\} \nonumber\\
 \Big\{\cos[\theta_1(x_1, x_2, t , 0)] - \sin[\theta_1(x_1, x_2, t , 0)] 
 \Big(- a  \partial_{x_1} \theta_1(x_1, x_2, t , 0) + a  \partial_{x_2} \theta_1(x_1, x_2, t , 0)
 + \tau \vartheta_1(x_1, x_2, t, 0) \Big)  \Big\}\ket{x_1, x_2}\bra{x_1, x_2} \nonumber\\
 - \Big\{ \sin[\theta_2(x_1, x_2, t, 0)] + \cos[\theta_2(x_1, x_2, t, 0)]
 \Big( - a \partial_{x_1} \theta_2(x_1, x_2, t, 0) + \tau \vartheta_2(x_1, x_2, t, 0)\Big) \Big\} \nonumber\\
\Big\{ \sin[\theta_1(x_1, x_2, t , 0)] + \tau \vartheta_1(x_1, x_2, t , 0) \cos[\theta_1(x_1, x_2, t , 0)] \Big\} 
\ket{x_1, x_2}\bra{x_1, x_2} + \mathcal{O}(\tau^2).
 \end{align}
 \subsubsection{\bf Second row third column term} 
  
\begin{align}
 \bigg( \big[S^1_+ \otimes S^2_+ \big] \cdot C_2(t, \tau) \cdot \big[S^1_- \otimes S^2_- \big] \cdot C_1(t, \tau)\bigg)_{i=2,j=3} \nonumber\\
 =  - i \sum_{x_1, x_2} \cos[\theta_2(x_1, x_2-a, t, \tau)] \sin[\theta_1(x_1, x_2, t, \tau)] \ket{x_1 + a, x_2 -a}\bra{x_1, x_2} 
 +   \sin[\theta_2(x_1-a, x_2, t, \tau)] \cos[\theta_1(x_1, x_2, t, \tau) ]\ket{x_1, x_2}\bra{x_1, x_2} \nonumber\\
 = - i \sum_{x_1, x_2} \cos[\theta_2(x_1-a, x_2, t, \tau)] \sin[\theta_1(x_1-a, x_2+a, t, \tau)] \Big]
 \ket{x_1, x_2}\bra{x_1, x_2} e^{-\frac{i (\hat{p}_1 \otimes \mathbb{I}_2 -  \mathbb{I}_1 \otimes \hat{p}_2)a}{\hbar}} \nonumber\\
  +   \sin[\theta_2(x_1-a, x_2, t, \tau)] \cos[\theta_1(x_1, x_2, t, \tau) ]\ket{x_1, x_2}\bra{x_1, x_2}
\end{align}

\begin{align}
  =  - \frac{a}{\hbar} \sum_{x_1, x_2} \cos[\theta_2(x_1, x_2, t, 0)] \sin[\theta_1(x_1, x_2, t, 0)] \Big]
 \ket{x_1, x_2}\bra{x_1, x_2}  (\hat{p}_1 \otimes \mathbb{I}_2 -  \mathbb{I}_1 \otimes \hat{p}_2) \nonumber\\
 - i \sum_{x_1, x_2} \Big\{ \cos[\theta_2(x_1, x_2, t, 0)] - \sin[\theta_2(x_1, x_2, t, 0)]
  \Big( - a \partial_{x_1}\theta_2(x_1, x_2, t, 0) + \tau \vartheta_2(x_1, x_2, t, 0) \Big) \Big\} \nonumber\\
  \Big\{ \sin[\theta_1(x_1, x_2, t, 0)] + \cos[\theta_1(x_1, x_2, t, 0)] 
  \Big( - a \partial_{x_1}\theta_1(x_1, x_2, t, 0) + a \partial_{x_2}\theta_1(x_1, x_2, t,0)
  + \tau \vartheta_1(x_1, x_2, t, 0) \Big) \Big\} \ket{x_1, x_2}\bra{x_1, x_2} \nonumber\\
  +   \Big\{ \sin[\theta_2(x_1, x_2, t, 0)] + \cos[\theta_2(x_1, x_2, t, 0)]
  \Big( - a \partial_{x_1}\theta_2(x_1, x_2, t, 0) + \tau \vartheta_2(x_1, x_2, t, 0) \Big) \Big\} \nonumber\\
  \Big\{ \cos[\theta_1(x_1, x_2, t, 0) ] 
  - \tau \vartheta_1(x_1, x_2, t, 0) \sin[\theta_1(x_1, x_2, t, 0)]\Big\} \ket{x_1, x_2}\bra{x_1, x_2} + \mathcal{O}(\tau^2).
\end{align}

\subsubsection{\bf Third row second column term}

\begin{align}
 \bigg( \big[S^1_+ \otimes S^2_+ \big] \cdot C_2(t, \tau) \cdot \big[S^1_- \otimes S^2_- \big] \cdot C_1(t, \tau)\bigg)_{i=3,j=2} \nonumber\\
 = - i \sum_{x_1, x_2} \sin[\theta_2(x_1, x_2-a, t, \tau)] \cos[\theta_1(x_1, x_2, t, \tau)]\ket{x_1, x_2}\bra{x_1, x_2} 
  + \cos[\theta_2(x_1-a, x_2, t, \tau)] \sin[\theta_1(x_1, x_2, t, \tau)] \ket{x_1-a, x_2 + a}\bra{x_1, x_2}\nonumber\\
 =  - i \sum_{x_1, x_2} \sin[\theta_2(x_1, x_2-a, t, \tau)] \cos[\theta_1(x_1, x_2, t, \tau)]\ket{x_1, x_2}\bra{x_1, x_2} \nonumber\\
  + \cos[\theta_2(x_1, x_2-a, t, \tau)] \sin[\theta_1(x_1+a, x_2-a, t, \tau)] 
 \ket{x_1, x_2}\bra{x_1, x_2}e^{\frac{i (\hat{p}_1 \otimes \mathbb{I}_2 -  \mathbb{I}_1 \otimes \hat{p}_2)a}{\hbar}}
 \end{align}
 \begin{align}
 =  - i \sum_{x_1, x_2} \Big\{ \sin[\theta_2(x_1, x_2, t, 0)] + \cos[\theta_2(x_1, x_2, t, 0)]
 \Big( - a \partial_{x_2} \theta_2(x_1, x_2, t, 0) + \tau \vartheta_2(x_1, x_2, t, 0) \Big) \Big\} \nonumber\\
 \Big\{ \cos[\theta_1(x_1, x_2, t, 0)] - \tau \vartheta_1(x_1, x_2, t, 0) \sin[\theta_1(x_1, x_2, t, 0)] \Big\}  \ket{x_1, x_2}\bra{x_1, x_2} \nonumber\\
  + \Big\{ \cos[\theta_2(x_1, x_2, t, 0)] - \sin[\theta_2(x_1, x_2, t, 0)]\Big( - a \partial_{x_2} \theta_2(x_1, x_2, t, 0)
  + \tau \vartheta_2(x_1, x_2, t, 0)\Big) \Big\} \nonumber\\
  \Big\{ \sin[\theta_1(x_1, x_2, t, 0)] + \cos[\theta_1(x_1, x_2, t, 0)] 
  \Big( a \partial_{x_1}\theta_1(x_1, x_2, t, 0) - a \partial_{x_2}\theta_1(x_1+a, x_2, t, 0) 
  + \tau \vartheta_1(x_1, x_2, t, 0) \Big) \Big\} \ket{x_1, x_2}\bra{x_1, x_2} \nonumber\\
 + \frac{a}{\hbar}\sum_{x_1, x_2}\cos[\theta_2(x_1, x_2, t, 0)] \sin[\theta_1(x_1, x_2, t, 0)] 
 \ket{x_1, x_2}\bra{x_1, x_2}(\hat{p}_1 \otimes \mathbb{I}_2 -  \mathbb{I}_1 \otimes \hat{p}_2) + \mathcal{O}(\tau^2).
\end{align}
\subsubsection{\bf Third row third column element}  
  \begin{align}
 \bigg( \big[S^1_+ \otimes S^2_+ \big] \cdot C_2(t, \tau) \cdot \big[S^1_- \otimes S^2_- \big] \cdot C_1(t, \tau)\bigg)_{i=3,j=3} \nonumber\\
 = - \sum_{x_1, x_2}  \sin[\theta_2(x_1, x_2-a, t, \tau)] \sin[\theta_1(x_1, x_2, t, \tau)] \ket{x_1, x_2}\bra{x_1, x_2} 
 -  \cos[\theta_2(x_1-a, x_2, t, \tau)] \cos[\theta_1(x_1, x_2, t, \tau)]  \ket{x_1-a, x_2 + a}\bra{x_1, x_2}\nonumber\\
  = - \sum_{x_1, x_2}  \sin[\theta_2(x_1, x_2-a, t, \tau)] \sin[\theta_1(x_1, x_2, t, \tau)] \ket{x_1, x_2}\bra{x_1, x_2}  \nonumber\\
 +  \sum_{x_1, x_2}  \cos[\theta_2(x_1, x_2-a, t, \tau)] \cos[\theta_1(x_1+a, x_2-a, t, \tau)]  
  \ket{x_1, x_2}\bra{x_1, x_2} e^{\frac{i (\hat{p}_1 \otimes \mathbb{I}_2 -  \mathbb{I}_1 \otimes \hat{p}_2)a}{\hbar}}
 \end{align}
 \begin{align}
  = - \sum_{x_1, x_2}  \Big\{ \sin[\theta_2(x_1, x_2, t, 0)] + \cos[\theta_2(x_1, x_2, t, 0)] 
  \Big( - a \partial_{x_2}\theta_2(x_1, x_2, t, 0) + \tau \vartheta_2(x_1, x_2, t, 0)\Big)\Big\} \nonumber\\
   \Big\{ \sin[\theta_1(x_1, x_2, t, 0)]  + \tau \vartheta_1(x_1, x_2, t, 0) \cos[\theta_1(x_1, x_2, t, 0)] \Big\} \ket{x_1, x_2}\bra{x_1, x_2}  \nonumber\\
 +  \sum_{x_1, x_2}  \Big\{ \cos[\theta_2(x_1, x_2, t, 0)] -  \sin[\theta_2(x_1, x_2, t, 0)] 
 \Big( - a \partial_{x_2} \theta_2(x_1, x_2, t, 0) + \tau \vartheta_2(x_1, x_2, t, 0) \Big)\Big\} \nonumber\\
 \Big\{ \cos[\theta_1(x_1, x_2, t, 0)] - \sin[\theta_1(x_1, x_2, t, 0)]
 \Big( a \partial_{x_1}\theta_1(x_1, x_2, t, 0) - a \partial_{x_2}\theta_1(x_1, x_2, t, 0) + \tau \vartheta_1(x_1, x_2, t, 0) \Big)\Big\}  
  \ket{x_1, x_2}\bra{x_1, x_2}  \nonumber\\
  + \frac{i a}{\hbar} \sum_{x_1, x_2}  \cos[\theta_2(x_1, x_2, t, 0)] \cos[\theta_1(x_1, x_2, t,0)]  
  \ket{x_1, x_2}\bra{x_1, x_2} (\hat{p}_1 \otimes \mathbb{I}_2 -  \mathbb{I}_1 \otimes \hat{p}_2) + \mathcal{O}(\tau^2) .
 \end{align}

 \subsubsection{\bf Fourth row first column term}

   \begin{align}
 \bigg( \big[S^1_+ \otimes S^2_+ \big] \cdot C_2(t, \tau) \cdot \big[S^1_- \otimes S^2_- \big] \cdot C_1(t, \tau)\bigg)_{i=4,j=1} \nonumber\\
 = - i \sum_{x_1, x_2} \sin[\theta_2(x_1, x_2, t, \tau)] \cos[\theta_1(x_1, x_2, t, \tau)] \ket{x_1, x_2}\bra{x_1, x_2}
 +  \cos[\theta_2(x_1-a, x_2-a, t, \tau)] \sin[\theta_1(x_1, x_2, t, \tau)] \ket{x_1-a, x_2-a}\bra{x_1, x_2} \nonumber\\
 = - i \sum_{x_1, x_2} \sin[\theta_2(x_1, x_2, t, \tau)] \cos[\theta_1(x_1, x_2, t, \tau)] \ket{x_1, x_2}\bra{x_1, x_2} \nonumber\\
 +  \cos[\theta_2(x_1, x_2, t, \tau)] \sin[\theta_1(x_1 + a, x_2 + a, t, \tau)] \ket{x_1, x_2}\bra{x_1, x_2}
 e^{\frac{i (\hat{p}_1 \otimes \mathbb{I}_2 +  \mathbb{I}_1 \otimes \hat{p}_2)a}{\hbar}}
 \end{align}
  \begin{align}
  = - i \sum_{x_1, x_2} \Big\{  \sin[\theta_2(x_1, x_2, t, 0)] + \tau \vartheta_2(x_1, x_2, t, 0) \cos[\theta_2(x_1, x_2, t, 0)] \Big\} \nonumber\\
 \Big\{ \cos[\theta_1(x_1, x_2, t, 0)] - \tau \vartheta_1(x_1, x_2, t, 0) \sin[\theta_1(x_1, x_2, t, 0)]  \Big\} \ket{x_1, x_2}\bra{x_1, x_2} \nonumber\\
 +  \Big\{ \cos[\theta_2(x_1, x_2, t, 0)] - \tau \vartheta_2(x_1, x_2, t, 0) \sin[\theta_2(x_1, x_2, t, 0)] \Big\} \nonumber\\
 \Big\{ \sin[\theta_1(x_1, x_2, t, 0)]  + \cos[\theta_1(x_1, x_2, t, 0)] 
 \Big(a \partial_{x_1}\theta_1(x_1, x_2, t, 0) + a \partial_{x_2} \theta_1(x_1, x_2, t, 0)
 + \tau \vartheta_1(x_1, x_2, t, 0) \Big)\Big\} \ket{x_1, x_2}\bra{x_1, x_2} \nonumber\\
 + \frac{a}{\hbar} \sum_{x_1, x_2} \cos[\theta_2(x_1, x_2, t, 0)] \sin[\theta_1(x_1, x_2, t, 0)] \ket{x_1, x_2}\bra{x_1, x_2}
  (\hat{p}_1 \otimes \mathbb{I}_2 +  \mathbb{I}_1 \otimes \hat{p}_2) + \mathcal{O}(\tau^2).
 \end{align}

\subsubsection{\bf Fourth row fourth column term}

   \begin{align}
 \bigg( \big[S^1_+ \otimes S^2_+ \big] \cdot C_2(t, \tau) \cdot \big[S^1_- \otimes S^2_- \big] \cdot C_1(t, \tau)\bigg)_{i=4,j=4} \nonumber\\
 =  - \sum_{x_1, x_2} \sin[\theta_2(x_1, x_2, t, \tau)]  \sin[\theta_1(x_1, x_2, t, \tau)] \ket{x_1, x_2}\bra{x_1, x_2} 
 - \cos[\theta_2(x_1-a, x_2-a, t, \tau)]  \cos[\theta_1(x_1, x_2, t, \tau)] \ket{x_1-a, x_2-a}\bra{x_1, x_2} \nonumber\\
 =  - \sum_{x_1, x_2} \sin[\theta_2(x_1, x_2, t, \tau)]  \sin[\theta_1(x_1, x_2, t, \tau)] \ket{x_1, x_2}\bra{x_1, x_2}  \nonumber\\
 - \cos[\theta_2(x_1, x_2, t, \tau)]  \cos[\theta_1(x_1+a, x_2+a, t, \tau)]
 \ket{x_1, x_2}\bra{x_1, x_2} e^{\frac{i (\hat{p}_1 \otimes \mathbb{I}_2 +  \mathbb{I}_1 \otimes \hat{p}_2)a}{\hbar}}
 \end{align}
    \begin{align}
  =  - \sum_{x_1, x_2} \Big\{ \sin[\theta_2(x_1, x_2, t, 0)] + \tau \vartheta_2(x_1, x_2, t, 0) \cos[\theta_2(x_1, x_2, t, 0)] \Big\} \nonumber\\
  \Big\{ \sin[\theta_1(x_1, x_2, t, 0)] + \tau \vartheta_1(x_1, x_2, t, 0) \cos[\theta_1(x_1, x_2, t, 0)] \Big\} \ket{x_1, x_2}\bra{x_1, x_2}  \nonumber\\
 -  \Big\{ \cos[\theta_2(x_1, x_2, t, 0)] - \tau \vartheta_2(x_1, x_2, t, 0) \sin[\theta_2(x_1, x_2, t, 0)] \Big\} \nonumber\\ 
 \Big\{ \cos[\theta_1(x_1, x_2, t, 0)] - \sin[\theta_1(x_1, x_2, t, 0)]\Big( a \partial_{x_1}\theta_1(x_1, x_2, t, 0)
 + a \partial_{x_2}\theta_1(x_1, x_2, t, 0) + \tau\vartheta_1(x_1, x_2, t, 0)  \Big) \Big\}
 \ket{x_1, x_2}\bra{x_1, x_2}  \nonumber\\
 + \frac{i a}{\hbar} \sum_{x_1, x_2} \cos[\theta_2(x_1, x_2, t, 0)]  \cos[\theta_1(x_1, x_2, t, 0)]
 \ket{x_1, x_2}\bra{x_1, x_2} (\hat{p}_1 \otimes \mathbb{I}_2 +  \mathbb{I}_1 \otimes \hat{p}_2) + \mathcal{O}(\tau^2).
 \end{align}
 
 \subsubsection*{}

 All the other matrix elements of the operator $\big[S^1_+ \otimes S^2_+ \big] \cdot C_2(t, \tau) \cdot \big[S^1_- \otimes S^2_- \big] \cdot C_1(t, \tau)$ are zeros.

\subsection{The matrix elements of the operator $[ U^\text{two}(t, 0)]^\dagger = C_1^\dagger(t, 0) \cdot C^\dagger_2(t, 0)$ in coin basis}
Here we have considered that the shift operators become the identity operator when $\tau$ goes to zero.

\begin{align}
 \bigg( C_1^\dagger(t, 0) \cdot C^\dagger_2(t, 0)\bigg)
 = \sum_{x_1, x_2} \Big\{ \cos[\theta_2(x_1, x_2, t, 0)] \sigma_0 \otimes \sigma_0
 + i \sin[\theta_2(x_1, x_2, t, 0)] \sigma_1 \otimes \sigma_1 \Big\}  \nonumber\\
 \Big\{ \cos[\theta_1(x_1, x_2, t, 0)] \sigma_0 \otimes \sigma_0
 + i \sin[\theta_1(x_1, x_2, t, 0)] \sigma_1 \otimes \sigma_1 \Big\} \ket{x_1, x_2}\bra{x_1, x_2} \nonumber\\
 =  \sum_{x_1, x_2} \Big\{ \cos[\theta_2(x_1, x_2, t, 0) +\theta_1(x_1, x_2, t, 0) ] \sigma_0 \otimes \sigma_0
 + i \sin[\theta_1(x_1, x_2, t, 0) + \theta_2(x_1, x_2, t, 0)] \sigma_1 \otimes \sigma_1 \Big\} \ket{x_1, x_2}\bra{x_1, x_2}.
\end{align}

 \subsection{The matrix elements of the operator $\mathscr{U}^\text{two}(t, \tau)$ in the coin-basis}
 
From this section, for notational convenience, we will denote
$\theta_j(x_1, x_2, t, 0)$ by $\theta_j$ and 
$\vartheta_j(x_1, x_2, t)$ by $\vartheta_j$ for all $j = 1, 2$.

\subsubsection{\bf First row first column element in coin-basis}

\begin{align}
 \mathscr{U}^\text{two}_{00}(t, \tau) = \bigg(C_1^\dagger(t, 0) \cdot C^\dagger_2(t, 0)\bigg)_{i=1,j=1} \cdot
 \bigg( \big[S^1_+ \otimes S^2_+ \big] \cdot C_2(t, \tau) \cdot \big[S^1_- \otimes S^2_- \big] \cdot C_1(t, \tau)\bigg)_{i=1,j=1} \nonumber\\
 + \bigg(C_1^\dagger(t, 0) \cdot C^\dagger_2(t, 0)\bigg)_{i=1,j=4} \cdot
 \bigg( \big[S^1_+ \otimes S^2_+ \big] \cdot C_2(t, \tau) \cdot \big[S^1_- \otimes S^2_- \big] \cdot C_1(t, \tau)\bigg)_{i=4,j=1}
 \end{align} \begin{align}
 = \sum_{x_1, x_2} \cos[\theta_2 +\theta_1 ] \ket{x_1, x_2}\bra{x_1, x_2} \nonumber\\
 \Bigg[- \frac{i a}{\hbar} \sum_{x_1, x_2}  \cos[\theta_2] \cos[\theta_1] 
  \ket{x_1, x_2}\bra{x_1, x_2}  (\hat{p}_1 \otimes \mathbb{I}_2 +  \mathbb{I}_1 \otimes \hat{p}_2) 
  +  \sum_{x_1, x_2} \Big\{ \cos[\theta_2] - \sin[\theta_2] \Big(- a \partial_{x_1} \theta_2
  - a \partial_{x_2} \theta_2 + \tau \vartheta_2\Big) \Big\} \nonumber\\
  \Big\{ \cos[\theta_1] - \sin[\theta_1] \Big(-a \partial_{x_1} \theta_1
  - a \partial_{x_2} \theta_1 + \tau \vartheta_1\Big) \Big\}
  \ket{x_1, x_2}\bra{x_1, x_2} 
  - \Big\{ \sin[\theta_2] + \cos[\theta_2] \Big( - a \partial_{x_1} \theta_2
  - a \partial_{x_2} \theta_2 + \tau \vartheta_2\Big) \Big\}  \nonumber\\
  \Big\{ \sin[\theta_1] 
  + \tau \vartheta_1\cos[\theta_1]\Big\}\ket{x_1, x_2}\bra{x_1, x_2} \Bigg] 
  +  i\sum_{x_1, x_2} \sin[\theta_1 + \theta_2] \ket{x_1, x_2}\bra{x_1, x_2} 
  \Bigg[  - i \sum_{x_1, x_2} \Big\{  \sin[\theta_2] + \tau \vartheta_2 \cos[\theta_2] \Big\} \nonumber\\
 \Big\{ \cos[\theta_1] - \tau \vartheta_1 \sin[\theta_1]  \Big\} \ket{x_1, x_2}\bra{x_1, x_2} 
 +  \Big\{ \cos[\theta_2] - \tau \vartheta_2 \sin[\theta_2] \Big\}
 \Big\{ \sin[\theta_1]  + \cos[\theta_1] 
 \Big(a \partial_{x_1}\theta_1 + a \partial_{x_2} \theta_1
 + \tau \vartheta_1 \Big)\Big\} \ket{x_1, x_2}\bra{x_1, x_2} \nonumber\\
 + \frac{a}{\hbar} \sum_{x_1, x_2} \cos[\theta_2] \sin[\theta_1] \ket{x_1, x_2}\bra{x_1, x_2}
  (\hat{p}_1 \otimes \mathbb{I}_2 +  \mathbb{I}_1 \otimes \hat{p}_2) \Bigg] + \mathcal{O}(\tau^2)
\end{align}
 
 $\Longrightarrow$
 
\begin{align}
 \mathscr{U}^\text{two}_{00}(t, \tau) - \sum_{x_1, x_2} \ket{x_1, x_2}\bra{x_1, x_2} 
 = - \frac{i a}{\hbar} \sum_{x_1, x_2} \cos[\theta_2] \cos[2 \theta_1 + \theta_2]
 \ket{x_1, x_2}\bra{x_1, x_2}  (\hat{p}_1 \otimes \mathbb{I}_2 +  \mathbb{I}_1 \otimes \hat{p}_2) \nonumber\\
 + \frac{a}{2} \sum_{x_1, x_2} \sin[2\theta_1 + 2\theta_2]~
 \Big[\partial_{x_1} \theta_2 + \partial_{x_2} \theta_2 \Big]\ket{x_1, x_2}\bra{x_1, x_2} 
 + a \sum_{x_1, x_2} \cos[\theta_2] \sin[2\theta_1 + \theta_2]~
 \Big[\partial_{x_1} \theta_1 + \partial_{x_2} \theta_1 \Big]\ket{x_1, x_2}\bra{x_1, x_2} + \mathcal{O}(\tau^2).
\end{align}

\subsubsection{\bf First row fourth column element in coin-basis}

\begin{align}
 \mathscr{U}^\text{two}_{03}(t, \tau) = \bigg(C_1^\dagger(t, 0) \cdot C^\dagger_2(t, 0)\bigg)_{i=1,j=1} \cdot
 \bigg( \big[S^1_+ \otimes S^2_+ \big] \cdot C_2(t, \tau) \cdot \big[S^1_- \otimes S^2_- \big] \cdot C_1(t, \tau)\bigg)_{i=1,j=4} \nonumber\\
 + \bigg(C_1^\dagger(t, 0) \cdot C^\dagger_2(t, 0)\bigg)_{i=1,j=4} \cdot
 \bigg( \big[S^1_+ \otimes S^2_+ \big] \cdot C_2(t, \tau) \cdot \big[S^1_- \otimes S^2_- \big] \cdot C_1(t, \tau)\bigg)_{i=4,j=4} 
 \end{align} \begin{align}
 = \sum_{x_1, x_2}\cos[\theta_2 +\theta_1 ] \ket{x_1, x_2} \bra{x_1, x_2} 
 \Bigg[- \frac{a}{\hbar} \sum_{x_1, x_2} \cos[\theta_2] \sin[\theta_1] \ket{x_1, x_2}\bra{x_1, x_2}
   (\hat{p}_1 \otimes \mathbb{I}_2 +  \mathbb{I}_1 \otimes \hat{p}_2) \nonumber\\
   - i \sum_{x_1, x_2}  \Big\{ \cos[\theta_2] - \sin[\theta_2]
   \Big( - a \partial_{x_1}\theta_2 
   - a \partial_{x_2}\theta_2 + \tau \vartheta_2\Big)\Big\} 
  \Big\{ \sin[\theta_1] +\cos[\theta_1]
   \Big( - a \partial_{x_1}\theta_1 
   - a \partial_{x_2}\theta_1 + \tau \vartheta_1\Big) \Big\}\ket{x_1, x_2}\bra{x_1, x_2} \nonumber\\
   + \Big\{ \sin[\theta_2] + \cos[\theta_2]
   \Big( - a \partial_{x_1}\theta_2 
   - a \partial_{x_2}\theta_2 + \tau \vartheta_2\Big)\Big\} 
   \Big\{  \cos[\theta_1] - \tau \vartheta_1 \sin[\theta_1] \Big\}\ket{x_1, x_2}\bra{x_1, x_2} \Bigg] 
   + i \sum_{x_1, x_2} \sin[\theta_2 +\theta_1 ] \ket{x_1, x_2} \bra{x_1, x_2} \nonumber\\
   \Bigg[- \sum_{x_1, x_2} \Big\{ \sin[\theta_2] + \tau \vartheta_2 \cos[\theta_2] \Big\}
  \Big\{ \sin[\theta_1] + \tau \vartheta_1 \cos[\theta_1] \Big\} \ket{x_1, x_2}\bra{x_1, x_2} 
 -  \Big\{ \cos[\theta_2] - \tau \vartheta_2 \sin[\theta_2] \Big\} \nonumber\\ 
 \Big\{ \cos[\theta_1] - \sin[\theta_1]\Big( a \partial_{x_1}\theta_1
 + a \partial_{x_2}\theta_1 + \tau\vartheta_1  \Big) \Big\}
 \ket{x_1, x_2}\bra{x_1, x_2} 
 + \frac{i a}{\hbar} \sum_{x_1, x_2} \cos[\theta_2]  \cos[\theta_1]
 \ket{x_1, x_2}\bra{x_1, x_2} (\hat{p}_1 \otimes \mathbb{I}_2 +  \mathbb{I}_1 \otimes \hat{p}_2)  \Bigg] + \mathcal{O}(\tau^2)
 \end{align}

$\Longrightarrow$

\begin{align}
 \mathscr{U}^\text{two}_{03}(t, \tau) = - \frac{a}{ \hbar} \sum_{x_1, x_2}  \cos[\theta_2]
 \sin[2 \theta_1 + \theta_2] \ket{x_1, x_2}\bra{x_1, x_2}
 (\hat{p}_1 \otimes \mathbb{I}_2 +  \mathbb{I}_1 \otimes \hat{p}_2) 
 + a i \sum_{x_1, x_2} \cos^2[\theta_1 + \theta_2]  
 \Big[\partial_{x_1} \theta_2 + \partial_{x_2} \theta_2 \Big]\ket{x_1, x_2}\bra{x_1, x_2}\nonumber\\
  + a i \sum_{x_1, x_2} \cos[ \theta_2] \cos[ 2\theta_1 +\theta_2 ]  
 \Big[\partial_{x_1} \theta_1 + \partial_{x_2} \theta_1 \Big]\ket{x_1, x_2}\bra{x_1, x_2}
 - i \tau \sum_{x_1, x_2} \Big[\vartheta_2 + \vartheta_1  \Big] \ket{x_1, x_2}\bra{x_1, x_2} + \mathcal{O}(\tau^2).
\end{align}

\subsubsection{\bf Second row second column element in coin basis}

\begin{align}
 \mathscr{U}^\text{two}_{11}(t, \tau) = \bigg(C_1^\dagger(t, 0) \cdot C^\dagger_2(t, 0)\bigg)_{i=2,j=2} \cdot
 \bigg( \big[S^1_+ \otimes S^2_+ \big] \cdot C_2(t, \tau) \cdot \big[S^1_- \otimes S^2_- \big] \cdot C_1(t, \tau)\bigg)_{i=2,j=2} \nonumber\\
 + \bigg(C_1^\dagger(t, 0) \cdot C^\dagger_2(t, 0)\bigg)_{i=2,j=3} \cdot
 \bigg( \big[S^1_+ \otimes S^2_+ \big] \cdot C_2(t, \tau) \cdot \big[S^1_- \otimes S^2_- \big] \cdot C_1(t, \tau)\bigg)_{i=3,j=2} 
 \end{align}
 
 \begin{align}
  = \sum_{x_1, x_2}\cos[\theta_1 +  \theta_2] \ket{x_1, x_2} \bra{x_1, x_2} 
  \Bigg[-\frac{i a }{\hbar} \sum_{x_1, x_2}  \cos[\theta_2] \cos[\theta_1]
 \ket{x_1, x_2}\bra{x_1, x_2}(\hat{p}_1 \otimes \mathbb{I}_2 -  \mathbb{I}_1 \otimes \hat{p}_2)  \nonumber\\
 + \sum_{x_1, x_2} \Big\{\cos[\theta_2] - \sin[\theta_2] 
 \Big(- a  \partial_{x_1} \theta_2 + \tau \vartheta_2 \Big)  \Big\}
 \Big\{\cos[\theta_1] - \sin[\theta_1] 
 \Big(- a  \partial_{x_1} \theta_1 + a  \partial_{x_2} \theta_1
 + \tau \vartheta_1 \Big)  \Big\}\ket{x_1, x_2}\bra{x_1, x_2} \nonumber\\
 - \Big\{ \sin[\theta_2] + \cos[\theta_2]
 \Big( - a \partial_{x_1} \theta_2 + \tau \vartheta_2\Big) \Big\} 
\Big\{ \sin[\theta_1] + \tau \vartheta_1 \cos[\theta_1] \Big\} 
\ket{x_1, x_2}\bra{x_1, x_2} \Bigg] 
  + i\sum_{x_1, x_2} \sin[\theta_1 +  \theta_2]  \ket{x_1, x_2}\bra{x_1, x_2} \nonumber\\ 
  \Bigg[- i \sum_{x_1, x_2} \Big\{ \sin[\theta_2] + \cos[\theta_2]
 \Big( - a \partial_{x_2} \theta_2 + \tau \vartheta_2 \Big) \Big\} 
 \Big\{ \cos[\theta_1] - \tau \vartheta_1 \sin[\theta_1] \Big\}  \ket{x_1, x_2}\bra{x_1, x_2} 
  + \Big\{ \cos[\theta_2] - \sin[\theta_2]\Big( - a \partial_{x_2} \theta_2
  + \tau \vartheta_2\Big) \Big\} \nonumber\\
  \Big\{ \sin[\theta_1] + \cos[\theta_1] 
  \Big( a \partial_{x_1}\theta_1 - a \partial_{x_2}\theta_1 
  + \tau \vartheta_1 \Big) \Big\} \ket{x_1, x_2}\bra{x_1, x_2} 
 + \frac{a}{\hbar}\sum_{x_1, x_2}\cos[\theta_2] \sin[\theta_1] 
 \ket{x_1, x_2}\bra{x_1, x_2}(\hat{p}_1 \otimes \mathbb{I}_2 -  \mathbb{I}_1 \otimes \hat{p}_2) \Bigg] + \mathcal{O}(\tau^2)
 \end{align}
$\Longrightarrow$
 
\begin{align}
 \mathscr{U}^\text{two}_{11}(t, \tau) - \sum_{x_1, x_2}\ket{x_1, x_2}\bra{x_1, x_2}
 = - \frac{i a}{\hbar}  \sum_{x_1, x_2} \cos[\theta_2] 
 \cos[2\theta_1 + \theta_2 ]
 \ket{x_1, x_2}\bra{x_1, x_2}(\hat{p}_1 \otimes \mathbb{I}_2 -  \mathbb{I}_1 \otimes \hat{p}_2) \nonumber\\
+\frac{a}{2} \sum_{x_1, x_2}\sin[2\theta_1 + 2\theta_2 ]
  \Big[ \partial_{x_1} \theta_2 - \partial_{x_2} \theta_2) \Big]  \ket{x_1, x_2}\bra{x_1, x_2}
  + a \sum_{x_1, x_2}  \cos[\theta_2] \sin[2\theta_1 + \theta_2 ]
  \Big[ \partial_{x_1} \theta_1 - \partial_{x_2} \theta_1 \Big]  \ket{x_1, x_2}\bra{x_1, x_2} + \mathcal{O}(\tau^2).
\end{align}

\subsubsection{\bf Second row third column element in coin basis} 
 
\begin{align}
 \mathscr{U}^\text{two}_{12}(t, \tau) = \bigg(C_1^\dagger(t, 0) \cdot C^\dagger_2(t, 0)\bigg)_{i=2,j=2} \cdot
 \bigg( \big[S^1_+ \otimes S^2_+ \big] \cdot C_2(t, \tau) \cdot \big[S^1_- \otimes S^2_- \big] \cdot C_1(t, \tau)\bigg)_{i=2,j=3} \nonumber\\
 + \bigg(C_1^\dagger(t, 0) \cdot C^\dagger_2(t, 0)\bigg)_{i=2,j=3} \cdot
 \bigg( \big[S^1_+ \otimes S^2_+ \big] \cdot C_2(t, \tau) \cdot \big[S^1_- \otimes S^2_- \big] \cdot C_1(t, \tau)\bigg)_{i=3,j=3} 
 \end{align}

\begin{align}
 = \sum_{x_1, x_2} \cos[\theta_1 +  \theta_2]\ket{x_1, x_2}\bra{x_1, x_2}  
 \Bigg[  - \frac{a}{\hbar} \sum_{x_1, x_2}
 \cos[\theta_2] \sin[\theta_1] \Big]
 \ket{x_1, x_2}\bra{x_1, x_2}  (\hat{p}_1 \otimes \mathbb{I}_2 -  \mathbb{I}_1 \otimes \hat{p}_2) \nonumber\\
 - i \sum_{x_1, x_2} \Big\{ \cos[\theta_2] - \sin[\theta_2]
  \Big( - a \partial_{x_1}\theta_2 + \tau \vartheta_2 \Big) \Big\} 
  \Big\{ \sin[\theta_1] + \cos[\theta_1] 
  \Big( - a \partial_{x_1}\theta_1 + a \partial_{x_2}\theta_1
  + \tau \vartheta_1 \Big) \Big\} \ket{x_1, x_2}\bra{x_1, x_2} \nonumber\\
  +   \Big\{ \sin[\theta_2] + \cos[\theta_2]
  \Big( - a \partial_{x_1}\theta_2 + \tau \vartheta_2 \Big) \Big\} 
  \Big\{ \cos[\theta_1 ] 
  - \tau \vartheta_1 \sin[\theta_1]\Big\} \ket{x_1, x_2}\bra{x_1, x_2}\Bigg] \nonumber\\
 + i\sum_{x_1, x_2} \sin[\theta_1 +  \theta_2] \ket{x_1, x_2}\bra{x_1, x_2} 
 \Bigg[- \sum_{x_1, x_2}  \Big\{ \sin[\theta_2] + \cos[\theta_2] 
  \Big( - a \partial_{x_2}\theta_2 + \tau \vartheta_2\Big)\Big\} 
   \Big\{ \sin[\theta_1]  + \tau \vartheta_1 \cos[\theta_1] \Big\} \ket{x_1, x_2}\bra{x_1, x_2}  \nonumber\\
 +  \sum_{x_1, x_2}  \Big\{ \cos[\theta_2] -  \sin[\theta_2] 
 \Big( - a \partial_{x_2} \theta_2 + \tau \vartheta_2 \Big)\Big\}
 \Big\{ \cos[\theta_1] - \sin[\theta_1]
 \Big( a \partial_{x_1}\theta_1 - a \partial_{x_2}\theta_1 + \tau \vartheta_1 \Big)\Big\}  
  \ket{x_1, x_2}\bra{x_1, x_2}  \nonumber\\
  + \frac{i a}{\hbar} \sum_{x_1, x_2}  \cos[\theta_2] \cos[\theta_1]  
  \ket{x_1, x_2}\bra{x_1, x_2} (\hat{p}_1 \otimes \mathbb{I}_2 -  \mathbb{I}_1 \otimes \hat{p}_2) \Bigg] + \mathcal{O}(\tau^2)
\end{align}

$\Longrightarrow$

\begin{align}
  \mathscr{U}^\text{two}_{12}(t, \tau) = - \frac{a}{\hbar} \sum_{x_1, x_2}
 \cos[\theta_2] \sin[2 \theta_1 + \theta_2]
 \ket{x_1, x_2}\bra{x_1, x_2} (\hat{p}_1 \otimes \mathbb{I}_2 -  \mathbb{I}_1 \otimes \hat{p}_2) \nonumber\\
 + a i \sum_{x_1, x_2} \Big[ \cos^2[\theta_1 + \theta_2] \partial_{x_1}\theta_2
 +\sin^2[\theta_1 + \theta_2] \partial_{x_2}\theta_2 \Big]\ket{x_1, x_2}\bra{x_1, x_2} 
 + a i \sum_{x_1, x_2}  \cos[\theta_2] \cos[2 \theta_1 + \theta_2]
 \Big[\partial_{x_1}\theta_1 - \partial_{x_2}\theta_1 \Big]\ket{x_1, x_2}\bra{x_1, x_2} \nonumber\\
 - i \tau \sum_{x_1, x_2} \Big[ \vartheta_1 + \vartheta_2 \Big] \ket{x_1, x_2}\bra{x_1, x_2} + \mathcal{O}(\tau^2).
\end{align}

 \subsubsection{\bf Third row second column element in coin basis}
 
 \begin{align}
 \mathscr{U}^\text{two}_{21}(t, \tau) = \bigg(C_1^\dagger(t, 0) \cdot C^\dagger_2(t, 0)\bigg)_{i=3,j=2} \cdot
 \bigg( \big[S^1_+ \otimes S^2_+ \big] \cdot C_2(t, \tau) \cdot \big[S^1_- \otimes S^2_- \big] \cdot C_1(t, \tau)\bigg)_{i=2,j=2} \nonumber\\
 + \bigg(C_1^\dagger(t, 0) \cdot C^\dagger_2(t, 0)\bigg)_{i=3,j=3} \cdot
 \bigg( \big[S^1_+ \otimes S^2_+ \big] \cdot C_2(t, \tau) \cdot \big[S^1_- \otimes S^2_- \big] \cdot C_1(t, \tau)\bigg)_{i=3,j=2} 
 \end{align}
 
 \begin{align}
  = i\sum_{x_1, x_2} \sin[\theta_1 + \theta_2]\ket{x_1, x_2}\bra{x_1, x_2}
  \Bigg[-\frac{i a }{\hbar} \sum_{x_1, x_2}  \cos[\theta_2] \cos[\theta_1]
 \ket{x_1, x_2}\bra{x_1, x_2}(\hat{p}_1 \otimes \mathbb{I}_2 -  \mathbb{I}_1 \otimes \hat{p}_2)  \nonumber\\
 + \sum_{x_1, x_2} \Big\{\cos[\theta_2] - \sin[\theta_2] 
 \Big(- a  \partial_{x_1} \theta_2 + \tau \vartheta_2 \Big)  \Big\} 
 \Big\{\cos[\theta_1] - \sin[\theta_1] 
 \Big(- a  \partial_{x_1} \theta_1 + a  \partial_{x_2} \theta_1
 + \tau \vartheta_1 \Big)  \Big\}\ket{x_1, x_2}\bra{x_1, x_2} \nonumber\\
 - \Big\{ \sin[\theta_2] + \cos[\theta_2]
 \Big( - a \partial_{x_1} \theta_2 + \tau \vartheta_2\Big) \Big\} 
\Big\{ \sin[\theta_1] + \tau \vartheta_1 \cos[\theta_1] \Big\} 
\ket{x_1, x_2}\bra{x_1, x_2} \Bigg]
  + \sum_{x_1, x_2}\cos[\theta_1 + \theta_2] \ket{x_1, x_2}\bra{x_1, x_2} \nonumber\\
  \Bigg[- i \sum_{x_1, x_2} \Big\{ \sin[\theta_2] + \cos[\theta_2]
 \Big( - a \partial_{x_2} \theta_2 + \tau \vartheta_2 \Big) \Big\} 
 \Big\{ \cos[\theta_1] - \tau \vartheta_1 \sin[\theta_1] \Big\}  \ket{x_1, x_2}\bra{x_1, x_2} 
  + \Big\{ \cos[\theta_2] - \sin[\theta_2]\Big( - a \partial_{x_2} \theta_2
  + \tau \vartheta_2\Big) \Big\} \nonumber\\
  \Big\{ \sin[\theta_1] + \cos[\theta_1] 
  \Big( a \partial_{x_1}\theta_1 - a \partial_{x_2}\theta_1(x_1+a, x_2, t, 0) 
  + \tau \vartheta_1 \Big) \Big\} \ket{x_1, x_2}\bra{x_1, x_2}  \nonumber\\
 + \frac{a}{\hbar}\sum_{x_1, x_2}\cos[\theta_2] \sin[\theta_1] 
 \ket{x_1, x_2}\bra{x_1, x_2}(\hat{p}_1 \otimes \mathbb{I}_2 -  \mathbb{I}_1 \otimes \hat{p}_2) \Bigg]+ \mathcal{O}(\tau^2)
 \end{align}

$\Longrightarrow$

\begin{align}
\mathscr{U}^\text{two}_{21}(t, \tau) = \frac{a}{\hbar} \sum_{x_1, x_2} \cos[\theta_2]
\sin[2 \theta_1 + \theta_2]
 \ket{x_1, x_2}\bra{x_1, x_2} (\hat{p}_1 \otimes \mathbb{I}_2 -  \mathbb{I}_1 \otimes \hat{p}_2) \nonumber\\
 + a i\sum_{x_1, x_2} \Big[ \sin^2[\theta_1 + \theta_2] \partial_{x_1}\theta_2
 +\cos^2[\theta_1 + \theta_2] \partial_{x_2}\theta_2 \Big]\ket{x_1, x_2}\bra{x_1, x_2} 
 - i a \sum_{x_1, x_2}  \cos[\theta_2] \cos[2 \theta_1 + \theta_2]
 \Big[\partial_{x_1}\theta_1 - \partial_{x_2}\theta_1 \Big]\ket{x_1, x_2}\bra{x_1, x_2} \nonumber\\
  - i \tau \sum_{x_1, x_2} \Big[ \vartheta_1 + \vartheta_2 \Big] \ket{x_1, x_2}\bra{x_1, x_2} + \mathcal{O}(\tau^2).
\end{align}

 \subsubsection{\bf Third row third column element in coin basis}

 \begin{align}
 \mathscr{U}^\text{two}_{22}(t, \tau) = \bigg(C_1^\dagger(t, 0) \cdot C^\dagger_2(t, 0)\bigg)_{i=3,j=2} \cdot
 \bigg( \big[S^1_+ \otimes S^2_+ \big] \cdot C_2(t, \tau) \cdot \big[S^1_- \otimes S^2_- \big] \cdot C_1(t, \tau)\bigg)_{i=2,j=3} \nonumber\\
 + \bigg(C_1^\dagger(t, 0) \cdot C^\dagger_2(t, 0)\bigg)_{i=3,j=3} \cdot
 \bigg( \big[S^1_+ \otimes S^2_+ \big] \cdot C_2(t, \tau) \cdot \big[S^1_- \otimes S^2_- \big] \cdot C_1(t, \tau)\bigg)_{i=3,j=3} 
 \end{align}

 \begin{align}
  = i \sum_{x_1, x_2} \sin[\theta_1 + \theta_2] \ket{x_1, x_2}\bra{x_1, x_2} 
  \Bigg[- \frac{a}{\hbar} \sum_{x_1, x_2} \cos[\theta_2] \sin[\theta_1] \Big]
 \ket{x_1, x_2}\bra{x_1, x_2}  (\hat{p}_1 \otimes \mathbb{I}_2 -  \mathbb{I}_1 \otimes \hat{p}_2) \nonumber\\
 - i \sum_{x_1, x_2} \Big\{ \cos[\theta_2] - \sin[\theta_2]
  \Big( - a \partial_{x_1}\theta_2 + \tau \vartheta_2 \Big) \Big\} 
  \Big\{ \sin[\theta_1] + \cos[\theta_1] 
  \Big( - a \partial_{x_1}\theta_1 + a \partial_{x_2}\theta_1
  + \tau \vartheta_1 \Big) \Big\} \ket{x_1, x_2}\bra{x_1, x_2} \nonumber\\
  +   \Big\{ \sin[\theta_2] + \cos[\theta_2]
  \Big( - a \partial_{x_1}\theta_2 + \tau \vartheta_2 \Big) \Big\}
  \Big\{ \cos[\theta_1 ] 
  - \tau \vartheta_1 \sin[\theta_1]\Big\} \ket{x_1, x_2}\bra{x_1, x_2}  \Bigg] 
  + \sum_{x_1, x_2} \cos[\theta_1 + \theta_2] \ket{x_1, x_2}\bra{x_1, x_2} \nonumber\\
  \Bigg[- \sum_{x_1, x_2}  \Big\{ \sin[\theta_2] + \cos[\theta_2] 
  \Big( - a \partial_{x_2}\theta_2 + \tau \vartheta_2\Big)\Big\} 
   \Big\{ \sin[\theta_1]  + \tau \vartheta_1 \cos[\theta_1] \Big\} \ket{x_1, x_2}\bra{x_1, x_2} 
 +  \sum_{x_1, x_2}  \Big\{ \cos[\theta_2] -  \sin[\theta_2] 
 \Big( - a \partial_{x_2} \theta_2 + \tau \vartheta_2 \Big)\Big\} \nonumber\\
 \Big\{ \cos[\theta_1] - \sin[\theta_1]
 \Big( a \partial_{x_1}\theta_1 - a \partial_{x_2}\theta_1 + \tau \vartheta_1 \Big)\Big\}  
  \ket{x_1, x_2}\bra{x_1, x_2}  
  + \frac{i a}{\hbar} \sum_{x_1, x_2}  \cos[\theta_2] \cos[\theta_1]  
  \ket{x_1, x_2}\bra{x_1, x_2} (\hat{p}_1 \otimes \mathbb{I}_2 -  \mathbb{I}_1 \otimes \hat{p}_2) \Bigg] + \mathcal{O}(\tau^2)
 \end{align}

$\Longrightarrow$

\begin{align}
  \mathscr{U}^\text{two}_{22}(t, \tau) - \sum_{x_1, x_2}  \ket{x_1, x_2}\bra{x_1, x_2}
  = \frac{i a}{\hbar} \sum_{x_1, x_2} \cos[\theta_2] \cos[2\theta_1 + \theta_2]
  \ket{x_1, x_2}\bra{x_1, x_2} (\hat{p}_1 \otimes \mathbb{I}_2 -  \mathbb{I}_1 \otimes \hat{p}_2) \nonumber\\
  + \frac{a}{2} \sum_{x_1, x_2} \sin[2\theta_1 + 2\theta_2] 
  \Big[\partial_{x_2} \theta_2 - \partial_{x_1} \theta_2\Big]\ket{x_1, x_2}\bra{x_1, x_2}
  + a \sum_{x_1, x_2} \cos[\theta_2] \sin[2\theta_1 + \theta_2] 
  \Big[\partial_{x_2} \theta_1 - \partial_{x_1} \theta_1\Big]\ket{x_1, x_2}\bra{x_1, x_2} + \mathcal{O}(\tau^2).
\end{align}

  \subsubsection{\bf Fourth row first column element in coin basis}

 \begin{align}
 \mathscr{U}^\text{two}_{30}(t, \tau) = \bigg(C_1^\dagger(t, 0) \cdot C^\dagger_2(t, 0)\bigg)_{i=4,j=1} \cdot
 \bigg( \big[S^1_+ \otimes S^2_+ \big] \cdot C_2(t, \tau) \cdot \big[S^1_- \otimes S^2_- \big] \cdot C_1(t, \tau)\bigg)_{i=1,j=1} \nonumber\\
 + \bigg(C_1^\dagger(t, 0) \cdot C^\dagger_2(t, 0)\bigg)_{i=4,j=4} \cdot
 \bigg( \big[S^1_+ \otimes S^2_+ \big] \cdot C_2(t, \tau) \cdot \big[S^1_- \otimes S^2_- \big] \cdot C_1(t, \tau)\bigg)_{i=4,j=1} 
 \end{align}

\begin{align}
 =  i \sum_{x_1, x_2} \sin[\theta_1 + \theta_2] \ket{x_1, x_2}\bra{x_1, x_2}
 \Bigg[- \frac{i a}{\hbar} \sum_{x_1, x_2}  \cos[\theta_2] \cos[\theta_1] 
  \ket{x_1, x_2}\bra{x_1, x_2}  (\hat{p}_1 \otimes \mathbb{I}_2 +  \mathbb{I}_1 \otimes \hat{p}_2) \nonumber\\
  +  \sum_{x_1, x_2} \Big\{ \cos[\theta_2] - \sin[\theta_2] \Big(- a \partial_{x_1} \theta_2
  - a \partial_{x_2} \theta_2 + \tau \vartheta_2\Big) \Big\}
  \Big\{ \cos[\theta_1] - \sin[\theta_1] \Big(-a \partial_{x_1} \theta_1
  - a \partial_{x_2} \theta_1 + \tau \vartheta_1\Big) \Big\}
  \ket{x_1, x_2}\bra{x_1, x_2} \nonumber\\
  - \Big\{ \sin[\theta_2] + \cos[\theta_2] \Big( - a \partial_{x_1} \theta_2
  - a \partial_{x_2} \theta_2 + \tau \vartheta_2\Big) \Big\}  
  \Big\{ \sin[\theta_1] 
  + \tau \vartheta_1\cos[\theta_1]\Big\}\ket{x_1, x_2}\bra{x_1, x_2} \Bigg]\nonumber\\
 +  \sum_{x_1, x_2} \cos[\theta_1 + \theta_2] \ket{x_1, x_2}\bra{x_1, x_2}
\Bigg[ - i \sum_{x_1, x_2} \Big\{  \sin[\theta_2] + \tau \vartheta_2 \cos[\theta_2] \Big\} 
 \Big\{ \cos[\theta_1] - \tau \vartheta_1 \sin[\theta_1]  \Big\} \ket{x_1, x_2}\bra{x_1, x_2} \nonumber\\
 +  \Big\{ \cos[\theta_2] - \tau \vartheta_2 \sin[\theta_2] \Big\}
 \Big\{ \sin[\theta_1]  + \cos[\theta_1] 
 \Big(a \partial_{x_1}\theta_1 + a \partial_{x_2} \theta_1
 + \tau \vartheta_1 \Big)\Big\} \ket{x_1, x_2}\bra{x_1, x_2} \nonumber\\
 + \frac{a}{\hbar} \sum_{x_1, x_2} \cos[\theta_2] \sin[\theta_1] \ket{x_1, x_2}\bra{x_1, x_2}
  (\hat{p}_1 \otimes \mathbb{I}_2 +  \mathbb{I}_1 \otimes \hat{p}_2)\Bigg]+ \mathcal{O}(\tau^2)
 \end{align}

 $\Longrightarrow$
 
 \begin{align}
   \mathscr{U}^\text{two}_{30}(t, \tau) 
   = \frac{a}{\hbar}  \sum_{x_1, x_2} \cos[\theta_2] \sin[2\theta_1 + \theta_2]
  \ket{x_1, x_2}\bra{x_1, x_2} (\hat{p}_1 \otimes \mathbb{I}_2 + \mathbb{I}_1 \otimes \hat{p}_2) 
  + i a \sum_{x_1, x_2} \sin^2[\theta_1 + \theta_2] 
  \Big[ \partial_{x_1}  \theta_2 + \partial_{x_2} \theta_2 \Big] \ket{x_1, x_2}\bra{x_1, x_2} \nonumber\\
  -  i a \sum_{x_1, x_2} \cos[\theta_2] \cos[2\theta_1 + \theta_2] 
  \Big[ \partial_{x_1}  \theta_1 + \partial_{x_2} \theta_1 \Big] \ket{x_1, x_2}\bra{x_1, x_2} 
  - i \tau \sum_{x_1, x_2} \Big[ \vartheta_1 + \vartheta_2 \Big] \ket{x_1, x_2}\bra{x_1, x_2} + \mathcal{O}(\tau^2).
 \end{align}

 \subsubsection{\bf Fourth row fourth column element in coin basis}

 \begin{align}
 \mathscr{U}^\text{two}_{33}(t, \tau) = \bigg(C_1^\dagger(t, 0) \cdot C^\dagger_2(t, 0)\bigg)_{i=4,j=1} \cdot
 \bigg( \big[S^1_+ \otimes S^2_+ \big] \cdot C_2(t, \tau) \cdot \big[S^1_- \otimes S^2_- \big] \cdot C_1(t, \tau)\bigg)_{i=1,j=4} \nonumber\\
 + \bigg(C_1^\dagger(t, 0) \cdot C^\dagger_2(t, 0)\bigg)_{i=4,j=4} \cdot
 \bigg( \big[S^1_+ \otimes S^2_+ \big] \cdot C_2(t, \tau) \cdot \big[S^1_- \otimes S^2_- \big] \cdot C_1(t, \tau)\bigg)_{i=4,j=4} 
 \end{align}
 
 \begin{align}
 = i \sum_{x_1, x_2} \sin[\theta_1(x_1, x_2, t,0) +\theta_2(x_1, x_2, t,0) ] \ket{x_1, x_2} \bra{x_1, x_2}
 \Bigg[ - \frac{a}{\hbar} \sum_{x_1, x_2} \cos[\theta_2] \sin[\theta_1] \ket{x_1, x_2}\bra{x_1, x_2}
   (\hat{p}_1 \otimes \mathbb{I}_2 +  \mathbb{I}_1 \otimes \hat{p}_2) \nonumber\\
   - i \sum_{x_1, x_2}  \Big\{ \cos[\theta_2] - \sin[\theta_2]
   \Big( - a \partial_{x_1}\theta_2 
   - a \partial_{x_2}\theta_2 + \tau \vartheta_2\Big)\Big\}  
  \Big\{ \sin[\theta_1] +\cos[\theta_1]
   \Big( - a \partial_{x_1}\theta_1 
   - a \partial_{x_2}\theta_1 + \tau \vartheta_1\Big) \Big\}\ket{x_1, x_2}\bra{x_1, x_2} \nonumber\\
   + \Big\{ \sin[\theta_2] + \cos[\theta_2]
   \Big( - a \partial_{x_1}\theta_2 
   - a \partial_{x_2}\theta_2 + \tau \vartheta_2\Big)\Big\} 
   \Big\{  \cos[\theta_1] - \tau \vartheta_1 \sin[\theta_1] \Big\}\ket{x_1, x_2}\bra{x_1, x_2}\Bigg] \nonumber\\
 + \sum_{x_1, x_2} \cos[\theta_1 +\theta_2 ] \ket{x_1, x_2} \bra{x_1, x_2} 
 \Bigg[ - \sum_{x_1, x_2} \Big\{ \sin[\theta_2] + \tau \vartheta_2 \cos[\theta_2] \Big\} 
  \Big\{ \sin[\theta_1] + \tau \vartheta_1 \cos[\theta_1] \Big\} \ket{x_1, x_2}\bra{x_1, x_2}  \nonumber\\
 -  \Big\{ \cos[\theta_2] - \tau \vartheta_2 \sin[\theta_2] \Big\} 
 \Big\{ \cos[\theta_1] - \sin[\theta_1]\Big( a \partial_{x_1}\theta_1
 + a \partial_{x_2}\theta_1 + \tau\vartheta_1  \Big) \Big\}
 \ket{x_1, x_2}\bra{x_1, x_2}  \nonumber\\
 + \frac{i a}{\hbar} \sum_{x_1, x_2} \cos[\theta_2]  \cos[\theta_1]
 \ket{x_1, x_2}\bra{x_1, x_2} (\hat{p}_1 \otimes \mathbb{I}_2 +  \mathbb{I}_1 \otimes \hat{p}_2) \Bigg] + \mathcal{O}(\tau^2)
 \end{align}

 $\Longrightarrow$

 \begin{align}
  \mathscr{U}^\text{two}_{33}(t, \tau) - \sum_{x_1, x_2}  \ket{x_1, x_2}\bra{x_1, x_2}
  = \frac{i a }{\hbar} \sum_{x_1, x_2} \cos[\theta_2] \cos[2 \theta_1 +\theta_2 ]
  \ket{x_1, x_2}\bra{x_1, x_2} (\hat{p}_1 \otimes \mathbb{I}_2 +  \mathbb{I}_1 \otimes \hat{p}_2) \nonumber\\
  - \frac{a}{2} \sum_{x_1, x_2} \sin[2 \theta_1 + 2 \theta_2]
  \Big[\partial_{x_1}\theta_2  + \partial_{x_2}\theta_2 \Big] \ket{x_1, x_2}\bra{x_1, x_2}
  - a \sum_{x_1, x_2} \cos[\theta_2] \sin[2 \theta_1 +  \theta_2]
  \Big[\partial_{x_1}\theta_1  + \partial_{x_2}\theta_1 \Big] \ket{x_1, x_2}\bra{x_1, x_2}
  + \mathcal{O}(\tau^2).
 \end{align}

 \section{Deriving the Effective Two-particle Hamiltonian}\label{twoparham}

 Using the definition: \begin{align}
                        \mathscr{U}^\text{two}(t, \tau) = (\sigma_0 \otimes \sigma_0) \otimes \sum_{x_1, x_2} \ket{x_1, x_2}\bra{x_1, x_2}
                        - \frac{i \tau}{\hbar} H^\text{two}_\text{eff} + \mathcal{O}(\tau^2)
                        \end{align}
                        we get
      \begin{align}     
      H^\text{two}_\text{eff} = \lim_{\tau \to 0} \frac{i \hbar}{\tau}
                         \Big[ \mathscr{U}^\text{two}(t, \tau) - (\sigma_0 \otimes \sigma_0) \otimes \sum_{x_1, x_2} \ket{x_1, x_2}\bra{x_1, x_2}\Big]
                         \hspace{4.3cm}\nonumber\\ 
                         = \sum_{r_1,r_2 = 0}^3 (\sigma_{r_1} \otimes \sigma_{r_2}) \otimes \sum_{x_1, x_2} \Xi_{r_1 r_2}(x_1, x_2, t) \ket{x_1, x_2} \bra{x_1, x_2}
                          +~ c \sum_{r_1,r_2 = 0}^3 (\sigma_{r_1} \otimes \sigma_{r_2}) \otimes \sum_{x_1, x_2} 
                          \Theta^1_{r_1 r_2}(x_1, x_2, t) \ket{x_1, x_2} \bra{x_1, x_2} \hat{p}_1\otimes \mathbb{I}_2 \nonumber\\
                          + ~ c \sum_{r_1,r_2 = 0}^3 (\sigma_{r_1} \otimes \sigma_{r_2}) \otimes \sum_{x_1, x_2}
                          \Theta^2_{r_1 r_2}(x_1, x_2, t) \ket{x_1, x_2} \bra{x_1, x_2} \mathbb{I}_1 \otimes \hat{p}_2 .                          
                         \end{align}
                         where
                        
  \begin{align}
                               \sum_{x_1, x_2} \Xi_{0 0}(x_1, x_2, t) \ket{x_1, x_2} \bra{x_1, x_2} 
                               = \lim_{\tau \to 0} \frac{i \hbar}{4 \tau}
                         \Big[ \mathscr{U}_{00}^\text{two}(t, \tau) + \mathscr{U}_{11}^\text{two}(t, \tau) + 
                         \mathscr{U}_{22}^\text{two}(t, \tau) + \mathscr{U}_{33}^\text{two}(t, \tau)  -  4\sum_{x_1, x_2} \ket{x_1, x_2}\bra{x_1, x_2} \Big]
                            = 0,
                            \end{align}
      \begin{align}  \sum_{x_1, x_2} \Xi_{0 3}(x_1, x_2, t) \ket{x_1, x_2} \bra{x_1, x_2} + c \Theta^2_{0 3}(x_1, x_2, t) \ket{x_1, x_2} \bra{x_1, x_2} \mathbb{I}_1 \otimes \hat{p}_2
                               = \lim_{\tau \to 0} \frac{i \hbar}{4\tau}
                         \Big[ \mathscr{U}_{00}^\text{two}(t, \tau) - \mathscr{U}_{11}^\text{two}(t, \tau) + 
                         \mathscr{U}_{22}^\text{two}(t, \tau) - \mathscr{U}_{33}^\text{two}(t, \tau)\Big] \nonumber\\
                         = c \sum_{x_1, x_2} \cos[\theta_2] \cos[2 \theta_1 +\theta_2 ]
  \ket{x_1, x_2}\bra{x_1, x_2} \mathbb{I}_1 \otimes \hat{p}_2   
  + \frac{i \hbar c}{2}  \sum_{x_1, x_2} \sin[2 \theta_1 + 2 \theta_2]
  \partial_{x_2}\theta_2  \ket{x_1, x_2}\bra{x_1, x_2} \nonumber\\
  + i \hbar c \sum_{x_1, x_2} \cos[\theta_2] \sin[2 \theta_1 +  \theta_2]
  \partial_{x_2}\theta_1 \ket{x_1, x_2}\bra{x_1, x_2} ,
             \end{align}  
  \begin{align}  \sum_{x_1, x_2} \Xi_{30}(x_1, x_2, t) \ket{x_1, x_2} \bra{x_1, x_2} + c \Theta^1_{30}(x_1, x_2, t) \ket{x_1, x_2} \bra{x_1, x_2} 
   \hat{p}_1 \otimes \mathbb{I}_2
    =  \lim_{\tau \to 0} \frac{i \hbar}{4 \tau} \Big[ \mathscr{U}_{00}^\text{two}(t, \tau) + \mathscr{U}_{11}^\text{two}(t, \tau) - 
                         \mathscr{U}_{22}^\text{two}(t, \tau) - \mathscr{U}_{33}^\text{two}(t, \tau)\Big]\nonumber\\
                         = c \sum_{x_1, x_2} \cos[\theta_2] \cos[2 \theta_1 +\theta_2 ]
  \ket{x_1, x_2}\bra{x_1, x_2} \hat{p}_1 \otimes \mathbb{I}_2  
  + \frac{i \hbar c}{2}  \sum_{x_1, x_2} \sin[2 \theta_1 + 2 \theta_2]
  \partial_{x_1}\theta_2  \ket{x_1, x_2}\bra{x_1, x_2} \nonumber\\
  + i \hbar c \sum_{x_1, x_2} \cos[\theta_2] \sin[2 \theta_1 +  \theta_2]
  \partial_{x_1}\theta_1 \ket{x_1, x_2}\bra{x_1, x_2} ,
   \end{align}
          
       \begin{align}  \sum_{x_1, x_2} \Xi_{33}(x_1, x_2, t) \ket{x_1, x_2} \bra{x_1, x_2} 
    = \lim_{\tau \to 0} \frac{i \hbar}{4\tau}
                         \Big[ \mathscr{U}_{00}^\text{two}(t, \tau) - \mathscr{U}_{11}^\text{two}(t, \tau) - 
                         \mathscr{U}_{22}^\text{two}(t, \tau) + \mathscr{U}_{33}^\text{two}(t, \tau)\Big] = 0,
    \end{align}
             
      \begin{align}
         \sum_{x_1, x_2} \Xi_{11}(x_1, x_2, t) \ket{x_1, x_2} \bra{x_1, x_2} =  \lim_{\tau \to 0} \frac{i \hbar}{4\tau}
                         \Big[ \mathscr{U}_{03}^\text{two}(t, \tau) + \mathscr{U}_{12}^\text{two}(t, \tau) + 
                         \mathscr{U}_{21}^\text{two}(t, \tau) + \mathscr{U}_{30}^\text{two}(t, \tau)\Big] \nonumber\\
                         = - \frac{\hbar c}{2}  \sum_{x_1, x_2} \Big[ \partial_{x_1} \theta_2 
                         + \partial_{x_2} \theta_2 \Big] \ket{x_1, x_2} \bra{x_1, x_2}
                         + \hbar \sum_{x_1, x_2} \Big[\vartheta_1 
                         +  \vartheta_2 \Big] \ket{x_1, x_2} \bra{x_1, x_2},
        \end{align}
   \begin{align}  \sum_{x_1, x_2} \Xi_{12}(x_1, x_2, t) \ket{x_1, x_2} \bra{x_1, x_2} 
     + c \Theta^2_{12}(x_1, x_2, t) \ket{x_1, x_2} \bra{x_1, x_2} \mathbb{I}_1 \otimes \hat{p}_2 \nonumber\\
      =   -\lim_{\tau \to 0}  \frac{ \hbar}{4\tau}
                         \Big[ \mathscr{U}_{03}^\text{two}(t, \tau) - \mathscr{U}_{12}^\text{two}(t, \tau) + 
                         \mathscr{U}_{21}^\text{two}(t, \tau) - \mathscr{U}_{30}^\text{two}(t, \tau)\Big]
                  =   c  \sum_{x_1, x_2} \cos[\theta_2] \sin[2 \theta_1 + \theta_2] 
                  \ket{x_1, x_2}\bra{x_1, x_2} \mathbb{I}_1 \otimes \hat{p}_2   \nonumber\\
                  - \frac{i \hbar c}{2} \sum_{x_1, x_2} \cos[2 \theta_1 + 2\theta_2]
                  \partial_{x_2} \theta_2 \ket{x_1, x_2}\bra{x_1, x_2} 
                  - i \hbar c \sum_{x_1, x_2} \cos[\theta_2] \cos[2 \theta_1 + \theta_2]
                  \partial_{x_2} \theta_1 \ket{x_1, x_2}\bra{x_1, x_2},
     \end{align}

    \begin{align}
     =\sum_{x_1, x_2} \Xi_{21}(x_1, x_2, t) \ket{x_1, x_2} \bra{x_1, x_2} 
     + c \Theta^1_{21}(x_1, x_2, t) \ket{x_1, x_2} \bra{x_1, x_2} \hat{p}_1 \otimes \mathbb{I}_2 \nonumber\\
     = -\lim_{\tau \to 0} \frac{\hbar}{4\tau}
                         \Big[ \mathscr{U}_{03}^\text{two}(t, \tau) + \mathscr{U}_{12}^\text{two}(t, \tau) - 
                         \mathscr{U}_{21}^\text{two}(t, \tau) - \mathscr{U}_{30}^\text{two}(t, \tau)\Big]
                  =   c  \sum_{x_1, x_2} \cos[\theta_2] \sin[2 \theta_1 + \theta_2] 
                  \ket{x_1, x_2}\bra{x_1, x_2}\hat{p}_1 \otimes \mathbb{I}_2  \nonumber\\
                  - \frac{i \hbar c}{2} \sum_{x_1, x_2} \cos[2 \theta_1 + 2\theta_2]
                  \partial_{x_1} \theta_2 \ket{x_1, x_2}\bra{x_1, x_2} 
                  - i \hbar c \sum_{x_1, x_2} \cos[\theta_2] \cos[2 \theta_1 + \theta_2]
                  \partial_{x_1} \theta_1 \ket{x_1, x_2}\bra{x_1, x_2},
    \end{align}

   \begin{align}
      \sum_{x_1, x_2} \Xi_{22}(x_1, x_2, t) \ket{x_1, x_2} \bra{x_1, x_2} 
 = \lim_{\tau \to 0} \frac{i \hbar}{4\tau}
                         \Big[- \mathscr{U}_{03}^\text{two}(t, \tau) + \mathscr{U}_{12}^\text{two}(t, \tau) + 
                         \mathscr{U}_{21}^\text{two}(t, \tau) - \mathscr{U}_{30}^\text{two}(t, \tau)\Big] = 0.
                       \end{align}

 All other $\Xi_{r_1 r_2}(x_1, x_2, t),~ \Theta^1_{r_1 r_2}(x_1, x_2, t),~ \Theta^2_{r_1 r_2}(x_1, x_2, t) $ are zero for all position and time steps 
 $x_1, x_2, t$.

\subsubsection*{} 
 
{\normalsize{\bf Therefore, in this case the effective two-particle Hamiltonian looks like}}
\begin{align}
  H^\text{two}_\text{eff}  = \sum_{x_1, x_2} (\sigma_{0} \otimes \sigma_{3}) \otimes  \Xi_{03}(x_1, x_2, t) \ket{x_1, x_2} \bra{x_1, x_2}
   +  c \Theta^2_{03}(x_1, x_2, t) \ket{x_1, x_2} \bra{x_1, x_2} \mathbb{I}_1 \otimes \hat{p}_2 
  + (\sigma_{3} \otimes \sigma_{0}) \otimes  \Xi_{30}(x_1, x_2, t) \ket{x_1, x_2} \bra{x_1, x_2}\nonumber\\
  + c \Theta^1_{30}(x_1, x_2, t) \ket{x_1, x_2} \bra{x_1, x_2} \hat{p}_1\otimes \mathbb{I}_2 
  + (\sigma_{1} \otimes \sigma_{2}) \otimes  \Xi_{12}(x_1, x_2, t) \ket{x_1, x_2} \bra{x_1, x_2}
   +  c \Theta^2_{12}(x_1, x_2, t) \ket{x_1, x_2} \bra{x_1, x_2} \mathbb{I}_1 \otimes \hat{p}_2 \nonumber\\
  + (\sigma_{2} \otimes \sigma_{1}) \otimes  \Xi_{21}(x_1, x_2, t) \ket{x_1, x_2} \bra{x_1, x_2}
  + c \Theta^1_{21}(x_1, x_2, t) \ket{x_1, x_2} \bra{x_1, x_2} \hat{p}_1\otimes \mathbb{I}_2 
  + (\sigma_{1} \otimes \sigma_{1}) \otimes  \Xi_{11}(x_1, x_2, t) \ket{x_1, x_2} \bra{x_1, x_2}. \nonumber\\
\end{align}

}

\end{document}